\documentclass[10pt,aps,prd,twocolumn,tightenlines,superscriptaddress, letterpaper, amsmath, amssymb, preprintnumbers, floatfix, longbibliography, nofootinbib]{revtex4-2}
\usepackage{graphicx}
\usepackage{amsmath}
\usepackage{physics}
\usepackage[dvipsnames]{xcolor}
\usepackage{wrapfig}
\usepackage{float}
\usepackage{tikz}
\usepackage{array}
\usepackage{qcircuit}
\usepackage{comment}
\usepackage{enumitem}
\usepackage{bm}
\usepackage{tabularx}
\usepackage{multirow}
\usepackage{soul}

\newcolumntype{Y}{>{\centering\arraybackslash}X}

\usepackage[hypertexnames=false]{hyperref}
\hypersetup{
    colorlinks=true,       % false: boxed links; true: coloorange links
    linkcolor=blue,          % color of internal links
    citecolor=blue,        % color of links to bibliography
    filecolor=blue,      % color of file links
    urlcolor=blue           % color of external links
}

\usepackage{orcidlink}

\begin{document}

\begin{figure}
\vskip -1.cm
\leftline{\includegraphics[width=0.15\textwidth]{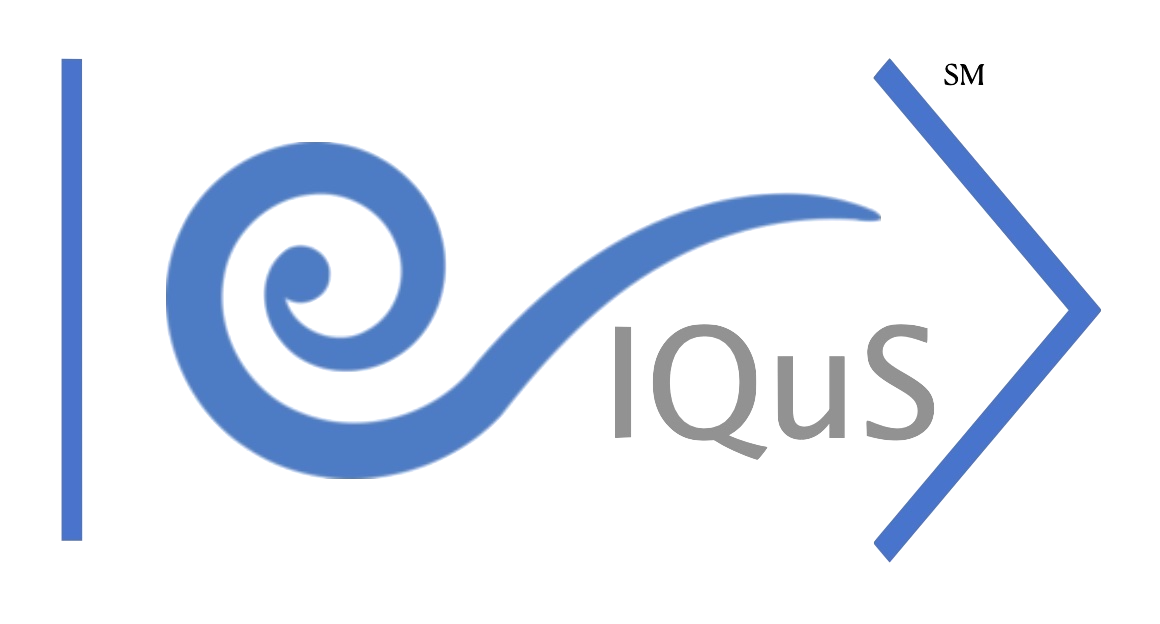}}
\vskip -1cm
\end{figure}

\title{Qutrit and Qubit Circuits for Three-Flavor Collective Neutrino Oscillations}

\author{Francesco Turro\,\orcidlink{0000-0002-1107-2873}}
\email[Corresponding author, ]{fturro@uw.edu}
\affiliation{InQubator for Quantum Simulation (IQuS), Department of Physics, University of Washington, Seattle, WA 98195, USA.}

\author{Ivan A.~Chernyshev\,\orcidlink{0000-0001-8289-1991}}
\email{ivanc3@uw.edu}
\affiliation{InQubator for Quantum Simulation (IQuS), Department of Physics, University of Washington, Seattle, WA 98195, USA.}

\author{Ramya Bhaskar\,\orcidlink{0000-0003-2148-2590}}
\email{rbhaskar@uw.edu}
\affiliation{InQubator for Quantum Simulation (IQuS), Department of Physics, University of Washington, Seattle, WA 98195, USA.}

\author{Marc Illa\,\orcidlink{0000-0003-3570-2849}}
\email{marcilla@uw.edu}
\affiliation{InQubator for Quantum Simulation (IQuS), Department of Physics, University of Washington, Seattle, WA 98195, USA.}

\preprint{IQuS@UW-21-082}
\date{\today}

\begin{abstract}

We explore the utility of qutrits and qubits for simulating the flavor dynamics of dense neutrino systems.
The evolution of such systems impacts some important astrophysical processes, such as core-collapse supernovae and the nucleosynthesis of heavy nuclei. 
Many-body simulations require classical resources beyond current computing capabilities for physically relevant system sizes. 
Quantum computers are therefore a promising candidate to efficiently simulate the many-body dynamics of collective neutrino oscillations.
Previous quantum simulation efforts have primarily focused on properties of the two-flavor approximation due to their direct mapping to qubits.
Here, we present new quantum circuits for simulating three-flavor neutrino systems on qutrit- and qubit-based platforms, and demonstrate their feasibility by simulating systems of two, four and eight neutrinos on IBM and Quantinuum quantum computers.

\end{abstract}

\maketitle

\section{Introduction }
\label{sec:intro}

The use of qudits~\cite{Gottesman:1998se} for simulating nuclear and high-energy physics systems has generated significant interest~\cite{Ciavarella:2021nmj,Gustafson:2021qbt,Calixto_2021,Gustafson:2022xlj,Gonzalez-Cuadra:2022hxt,Gustafson:2022xdt,Gustafson:2023swx,Zache:2023cfj,Illa:2023scc,Popov:2023xft,Meth:2023wzd,Calajo:2024qrc,Carena:2024dzu,Illa:2024kmf,Gustafson:2024kym}, as a result of recent advancements in experimental realizations of qudit-based platforms, including trapped-ion systems~\cite{Low:2019,Ringbauer:2021lhi,Low:2023dlg,Zalivako:2024bjm,Nikolaeva:2024wxl}, superconducting circuits~\cite{Blok:2020may,Seifert:2023ous,Nguyen:2023svc,Champion:2024ufp}, superconducting radio-frequency cavities~\cite{Roy:2024uro}, and photonic systems~\cite{Chi:2022}.
Multilevel quantum devices can efficiently map to high-dimensional systems, which is advantageous for the quantum simulation of such systems, as well as quantum algorithm performance~\cite{Cerf_2002,Gedik:2015,3307650,Baker:2020,Lim:2023hkb} (see Ref.~\cite{10.3389} for a review).
Three-level quantum systems (qutrits)~\cite{Blok:2020may,Yurtalan:2020,Morvan:2021qju,Cervera-Lierta:2021nhp,Ringbauer:2021lhi,Hrmo:2022bvo,Goss:2022bqd,Subramanian:2023xzi,Nikolaeva:2024wxl} are particularly attractive for simulating three-flavor neutrino systems.

In extreme astrophysical environments, neutrinos can reach high enough densities such that their flavor evolution can affect large scale dynamics. Examples include: core collapse supernovae (CCSNe) processes~\cite{Pantaleone:1992xh,Pantaleone:1992eq,Qian:1994wh,Pastor:2002we,Mirizzi:2015eza}, flavor transport in remnants of binary star mergers~\cite{Malkus:2012ts,Malkus:2015mda,Zhu:2016mwa,Frensel:2016fge,Chatelain:2016xva,Wu:2017qpc,Tian:2017xbr,Purcell:2024bim}, and nucleosynthesis~\cite{Fuller:1995ih,Balantekin:2004ug,Duan:2010af,Xiong:2020ntn,Balantekin:2023ayx} (see Refs.~\cite{Duan:2009cd,Duan:2010bg,Chakraborty:2016yeg,Tamborra:2020cul,Capozzi:2022slf,Richers:2022zug,Patwardhan:2022mxg,Volpe:2023met,Balantekin:2023qvm} for reviews on these topics).

At distance $\lesssim 100$ km away from the CCSNe center, self-interacting neutrino-neutrino currents \cite{Samuel:1993uw,Kostelecky:1993yt,Kostelecky:1993dm,Kostelecky:1993ys,Kostelecky:1995xc} predominate the region's dynamics, while at distance $\gtrsim 100$ km, neutrino-vacuum oscillations and the Mikeheyev-Smirnov-Wolfenstein (MSW) effect~\cite{Wolfenstein:1977ue,Wolfenstein:1979ni,Mikheyev:1985zog,Mikheyev:1989dy} become the primary mechanisms driving neutrino flavor evolution. 
The Hamiltonian describing the flavor dynamics of neutrinos propagating through a CCSN
environment therefore contains three terms: the one-body vacuum oscillation term, the one-body background matter interaction term modeled by the MSW effect, and the two-body neutrino-neutrino self-interaction term originating from the exchange of a $Z$ boson \cite{1987ApJ322795F,Notzold:1987ik,Savage:1990by,Pantaleone:1992xh,Pantaleone:1992eq,Sigl:1993ctk,Samuel:1995ri,Hoffman:1996aj,Balantekin:2006tg,Dasgupta:2017oko,Fiorillo:2024wej,Cirigliano:2024pnm} which gives rise to the quantum phenomenon of coherent collective flavor oscillations.

While mean-field studies have evidenced collective flavor dynamics 
\cite{Qian:1994wh,Duan:2006jv,Duan:2006an,Izaguirre:2016gsx,Capozzi:2020kge,Fiorillo:2023mze,Fiorillo:2023hlk,Fiorillo:2024fnl}, there is growing interest in collective dynamics beyond the mean-field approximation, such as when nontrivial neutrino-neutrino two-body correlations are taken into account~\cite{Roggero:2021asb,Xiong:2021evk,Roggero:2022hpy,Bhaskar:2023sta,Kost:2024esc}. 
For even modestly sized systems, the long-range entanglement immediately present in the all-to-all connected two-body operator necessitates exponentially-growing classical resources for simulating neutrino dynamics from the exact many-body Hamiltonian. Regardless of whether one works in position or momentum space, the computational complexity of the problem requires quantum computing resources at smaller volumes than what would have been required in systems with different connectivity structures. Operating in position space however requires from accounting for neutrinos' chirality on the lattice \cite{BORRELLI1990335, KAPLAN1992342, PhysRevLett.132.141603, PhysRevLett.132.141604}, thus we conduct our studies in momentum space.
\footnote{See Refs.~\cite{Shalgar:2023ooi,Johns:2023ewj} for discussions regarding the apparent differences between the mean-field and many-body approaches.}

Recent progress in quantum simulations of two flavor neutrino systems~\cite{Hall:2021rbv,Yeter-Aydeniz:2021olz,Illa:2022jqb,Amitrano:2022yyn,Illa:2022zgu,Siwach:2023wzy} has demonstrated quantum devices' potential~\cite{feynman} to efficiently capture the non-trivial entanglement structure present in many-body neutrino systems. 
Classical and quantum simulations of relatively small-sized two-flavor neutrino systems have uncovered a variety of uniquely quantum phenomena~\cite{Rrapaj:2019pxz,Patwardhan:2019zta,Patwardhan:2021rej,Roggero:2021asb,Xiong:2021evk,Illa:2022zgu,Bhaskar:2023sta,Martin:2023gbo,Neill:2024klc}, further motivating quantum simulations of three-flavor self-interacting neutrino systems~\cite{Siwach:2022xhx,Chernyshev:2024kpu}.

In this work, we introduce qubit and qutrit-based quantum circuits for simulating the time evolution of the three-flavor neutrino system. 
Simulation of the system dynamics for $N=2,4$ and $8$ neutrinos is demonstrated on the IBM \texttt{ibm\_torino}~\cite{ibm} and Quantinuum {\tt H1-1} qubit platforms~\cite{quantinuum} and time evolved observables, such as  single-neutrino flavor probabilities and entanglement entropy, are studied.

\section{Hamiltonian}
\label{sec:hamiltonian}

Within 100 km from the CCSN core, contributions from background matter (MSW effect) can be assumed negligible as similarly approximated in Refs.~\cite{Pehlivan:2011hp,Patwardhan:2019zta,Patwardhan:2021rej,Roggero:2021asb,Cervia:2022pro}), and antineutrinos are not considered in this work solely for reducing computational costs, as the size of the Hilbert space for a given system would increase by a factor of $3^N$, with $N$ being number of antineutrinos in the system.
The Hamiltonian governing the evolution of the neutrino system studied is then 
\begin{equation}
    H= H_{\nu}+H_{\nu\nu}\ ,
    \label{eq:simpleHam}
\end{equation}
where $H_{\nu}$  describes the one-body neutrino Hamiltonian given by the vacuum oscillation and $H_{\nu\nu}$ the  neutrino-neutrino interactions resulting from coherent forward scattering. 

This study follows the lead of Ref.~\cite{Balantekin:2006tg} in approximating neutrino-neutrino as forward scattering, which represents all neutrino-neutrino interactions as exchanges of momentum between neutrinos. Simulation of neutrino-neutrino interactions beyond the forward scattering approximation has been discussed in the literature \cite{PANTALEONE1992128, Pantaleone:1992xh, friedland2003manyparticle, Friedland2003Neutrino, Bell2003Speedup, Sawyer2004Instabilities, Friedland2006ManyBody, Balantekin:2006tg, Pehlivan:2011hp, birol2018neutrino, Cervia2019Entanglement, Roggero:2021asb, roggero2021dynamical, Martin2022Classical, Xiong:2021evk, Roggero:2022hpy, lacroix2022role, Cervia:2022pro, Martin2023ManyBody, Johns:2023ewj, Fiorillo:2024fnl} and implemented in Ref.~\cite{Cirigliano:2024pnm}. However, the forward scattering approximation enables all of the degrees of freedom of neutrinos to be expressed in terms of modes that each correspond to a neutrino momentum in the system’s initial state and for the Hamiltonian to be expressed in terms of these modes. The Hilbert space of each mode is spanned by the possible neutrino flavors and helicities, and the total Hilbert space of the system is a tensor product of the Hilbert spaces of the modes. For ultrarelativistic neutrinos, such as supernova neutrinos, the helicity can be eliminated as a degree of freedom as it is fixed to the chirality of the neutrinos. Thus, a three-flavor supernova neutrino in the forward scattering approximation can be mapped one-to-one to a spin-triplet, and the bilinear operators that the full collective neutrino oscillation operator Hamiltonian is built out of are in turn mapped to Gell-Mann matrices, which are a general-case set of generators for unitary operations on spin-triplets \cite{Balantekin:2006tg}.

The vacuum oscillations can be described in the mass basis as~\cite{Balantekin:2006tg}
\begin{equation}
H_{\nu}=\sum_i^N H^{(i)}_{\nu}=\sum_i^N -\frac{\omega}{2}\lambda^{(i)}_3 + \frac{\omega-2\Omega}{2\sqrt{3}} \lambda^{(i)}_8\,,
\label{eq:single_body_a}
\end{equation}
where the index $i$ sums over $N$ neutrinos in the system, $\lambda^{(i)}_n$ is the $n^{\rm th}$ Gell-Mann matrix acting on the $i^{\rm th}$ neutrino (the Gell-Mann matrices are detailed in App.~\ref{app:Gell_Mann}), and $\omega$ and $\Omega$ are the oscillation frequencies, defined as
\begin{equation}
\omega= \frac{1}{2E} \Delta m_{21}^2 \ , \quad \Omega=\frac{1}{2E} \Delta m_{31}^2=\omega \frac{\Delta m_{31}^2}{\Delta m_{21}^2} \ , 
\end{equation}
with $\Delta m^2_{ij}=m^2_i-m^2_j$, and $m^2_i$ being the squared mass of the $i^{\rm th}$ mass-eigenstate neutrino, with the neutrinos taken to have the same energy.
The $i^{\rm th}$ one-body Hamiltonian can also be written as
\begin{equation}
    H_\nu^{(i)}=\begin{pmatrix}
    0 & 0 & 0 \\
    0 & \omega & 0 \\
0 & 0 & \Omega 
\end{pmatrix} \ .
\label{eq:single_body_matrix}
\end{equation}
Here, identity contributions that correspond to global phases in the real-time evolution operator have been neglected.

To operate in the flavor basis, the Pontecorvo-Maki-Nakagawa-Sakata (PMNS) matrix is used to transform between mass and flavor basis,
\begin{align}
&\qquad\qquad\qquad\qquad\begin{pmatrix}
\nu_e \\ \nu_\mu \\ \nu_\tau
\end{pmatrix} = U_{\rm PMNS} \begin{pmatrix}
\nu_1 \\ \nu_2 \\ \nu_3
\end{pmatrix}, \\
    & U_{\rm PMNS} = \scalebox{0.83}{$\begin{pmatrix}
        1 & 0 & 0 \\
        0 & c_{23} & s_{23} \\
        0 & -s_{23} & c_{23}
    \end{pmatrix}\begin{pmatrix}
        c_{13} & 0 & s_{13}e^{-i\delta_{\rm CP}} \\
        0 & 1 & 0 \\
        -s_{13} & 0 & c_{13}
    \end{pmatrix}\begin{pmatrix}
        c_{12} & s_{12} & 0 \\
        -s_{12} & c_{12} & 0 \\
        0 & 0 & 1
    \end{pmatrix},$}
\end{align}
where $c_{ij}\equiv \cos(\theta_{ij})$ and $s_{ij}\equiv\sin(\theta_{ij})$, and the mixing angles $\theta_{ij}$ and phase $\delta_{\rm CP}$ are taken from NuFIT v5.3~\cite{Esteban:2020cvm,nufit} (with other groups recovering consistent results~\cite{deSalas:2020pgw,Capozzi:2021fjo}), and are tabulated in Table~\ref{tab:parameters}.
\begin{table}[t!]
    \centering
    \renewcommand{\arraystretch}{1.4}
    \begin{tabularx}{0.7\columnwidth}{|Y|Y|}
    \hline
        \multicolumn{2}{|c|}{PMNS parameters (deg.)}  \\
        \hline
        $\theta_{12}$ & $33.67^{+0.74}_{-0.71}$\\
         $\theta_{23}$ & $42.3^{+1.1}_{-0.9}$\\
         $\theta_{13}$ & $8.58^{+0.11}_{-0.11}$\\
         $\delta_{\rm CP}$ & $232^{+39}_{-25}$\\
        \hline \hline
        \multicolumn{2}{|c|}{Mass parameters ${\rm MeV}^2$}   \\
        \hline
        $\Delta m_{21}^2\; (\times 10^{17})$ & $7.41^{+0.21}_{-0.20}$ \\
        $\Delta m_{31}^2\; (\times 10^{15})$ & $2.505^{+0.024}_{-0.026}$ \\
        \hline
    \end{tabularx}
    \renewcommand{\arraystretch}{1.0}
    \caption{PNMS mixing parameters and mass differences taken from Refs.~\cite{Esteban:2020cvm,nufit}, assuming normal ordering.}
    \label{tab:parameters}
\end{table}

The three-flavor coherent neutrino-neutrino interaction can be described by the following Hamiltonian~\cite{Balantekin:2006tg},
\begin{equation}
  H_{\nu\nu} = 
  \sum_{i<j}  J_{ij}\bm{\lambda}^{(i)} \cdot \bm{\lambda}^{(j)} \ ,
  \label{eq:twobody}
\end{equation}
with $\bm{\lambda}^{(i)}=(\lambda^{(i)}_1,\lambda^{(i)}_2,\ldots,\lambda^{(i)}_8)$, and the coupling coefficient $J_{ij}$ is defined as 
\begin{equation}
    J_{ij}=\frac{G_F\rho_\nu}{\sqrt{2}N}(1-\cos\theta_{ij}) \ , 
\end{equation} 
with $G_F$ being Fermi's constant and $\rho_\nu$ the number-density of neutrinos.
The angle $\theta_{ij}$ is the angle between the momentum of the $i^{\rm th}$ and $j^{\rm th}$ neutrino, and for demonstration purposes, we use the simple model introduced in Ref.~\cite{Hall:2021rbv}, where it is sampled from a cone-shaped distribution, $\theta_{ij}=\frac{|i-j|}{N-1} \arccos(0.9)$, as done in Refs.~\cite{Hall:2021rbv,Illa:2022jqb,Amitrano:2022yyn,Illa:2022zgu,Chernyshev:2024kpu}.
The neutrino density coupling constant $\mu$ is defined such that the one-body and two-body terms have the same magnitude,

\begin{equation}
    \mu = \frac{G_F \rho_\nu}{\sqrt{2}}=\frac{\Delta m_{31}^2 N}{2E}\, ,
\end{equation}

 This choice of $\mu$ corresponds to a
simulation the most non-trivial part of the evolution. Close to a supernova’s proto-neutron star, the $H_{\nu \nu}$ interaction term dominates the collective neutrino oscillation Hamiltonian, while in the outer layers of the supernova the $H_{\nu}$ vacuum oscillation term dominates. The latter case is trivial: the oscillations of each neutrino can be treated separately. The former case is not trivial in general, but in the case where the forward scattering interaction between any two neutrino-modes is the same, $H_{\nu \nu}$ can be reduced to a single-neutrino time evolution \cite{friedland2003manyparticle, Friedland2006ManyBody}. Due in part to this, the transition between these two cases, where both $H_{\nu}$ and $H_{\nu \nu}$ matter, is the most non-trivial part of the neutrinos’ time evolution and is by extension the part where quantum computing would likely be the most useful. $\mu$ is also assumed to be time-dependent. However,  since the density of neutrinos decreases as a function of time, a realistic simulation should consider a time-dependent strength of the two-body term
(see, e.g., Refs.~\cite{Cervia:2022pro,Siwach:2022xhx,Siwach:2023wzy}).

The final Hamiltonian in the mass basis as a function of $\mu$ is then given by 
\begin{equation}
\begin{split}
    H & = \frac{\mu}{N}  \sum_i \left[-\frac{ \omega}{2\Omega} \lambda^{(i)}_3 +\frac{\omega-2\Omega}{2\sqrt{3}\Omega} \lambda^{(i)}_8 \right]\\ 
    &\qquad\qquad\qquad+\frac{\mu}{N} \sum_{i<j} [1-\cos(\theta_{ij})] \bm{\lambda}^{(i)} \cdot \bm{\lambda}^{(j)}\;.
\end{split}
\end{equation}
While the two-body term is basis-independent, to transform the one-body term into flavor space the PNMS matrix is applied to the one-body term (Eq.~\eqref{eq:single_body_a}) as follows
\begin{equation}
H^{(i)}_\nu|_{\rm flavor}= U_{\rm PNMS} \cdot H^{(i)}_\nu \cdot U^\dagger_{\rm PNMS}\,.
\label{eq:singleneut_flavor}
\end{equation}

\section{Qutrit mapping}
\label{sec:qutrit}

Motivated by neutrinos' natural three-level structure, we present qutrit-based quantum circuits that can be used for simulating the time evolution of a many-body three-flavored neutrino system.  Our proposed qutrit circuit follows the native qutrit gate set and the notation of the transmon qudits in Ref.~\cite{Goss:2022bqd} (see App.~\ref{app:qutrit_gate} for more details). 

The one-body term $H_{v}$, first implemented in Ref.~\cite{Nguyen:2022snr} on an IBM quantum computer, only involves single-qutrit gates. Equation~\eqref{eq:single_body_matrix} can be implemented with a single phase gate, 
\begin{equation}
e^{-itH^{(i)}_\nu}={\rm Ph}(0,-\omega\, t,-\Omega\, t)={\rm diag}(1,e^{-i\omega\,t},e^{-\Omega\,t})\,,    
\end{equation}
 and the PNMS matrix can be decomposed as
\begin{equation}
\begin{split}
    U_{\rm PNMS} \ = \ & R^{12}_y(\theta_{23})    
    R^{02}_Z\left(\frac{-\pi+\delta_{CP}}{2}\right) R^{02}_y(\theta_{13})\\ & \times R^{02}_Z\left(-\frac{-\pi+\delta_{CP}}{2}\right) R^{01}_y(\theta_{12}) \ ,    
\end{split}
\end{equation}
where App.~\ref{app:qutrit_gate} shows  our gate definitions.

\begin{figure}[!t]
\centering
$$\Qcircuit @C=1em @R=1em { & \ctrl{1}& \gate{X} & \qw &\gate{X^{\dagger}} &\qw & \ctrl{1} &  \qw \\
& \gate{X} & \ctrl{-1}  & \gate{X^{12}_{-2  t J_{ij}}} & \ctrl{-1} & \gate{{\rm Ph}(-2 t J_{ij},0,0)} & \gate{X^{\dagger}} &\qw\\
} $$
\caption{Quantum circuit implementing the term $e^{-i t J_{ij}\bm{\lambda}^{(i)}\cdot \bm{\lambda}^{(j)}}$ from the two-neutrino part $H_{\nu\nu}$. Definitions of the gates can be found in App. ~\ref{app:qutrit_gate}.}
\label{fig:qutrit_qc}
\end{figure}
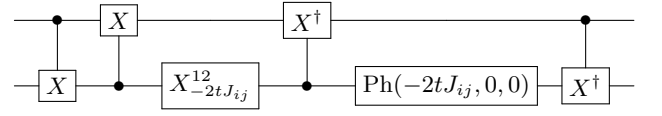

For the two-body term, while a numerical compilation requires at most 6 $CX$ (or $CX^\dagger$) gates~\cite{Goss:2022bqd}, two controlled-shift gates $CX$ and two $CX^\dagger$  were found to be sufficient to apply the specific SU(9) rotation described in Eq.~\eqref{eq:twobody}, as shown in Fig.~\ref{fig:qutrit_qc}.
The qutrit $CX$ gate is defined as $CX\ket{x,y}=\ket{x,\rm{mod}(x+y,3)}$, and the $X^{12}_{2tJ_{ij}}$ gate is given by

\begin{equation}
    X^{12}_{-2tJ_{ij}}=\begin{pmatrix}
        1 & 0 & 0\\
        0 & \cos(J_{ij}\,t) & -i\sin(J_{ij}\,t)\\
        0 & -i\sin(J_{ij}\,t) & \cos(J_{ij}\,t)
    \end{pmatrix}\ .
\end{equation}

\subsection{Swap network \label{sec:swap_network}}

\begin{figure}
    \raggedright
    (a)
    \includegraphics[width=0.95\columnwidth]{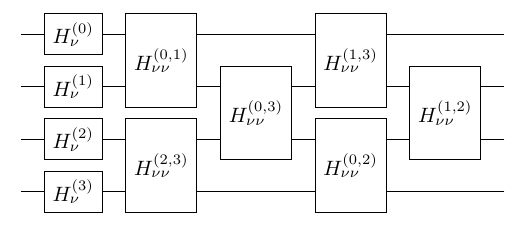}
    (b)
     \includegraphics[width=\columnwidth]{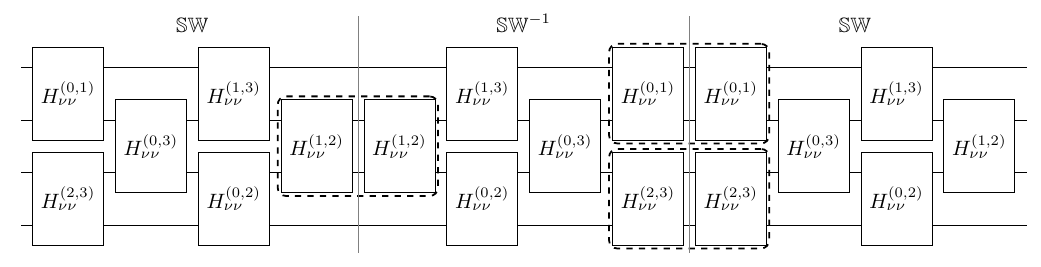}
    
    \caption{(a) Quantum circuit implementing a single LO Trotterized time evolution step via the swap network for four neutrinos. (b) Simplification when using three steps of the NLO$^*$ Trotterized time evolution operator for four neutrinos, where the highlighted operations can be simplified by a single two-qubit box with twice the time step.}
    \label{fig:circuitSWAP}
\end{figure}

Since strategic ordering of application of Eq.~\eqref{eq:twobody} is required to minmize circuit depth, we implement the swap network, $\mathbb{SW}$, which was first proposed in Ref.~\cite{Hall:2021rbv} and is similar to the fermionic swap  from Refs.~\cite{Kivlichan2018prl,OGorman:2019mll}. 
This network limits the depth to $N$ layers of Eq.~\eqref{eq:twobody} for a system with $N$ neutrinos (assuming the qubits are in a linear network and gates can be applied in parallel). 
In the original $\mathbb{SW}$, SWAP gates are needed between each layer to achieve the all-to-all connectivity. 
However, for our implementation, they can be absorbed into the two-body term,
\begin{align}
    {\rm SWAP}_{ij}\cdot e^{-it J_{ij}\bm{\lambda}^{(i)}\cdot \bm{\lambda}^{(j)}}&=e^{-i\frac{\pi}{4}\bm{\lambda}^{(i)}\cdot \bm{\lambda}^{(j)}}\cdot e^{-it J_{ij}\bm{\lambda}^{(i)}\cdot \bm{\lambda}^{(j)}} \nonumber \\
    &=e^{-i(t J_{ij}+\frac{\pi}{4})\bm{\lambda}^{(i)}\cdot \bm{\lambda}^{(j)}} \ ,
\end{align}
substantially reducing the number of entangling gates, particularly in hardware with limited connectivity such as superconducting systems.\footnote{The same simplification can be applied on the two-flavor case, where the SWAP gate can be written as $e^{-i\frac{\pi}{4}\bm{\sigma}^{(i)}\cdot \bm{\sigma}^{(j)}}$.} 
The schematic of such a network can be seen in Fig.~\ref{fig:circuitSWAP}a for four neutrinos.
While SWAP gates might not be required in devices with all-to-all connectivity (such as trapped ions), this new strategy 

can also improve the circuit fidelity in devices where qubits have to be physically moved around, such as in the Quantinuum devices.

The one- and two-body terms in Eqs.~\eqref{eq:single_body_a} and~\eqref{eq:twobody} commute, $[H_{\nu},H_{\nu\nu}]=0$, so they can be Trotterized independently. 
The leading-order (LO) Trotterized time evolution operator is therefore
\begin{equation}
    U(t)_{\rm LO}=e^{-itH_\nu}\prod_{i,j\, \in\, \mathbb{SW}}e^{-i t J_{ij} \bm{\lambda}^{(i)}\cdot \bm{\lambda}^{(j)}} \ .
\end{equation}
When applying multiple Trotter steps, it is more efficient to use the second order (NLO) Suzuki-Trotter formula~\cite{Suzuki1990,Suzuki1991}, as used in Ref.~\cite{Amitrano:2022yyn},
\begin{align}
    U(t)_{\rm NLO} \ = \ & e^{-itH_\nu} \prod_{i,j \, \in \, \mathbb{SW}} e^{-i \frac{t}{2} J_{ij} \bm{\lambda}^{(i)} \cdot \bm{\lambda}^{(j)}} \nonumber \\
    & \times \prod_{i,j \, \in \, \mathbb{SW}^{-1}} e^{-i \frac{t}{2} J_{ij} \bm{\lambda}^{(i)} \cdot  \bm{\lambda}^{(j)}} \ ,
\end{align}
with $\mathbb{SW}^{-1}$  applying the gates in reverse order. With this method, additional simplifications are possible. In particular, the last and first layer in $\mathbb{SW}$ and $\mathbb{SW}^{-1}$, respectively, can be merged into a single operation with twice the time step, as depicted in Fig.~\ref{fig:circuitSWAP}b.

We can generalize this to multiple Trotter steps by interleaving layers ordered in the $\mathbb{SW}$ and $\mathbb{SW}^{-1}$ network sequences, allowing for cancellations between these layers (referred as NLO$^*$).\footnote{This can be extended to higher-order Suzuki-Trotter formulas, although further simplifications are not expected to compensate for the increase in circuit depth.}
Then, the total number of $CX$ gates for $k$ number of Trotter steps (with $k \geq 2$) is
\begin{align}
    \left[U(\tfrac{t}{k})_{\rm LO}\right]^k\; & : \; N_{CX} k\frac{N(N-1)}{2}\ , \\
    U(t)_{{\rm NLO}^*_k}\; & : \; N_{CX}k\frac{N(N-1)}{2} -N_{CX}\left\lfloor \frac{k}{2} \right\rfloor \left(\left\lceil\frac{N}{2}\right\rceil-1\right) \nonumber\\
    & \ \ -N_{CX}\left\lfloor \frac{k-1}{2}\right\rfloor \cdot \left\lfloor\frac{N}{2}\right\rfloor     \ ,
\end{align}
where $\lceil \cdot \rceil$ ($\lfloor \cdot \rfloor$) is the ceiling (floor) function, $N$ is the number of neutrinos, and $N_{CX}$ is the number of $CX$ gates needed to compile a single neutrino-neutrino term (for the case in Fig.~\ref{fig:qutrit_qc}, $N_{CX}=4$).
A reduction in two-qubit gates is seen when using NLO$^*$ compared to LO with increasing number of Trotter steps, as well as improved convergence. 

\section{Qubit mapping}
\label{sec:qubit}

In the absence of qutrit-based platforms, an alternate approach involves mapping a three-flavor neutrino to two qubits~\cite{Arguelles:2019phs}
, with each flavor encoded as: $\ket{\nu_e} \rightleftharpoons \ket{00}$, $\ket{\nu_\mu} \rightleftharpoons \ket{01}$,  $\ket{\nu_\tau} \rightleftharpoons \ket{10}$, and the unassigned state $\ket{11}$ designated as the unphysical state (assuming no sterile neutrinos).
Therefore, no matrix elements in the unitaries that implement the one- and two-neutrino terms are allowed to mix states between the physical and the unphysical sub-spaces. However, we have the freedom to allow arbitrary mixing between unphysical states if the resulting quantum circuits are shallower.

For the one-neutrino term $H_{v}$, its time evolution operator in the mass basis is diagonal, and can be implemented with single-qubit $R_z$ gates,
\begin{equation}
    e^{-itH^{(i)}_{v}}=R_z(-\omega t)_{2i}\otimes R_z(-\Omega t)_{2i+1}\ .
\end{equation}

In the flavor basis, the $3\times 3$ matrix can be embedded into a $4\times 4$ one, and the resulting SU(4) matrix can be transpiled into the three-CNOT circuits from Refs.~\cite{Vatan:2004nmz,PhysRevA.69.010301,PhysRevA.77.066301} (or use the circuits from Refs.~\cite{Arguelles:2019phs,Molewski:2021ogs}).

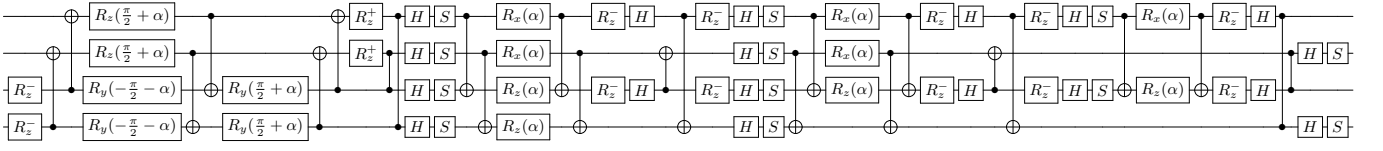
\begin{figure*}[ht]
    \centering
$\resizebox{\textwidth}{!}{
\Qcircuit @C=0.3em @R=0.6em {
    & \qw & \qw & \targ & \gate{R_z(\frac{\pi}{2}+\alpha)} & \qw & \ctrl{2} & \qw & \qw & \targ & \gate{R_z^+} & \qw & \ctrl{3} & \gate{H} & \gate{S} & \ctrl{2} & \qw & \gate{R_x(\alpha)} & \ctrl{2} & \qw & \gate{R_z^-} & \gate{H} & \qw & \ctrl{3} & \gate{R_z^-} & \gate{H} & \gate{S} & \qw & \ctrl{2} & \gate{ R_x(\alpha)} & \qw & \ctrl{2} & \gate{R_z^-} & \gate{H} & \qw & \ctrl{3} & \gate{R_z^-} & \gate{H} & \gate{S} & \ctrl{2} & \gate{R_x(\alpha)} & \ctrl{2} & \gate{R_z^-} & \gate{H} & \ctrl{3} & \qw & \qw & \qw & \qw \\
    & \qw & \targ & \qw & \gate{R_z(\frac{\pi}{2}+\alpha)} & \ctrl{2} & \qw & \qw & \targ & \qw & \gate{R_z^+} & \ctrl{1} & \qw & \gate{H} & \gate{S} & \qw & \ctrl{2} & \gate{R_x( {\alpha})} & \qw & \ctrl{2} & \qw & \qw & \targ & \qw  & \qw & \gate{H} & \gate{S} & \ctrl{2} & \qw & \gate{R_x(\alpha)} & \ctrl{2} & \qw & \qw & \qw & \targ & \qw & \qw & \qw & \qw & \qw & \qw & \qw & \qw & \qw & \qw & \ctrl{1} & \gate{H} & \gate{S} & \qw \\
    & \gate{R_z^-} & \qw & \ctrl{-2} & \gate{R_y(-\frac{\pi}{2}-\alpha)} & \qw & \targ & \gate{ R_y(\frac{\pi}{2}+\alpha)} & \qw & \ctrl{-2} & \qw & \control\qw & \qw & \gate{H} & \gate{S} & \targ & \qw & \gate{R_z(\alpha)} & \targ & \qw & \gate{R_z^-} & \gate{H} & \ctrl{-1} & \qw & \gate{R_z^-} & \gate{H} & \gate{S} & \qw & \targ & \gate{R_z( {\alpha})} & \qw & \targ & \gate{R_z^-} & \gate{H} & \ctrl{-1} & \qw & \gate{R_z^-} & \gate{H} & \gate{S} & \targ & \gate{R_z(\alpha)} & \targ & \gate{R_z^-} & \gate{H} & \qw & \control\qw & \qw & \qw & \qw \\
    & \gate{R_z^-} & \ctrl{-2} & \qw & \gate{R_y(-\frac{\pi}{2}-\alpha)}  & \targ & \qw & \gate{ R_y(\frac{\pi}{2}+\alpha)} & \ctrl{-2} & \qw & \qw & \qw & \control\qw & \gate{H} & \gate{S} & \qw & \targ & \gate{R_z(\alpha)} & \qw & \targ & \qw & \qw & \qw & \targ & \qw & \gate{H} & \gate{S} & \targ & \qw & \qw & \targ & \qw & \qw & \qw & \qw & \targ & \qw & \qw & \qw & \qw & \qw & \qw & \qw & \qw & \control\qw & \qw & \gate{H} & \gate{S} & \qw \\}
}
$
\caption{Circuit A implementing $e^{-i\alpha \bm{\lambda}^{(i)} \cdot \bm{\lambda}^{(j)}}$ in the physical subspace, using 24 CNOTs. The gates $R_z^\pm$ represents the short-hand version of $R_z(\pm\frac{\pi}{2})$.}
\label{fig:2neutA}
\end{figure*}
\begin{figure*}[ht]
    \centering 
$\resizebox{0.7\textwidth}{!}{
\Qcircuit @C=0.3em @R=0.6em { \\
	& \ctrl{2} & \qw & \gate{R_y(\frac{\pi}{4})} & \targ & \qw & \gate{R_y(-\frac{\pi}{4})}  & \ctrl{2} & \qw & \ctrl{1} & \qw & \ctrl{1} & \qw & \qw & \ctrl{2} & \qw & \ctrl{3} & \qw & \qw & \qw & \qw & \ctrl{3} & \gate{R_y(\frac{\pi}{4})} & \qw & \targ & \gate{R_y(-\frac{\pi}{4})} & \qw & \ctrl{2} & \qw \\
	& \qw & \ctrl{2} & \gate{R_y(\frac{\pi}{4})} & \qw & \targ & \gate{R_y(-\frac{\pi}{4})} & \qw & \ctrl{2} & \targ & \gate{R_z(-2\,\alpha)} & \targ & \ctrl{1} & \qw & \qw & \qw & \qw & \ctrl{1} & \qw & \ctrl{2} & \qw & \qw & \gate{R_y(\frac{\pi}{4})} & \targ & \qw & \gate{R_y(-\frac{\pi}{4})} & \ctrl{2} & \qw & \qw \\
	& \targ & \qw & \qw & \ctrl{-2} & \qw & \gate{R_z(-\alpha)} & \targ & \qw & \qw & \gate{R_z(\alpha)} & \qw & \targ & \gate{R_z(\alpha)} & \targ & \gate{R_z(-\alpha)} & \qw & \targ & \qw & \qw & \qw & \qw & \qw & \qw & \ctrl{-2} & \qw & \qw & \targ & \qw \\
	& \qw & \targ & \qw & \qw & \ctrl{-2} & \gate{ R_z(-\alpha)} & \qw & \targ & \qw & \gate{ R_z(\alpha)} & \qw & \qw & \qw & \qw & \qw & \targ & \qw & \gate{ R_z(\alpha)} & \targ & \gate{ R_z (-\alpha)} & \targ  & \qw & \ctrl{-2} & \qw & \qw & \targ & \qw & \qw \\
}
}$
    \caption{Circuit B implementing $e^{-i\alpha \bm{\lambda}^{(i)} \cdot \bm{\lambda}^{(j)}}$ in the physical subspace, using 18 CNOTs.}
    \label{fig:2neutB}
\end{figure*}
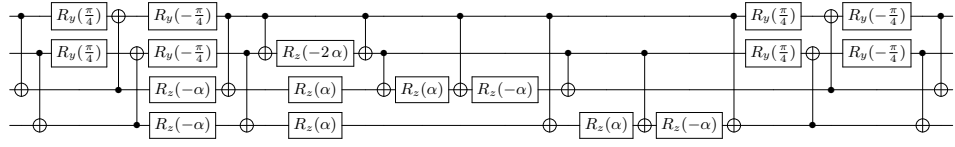
The two-neutrino term $H_{\nu\nu}$ is more delicate in this case, compared to the qutrit implementation. As mentioned, while the physical subspace is fixed, we have the freedom on the unphysical one to implement any rotation, as long as the two subspaces do not get mixed. Here we propose two circuits, A (shown in Fig.~\ref{fig:2neutA}) and B (shown in Fig.~\ref{fig:2neutB}), both using qiskit conventions~\cite{qiskit} for the gate definitions. The difference between these two circuits can be seen by looking at the unitary they implement,
\begin{align}
    & \left. e^{-i\alpha \bm{\lambda}^{(i)} \cdot \bm{\lambda}^{(j)}}\right|_A=\left(\vcenter{\hbox{\includegraphics[width=0.4\columnwidth]{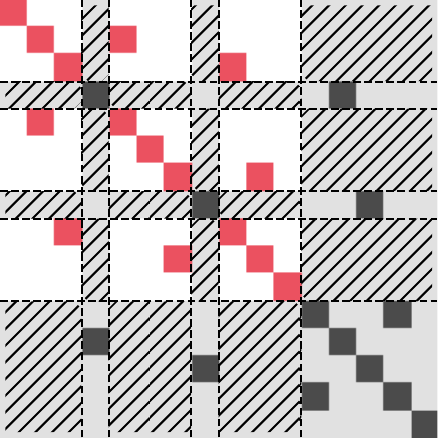}}}\right) \ , \label{eq:circA}\\[1em]
    & \left. e^{-i\alpha \bm{\lambda}^{(i)} \cdot \bm{\lambda}^{(j)}}\right|_B=\left(\vcenter{\hbox{\includegraphics[width=0.4\columnwidth]{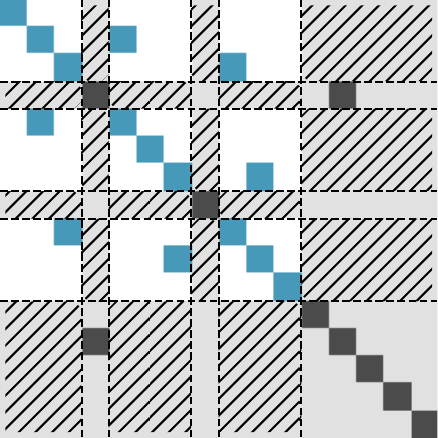}}}\right) \label{eq:circB} \ .
\end{align}
While both circuits have the same effect in the physical subspace, shown with white background in Eqs.~\eqref{eq:circA}-\eqref{eq:circB}, the rotations in the unphysical subspace, shown with a gray background, are different (with the dashed region being the transition between the two subspaces). 
For example, circuit A, in the full four-qubit space, implements the more general gate $e^{-i\alpha \tilde{\bm{\lambda}}^{(i)} \cdot \tilde{\bm{\lambda}}^{(j)}}$ (with $\tilde{\bm{\lambda}}^{(i)}$ being the set of SU(4) generators that act on the $i^{\rm th}$ neutrino), which can also be written as $e^{-i\frac{\alpha}{2}\sum_{a,b}(\sigma_a\otimes \sigma_b)^i\otimes (\sigma_a\otimes \sigma_b)^j}$ (with $\sigma_a$ being all possible elements of the Pauli group and the superscripts denoting neutrino-indices). 
While circuit A does not perform the most general SU(16) gate, it contains significantly less CNOT gates than what one would obtain for an SU(16) gate using currently available operator-to-circuit transpilers such as qiskit~\cite{qiskit,Javadi-Abhari:2024kbf}, tket~\cite{Sivarajah:2020lfo}, or other decomposition approaches~\cite{Shende:2006,Mottonen:2006,Mansky:2022bai,Krol:2024taf}, which result in circuits containing on the order of $\mathcal{O}(100)$ CNOTs. In the noiseless case, both circuits yield the same results. However, we expect that the results from running the circuit in Fig.~\ref{fig:2neutB} will be cleaner than those from Fig.~\ref{fig:2neutA}, as the computational time is shorter, leading to smaller decoherence effects.

In both circuits A and B, the method discussed in Sec.~\ref{sec:swap_network} can absorb the SWAP operation into the two-body term, modifying $\alpha\rightarrow\alpha+\frac{\pi}{4}$.
For this case, where each neutrino is composed of two qubits, the action of the SWAP gate is ${\rm SWAP}_{ij} \ket{ab}_i\otimes\ket{cd}_j=\ket{cd}_i\otimes\ket{ab}_j$.
Table~\ref{tab:cnot_counts} reports the number of required $CX$ or CNOT gates and its corresponding depth for each quantum circuit (both qubit and qutrit) after compilation in both an-all-to-all and linear-chain architecture.
\begin{table}[h!]
\centering
\renewcommand{\arraystretch}{1.2}
\begin{tabularx}{\columnwidth}{|Y|c|YY|YY|}
\hline
\multirow{2}{*}[-0.5em]{Qudit} &\multirow{2}{*}[-0.5em]{Circuit} & \multicolumn{2}{c|}{All-to-all} & \multicolumn{2}{c|}{Linear chain}  \\
& & 2-q gate count & 2-q gate depth & 2-q gate count & 2-q gate depth \\
\hline \hline
Qutrit & Fig.~\ref{fig:qutrit_qc} & 4 & 4 & 4 & 4 \\ \hline
\multirow{2}{*}{Qubit} & A (Fig.~\ref{fig:2neutA}) & 24 & 13 & 42 & 31 \\ 
& B (Fig.~\ref{fig:2neutB})  & 18 & 12 & 30 & 25 \\
\hline
\end{tabularx}
\renewcommand{\arraystretch}{1.0}
\caption{The two-qudit entangling gate count and depth for the two-neutrino quantum circuits proposed, involving two qutrits or four qubits.}
\label{tab:cnot_counts}
\end{table}

\section{Results}
\label{sec:results}

Dynamics for $N=\{2,4,8\}$ neutrinos were simulated with the \texttt{H1-1} Quantinuum trapped-ion and \texttt{ibm\_torino} IBM superconducting quantum computers (device parameters can be found in App.~\ref{app:device_parameters}). We use the circuits described in Sec.~\ref{sec:qubit}, in particular circuit B in Fig.~\ref{fig:2neutB}, to implement the two-neutrino interaction. For two and four neutrinos, the time step was fixed (while increasing the number of Trotter steps), while for eight neutrinos number of Trotter steps was fixed (while increasing the time step).

For current noisy intermediate-scale quantum (NISQ) devices~\cite{Preskill:2018jim} error mitigation techniques are critical for reliable results.
Well-known techniques, such as zero-noise extrapolation~\cite{Li:2016vmf,Temme:2016vkz,2020arXiv200510921G} and probabilistic error cancellation~\cite{Temme:2016vkz,Berg:2022ugn}, require a large overhead in circuit sampling or an increase in circuit depth. 
For this study, we decided to use a more resource-friendly algorithm, decoherence renormalization (DR), first introduced in Refs.~\cite{Urbanek:2021oej,ARahman:2022tkr}, and later implemented in increasingly larger simulations in Refs.~\cite{Farrell:2022wyt,Ciavarella:2023mfc,Farrell:2023fgd,Farrell:2024fit,Ciavarella:2024fzw}, to mitigate decoherent errors. 
For each time step, two different quantum circuits were ran: the first one (which we call the \textit{physics} quantum circuit)
implements the correct dynamics, and the second one (which we call the \textit{identity} quantum circuit) runs a quantum circuit with the same structure as the \textit{physics} circuit, except its action on the initial state is the identity. For first-order Trotter, this can be achieved by setting the time step to zero. For second-order Trotter, the sign in the time step for the second half of the circuit is flipped. 
Through the following rescaling formula, the experimental noiseless probability $P_{phys}^{\rm ex}(t)$ can be computed as:
\begin{equation}
    P_{phys}^{\rm ex}(t)-d_n = \frac{ P_{id}^{\rm ex}-d_n}{    P_{id}^{\rm noisy}-d_n
} \left( P_{phys}^{\rm noisy}(t)-d_n\right) \,,
\label{eq:DR}
\end{equation}
where $P_{phys}^{\rm noisy}(t)$ indicates the obtained probability value from the \textit{physics} quantum circuit and $P_{id}^{\rm noisy}$ indicates the obtained probability from the \textit{identity} quantum circuit. $P_{id}^{\rm ex}$ corresponds to the noiseless result of the \textit{identity} quantum circuit. $d_n$ represents the decoherence value of the quantity computed $P$, which, for a generic $n$-neutrino measurement probability, is $d_n=1/4^n$.

Equation~\eqref{eq:DR} assumes all noise from the device is depolarizing. While this seems a reasonable assumption for the Quantinuum device, as observed in other trapped ion devices~\cite{Nguyen:2021hyk}, the IBM quantum computer requires additional steps. To ensure all noise to be depolarizing, Pauli twirling~\cite{Wallman:2015uzh,Hashim:2020cop} was applied to transform coherent noise into incoherent noise, as well as dynamical decoupling~\cite{Viola:1998dsd,2012RSPTA.370.4748S,Ezzell:2022uat} to suppress cross-talk and idling errors. Moreover, the matrix-free measurement mitigation (M3)~\cite{Nation:2021kye} provided by the {\tt Sampler} function from {\tt qiskit}~\cite{qiskit} was used to correct readout errors.

\begin{figure}[t]
    \centering
    \includegraphics[width=\columnwidth]{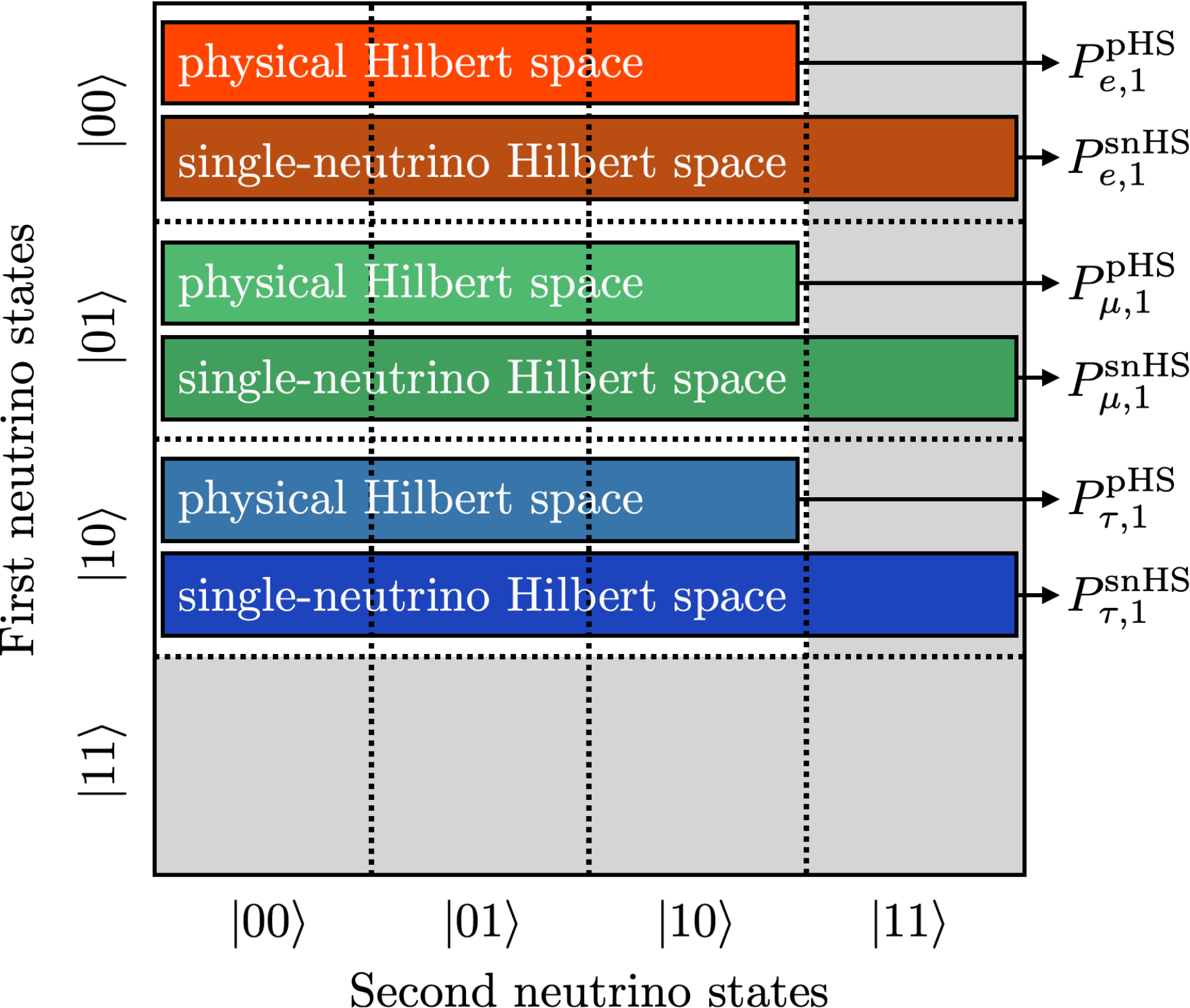}
    \caption{Different post-selecting procedures for computing the single-neutrino flavor probability, for a system of two neutrinos. The physical Hilbert space (pHS) approach only accounts for the physical flavor states; the single neutrino Hilbert space (snHS) approach keeps all contributions from the other neutrino states (both physical and unphysical).}
    \label{fig:postselecting_scheme}
\end{figure}

Two observables for each system were computed: the probability of a single neutrino in a particular flavor state $P_{\nu}$ and the persistence probability of the initial state, $|\langle \psi(0)|\psi(t)\rangle|^2$. Since device errors will populate the unphysical Hilbert space, we implement two different strategies for computing the single-neutrino probabilities. The first method uses the full physical Hilbert space (pHS) where all nonphysical states $\ket{11}$ are discarded. The second method involves summing over the remaining states in the single-neutrino Hilbert space (snHS), mimicking a single measurement of the qubits representing the $i^{\rm th}$ neutrino. While unphysical states for other neutrinos contribute to the probability, after being corrected with DR their contribution can be small.\footnote{One should notice that the two methods have two different decoherence values in Eq.~\eqref{eq:DR}. When computing $P_\nu$, $d^{\rm pHS}_{1} = 3^{N-1}/{4^N}$ (with $N$ the total number of neutrinos) and $d^{\rm snHS}_{1} = 1/4$.}
The two methods are illustrated in Fig.~\ref{fig:postselecting_scheme} for the two neutrino case, where the big squares represent all possible 16 states.

\subsection{Quantinuum}
\label{sec:quantinuum}

Due to the all-to-all architecture of the Quantinuum device~\cite{quantinuum}, the circuit in Fig.~\ref{fig:2neutB} for implementing the neutrino-neutrino term can be used without having to rewrite the CNOTs connecting distant qubits. 
After transpiling the circuit to the native gate-set (single-qubit gates and ZZ$(\theta)=e^{-i(\theta/2) Z\otimes Z}$), the entangling gate count and depth gets reduced by one unit, compared to the numbers in Table~\ref{tab:cnot_counts}.\footnote{The simplification occurs in the $R_z(-2\alpha)$ rotation and its two neighbouring CNOTs in Fig.~\ref{fig:2neutB}, which get transformed into a single ZZ$(-2\alpha)$ gate.}
We use the trick of adding a $\pi/4$ phase to the neutrino-neutrino terms to incorporate the SWAP gate into the latter. This is because although it is not necessary for this hardware, as mentioned in Sec.~\ref{sec:swap_network}, it should reduce the shuffling of trapped ions.

\begin{figure*}[htb!]
    \centering
    \includegraphics[width=\textwidth]{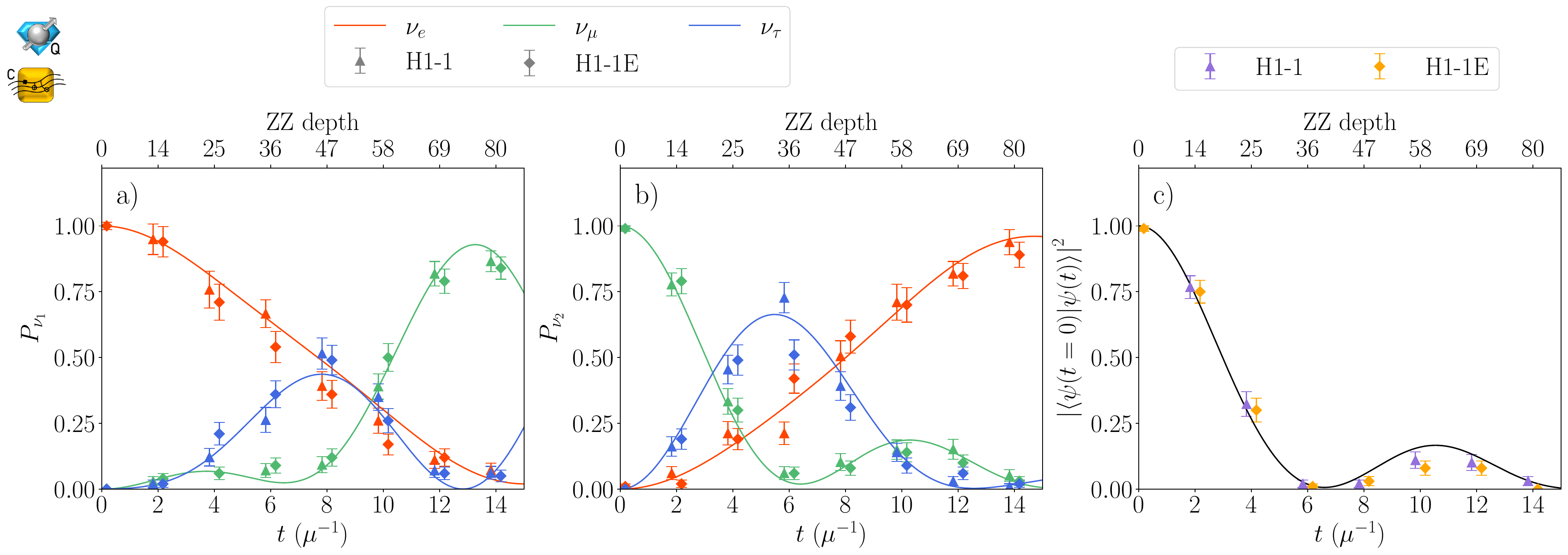}
    \caption{Flavor evolution of a two-neutrino system as a function of time. Panels (a) and (b) show the flavor evolution of the first and second neutrinos, respectively. Panel (c) shows the persistence probability of the initial state. Triangles and diamonds indicate results from the device (\texttt{H1-1}) and its emulator (\texttt{H1-1E}), respectively, slightly shifting the points for ease of readability; solid lines show the exact result. The top axis measures the ZZ depth for each point}
    \label{fig:quantinnum_2_neutrinos}
\end{figure*}
First, the evolution of two neutrinos was simulated, using multiple numbers of Trotter time steps to study the propagation at long times and the noise sources for deep quantum circuits.\footnote{Notice that for two neutrinos, there are no Trotter errors in the decomposition of the time evolution circuits, therefore the multiple Trotter steps are a way to increase the circuit depth and benchmark the quantum computer.} 
Figure~\ref{fig:quantinnum_2_neutrinos} shows the results when we start from the $\ket{\nu_e \nu_\mu}$ state from the device  \texttt{H1-1} and the emulator \texttt{H1-1E}, and use 100 shots per circuit. In this simulation, no error-mitigation techniques were implemented. Also, a 1-3\% contribution from unphysical states was found, supporting the pHS approach to be a good approximation. 
The results, even for longer times, ($t=14 \mu^{-1}$), are compatible with the exact evolution (represented with solid lines). The icons in the top-left corner that appear in this and subsequent plots, identify if the noisy simulator (yellow icon) or quantum device (blue icon) was used~\cite{Klco:2019xro}.

\begin{figure*}[htb!]
\centering
\includegraphics[width=\textwidth]{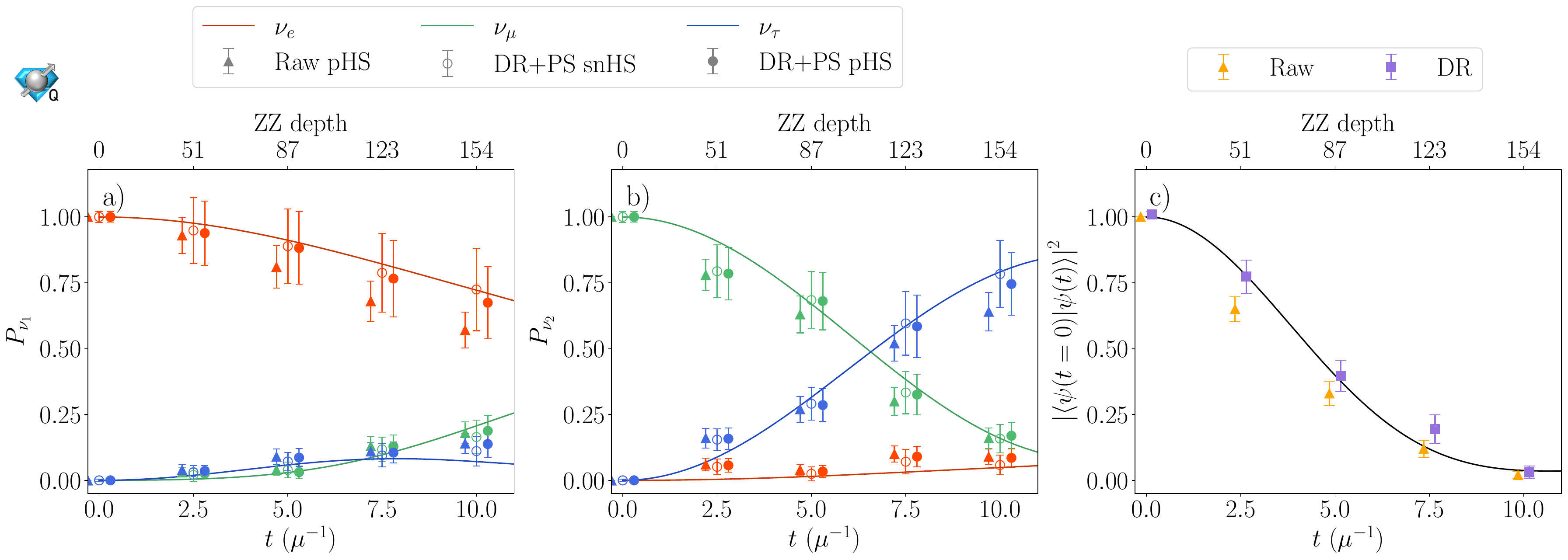}
\caption{Flavor evolution of a four-neutrino system as a function of time, using the \texttt{H1-1} device.  Panels (a) and (b) show the flavor evolution of the first and second neutrinos, respectively. Triangles indicate the raw pHS results. Empty and solid circles represent the results of applying DR,  snHS, and pHS post-selecting procedures, respectively.  Panel (c) shows the persistence probability of the initial state. The triangle and square markers represent the results without and with applying DR, respectively. 
In all panels, the solid lines show the exact evolution, and the points have been slightly shifted to ease the readability.
The top axis measures the ZZ depth for each point.}
    \label{fig:quantinuum_4_neutrinos}
\end{figure*}
The dynamics of four neutrinos starting from the $\ket{\nu_e \nu_\mu \nu_e \nu_\tau
}$ state on the \texttt{H1-1} device was then simulated.
Figure~\ref{fig:quantinuum_4_neutrinos} shows the obtained results, using 100 shots per circuit. Like in Fig.~\ref{fig:quantinnum_2_neutrinos}, panels (a) and (b) illustrate the flavor evolution of the first (the neutrino starting as $\nu_e$) and second neutrino (the neutrino starting as $\nu_\mu$), respectively. In this case DR was used for error-mitigation, and after post-selection via snHS and pHS, the probabilities were normalized to sum to $1$.
Panel (c) shows the persistence probability of the initial state. For this larger system, an improvement after performing error mitigation was observed.

In App.~\ref{app:4neutrinos} the emulator \texttt{H1-1E} and device \texttt{H1-1} results is compared. Generally the emulator and device results are observed to be compatible within reasonable uncertainties. Nevertheless, we observe that the \texttt{H1-1E} emulator gives more pessimistic results than the actual machine.  
\begin{figure*}[ht]
    \centering
    \includegraphics[width=\textwidth]{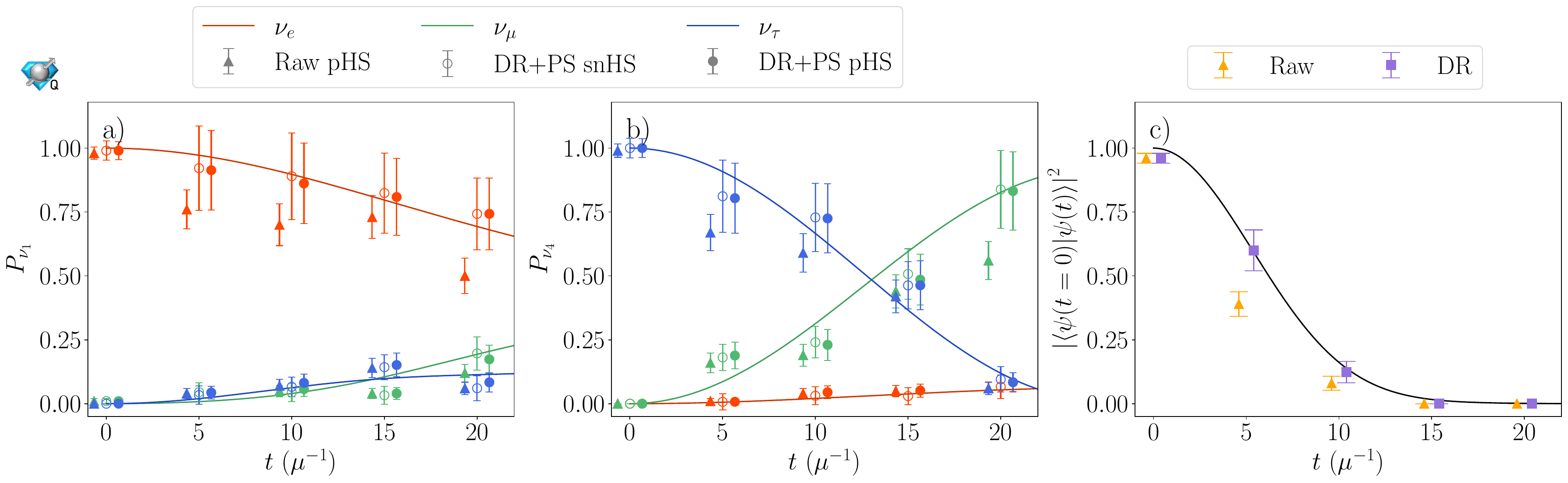}
\caption{Flavor evolution for an eight-neutrino system as a function of time, using the \texttt{H1-1} device. Panels (a) and (b) show the flavor evolution of the first and fourth neutrinos, respectively. Panel (c) shows the persistence probability of the initial state. 
Details are as in Fig.~\ref{fig:quantinuum_4_neutrinos}. 
}
\label{fig:quantinuum_8_neutrinos}
\end{figure*}
Figure~\ref{fig:quantinuum_8_neutrinos} shows the results for an eight-neutrino system, starting from $\ket{\nu_{e}\nu_{\mu}\nu_{e}\nu_{\tau}\nu_{e}\nu_{\mu}\nu_{e}\nu_{\tau}}$. In this case only a single Trotter time step was performed, increasing the time step. 
The implemented quantum circuits have a ZZ depth of 91, and 100 shots were used (except for $t=10\mu^{-1}$, which used 79 shots).
All results are compatible within $2\sigma$ with the exact evolution, albeit with larger uncertainties than the cases for two and four neutrinos.

\subsection{IBM}

Compared to Quantinuum, the IBM hardware has the constraint of linear connectivity between qubits (a one-dimensional chain was selected from the heavy-hex lattice), necessitating the circuit in Fig.~\ref{fig:2neutB} to be compiled into a linear chain architecture. All results were obtained from implementations on \texttt{ibm\_torino}, where the native entangling gate is the controlled-Z (CZ) gate, leading to the same depth and number of gates as in Table~\ref{tab:cnot_counts}.

As discussed at the beginning of 
in Sec.~\ref{sec:results}, $10\times N$ different Pauli-twirled quantum circuits were ran for each time step in order to average out the coherence noise (with $N$ being the number of neutrinos), using 8000 shots per circuit. 
After applying DR and the post-selecting procedures it was noted that the resulting single-neutrino probabilities could become unphysical (either negative or greater than 1), an issue not encountered when using the Quantinuum device in Sec.~\ref{sec:quantinuum}. To fix this, the algorithm of 
Ref.~\cite{PhysRevLett.108.070502} was applied to find the closest probability distribution. 
The uncertainties from combining the different twirled circuits were computed via bootstrap resampling.

\begin{figure*}[ht]
    \centering
    \includegraphics[width=\textwidth]{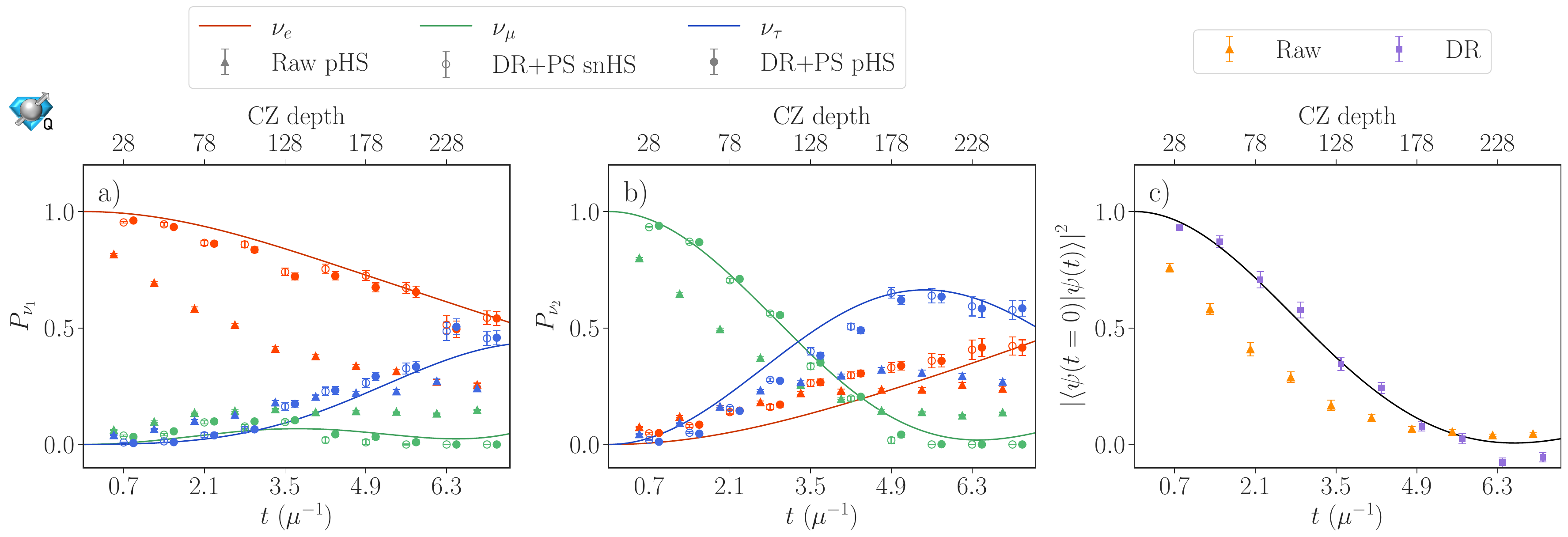}
\caption{Flavor evolution for two neutrinos as a function of time obtained from \texttt{ibm\_torino} device.  
Panels (a) and (b) show the flavor evolution of the first and second neutrinos, respectively. 
Panel (c) shows the persistence probability of the initial state.
Details are as in Fig.~\ref{fig:quantinuum_4_neutrinos}. 
The top axis measures the CZ depth at every other point.}
    \label{fig:ibm_2_neutrinos}
\end{figure*}
Figure~\ref{fig:ibm_2_neutrinos} depicts the evolution for a two-neutrino system, starting from the $\ket{\nu_{e}\nu_{\mu}}$ state. Unlike for Quantinuum (Fig.~\ref{fig:quantinnum_2_neutrinos}), in this case error mitigation is essential for the results to be compatible with the exact evolution.
While most points are within $1\sigma$ and $3\sigma$, it seems that the initial state persistence is more robust against errors than the single-neutrino probabilities.

\begin{figure*}[ht]
    \centering
    \includegraphics[width=\textwidth]{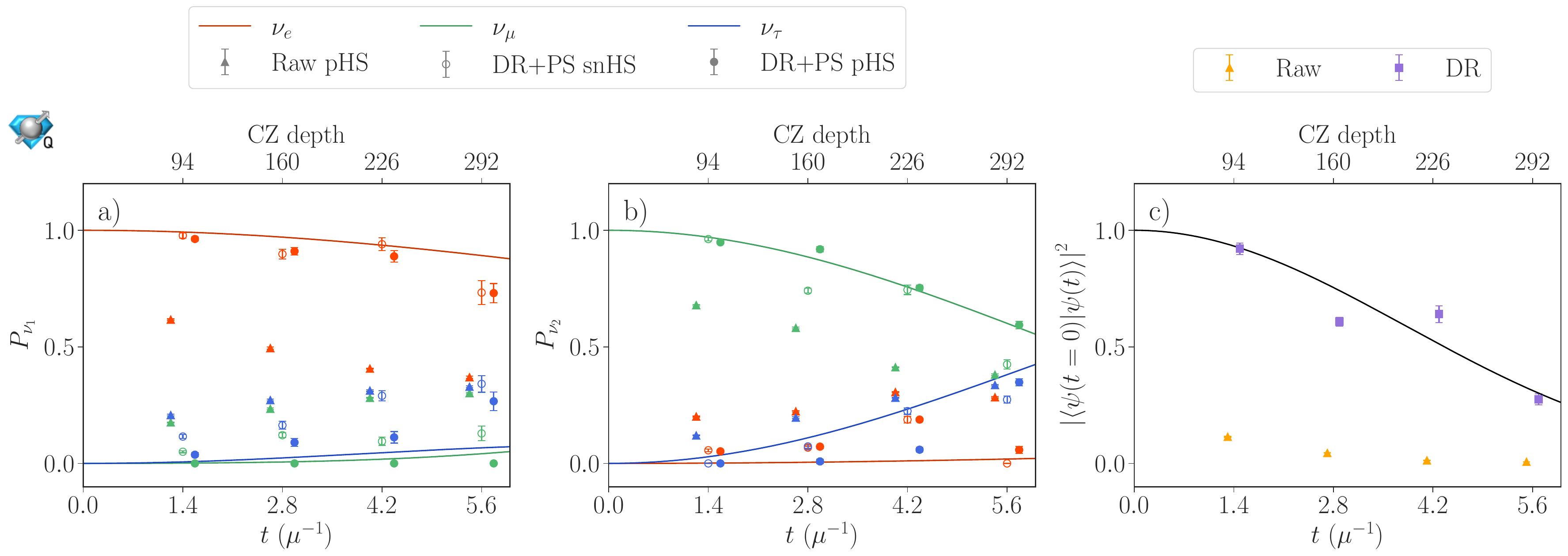}
\caption{Flavor evolution for four neutrinos as a function of time obtained from the \texttt{ibm\_torino} device.  
Panels (a) and (b) show the flavor evolution of the first and second neutrinos, respectively. 
Panel (c) shows the persistence probability of the initial state.
Details are as in Fig.~\ref{fig:quantinuum_4_neutrinos}. 
The top axis measures the CZ depth at each point.}
\label{fig:ibm_4_neutrinos}
\end{figure*}
Figure~\ref{fig:ibm_4_neutrinos} shows the evolution for the four-neutrino system, starting from the $\ket{\nu_{e}\nu_{\mu}\nu_{e}\nu_{\tau}}$ state. 
Like in the two-neutrino case, the obtained results follow the analytical evolution, although in some cases there is a difference of more than $5\sigma$.
This growing tension is investigated further in the eight-neutrino system.
\begin{figure*}[ht]
    \centering
\includegraphics[width=\textwidth]{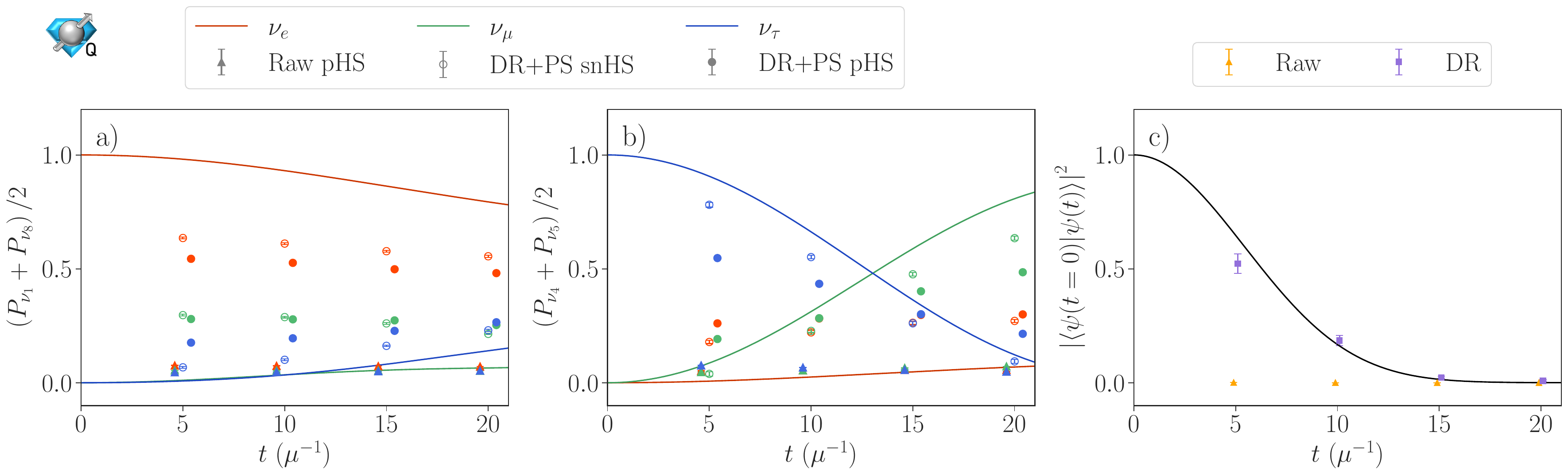}
\caption{Flavor evolution for eight neutrinos as a function of time obtained from the \texttt{ibm\_torino} device.  Panels (a) and (b) show the flavor evolution of the (symmetrized) first and fourth neutrinos, respectively. 
Panel (c) shows the persistence probability of the initial state.
Details are as in Fig.~\ref{fig:quantinuum_4_neutrinos}.
}
\label{fig:ibm_8_neutrinos}
\end{figure*}
Figure~\ref{fig:ibm_8_neutrinos} shows the evolution for the eight-neutrino system, with quantum circuits that have a CZ depth of 182.
In contrast to the previous results, here the initial state was the symmetric state $\ket{\nu_{e}\nu_{\mu}\nu_{e}\nu_{\tau}\nu_{\tau}\nu_{e}\nu_{\mu}\nu_{e}}$. 
This change enables averaging of the single-neutrino probability between the $i^{\rm th}$ and $(N+1-i)^{\rm th}$ neutrino, improving the quality of the obtained results (as seen by the degradation in the four neutrino system in Fig.~\ref{fig:ibm_4_neutrinos}).\footnote{A similar exchange symmetry has been observed for the two-flavor case~\cite{Hall:2021rbv,Amitrano:2022yyn,Illa:2022zgu}, although in the three-flavor case it is only manifested for symmetric initial states.} 
After performing the symmetrization, the noise contributions can be averaged out and reduced (for the non-symmetric initial state, see App.~\ref{app:8ibm}). 
Despite performing error mitigation,  the results in Fig.~\ref{fig:ibm_8_neutrinos} for the single-neutrino probabilities have large deviations from the expected values (more noticeable with pHS post-selection than with snHS). 
As before, the initial state persistence seems to be more effectively recovered after DR.

\begin{figure*}[ht]
    \centering
\includegraphics[width=\textwidth]{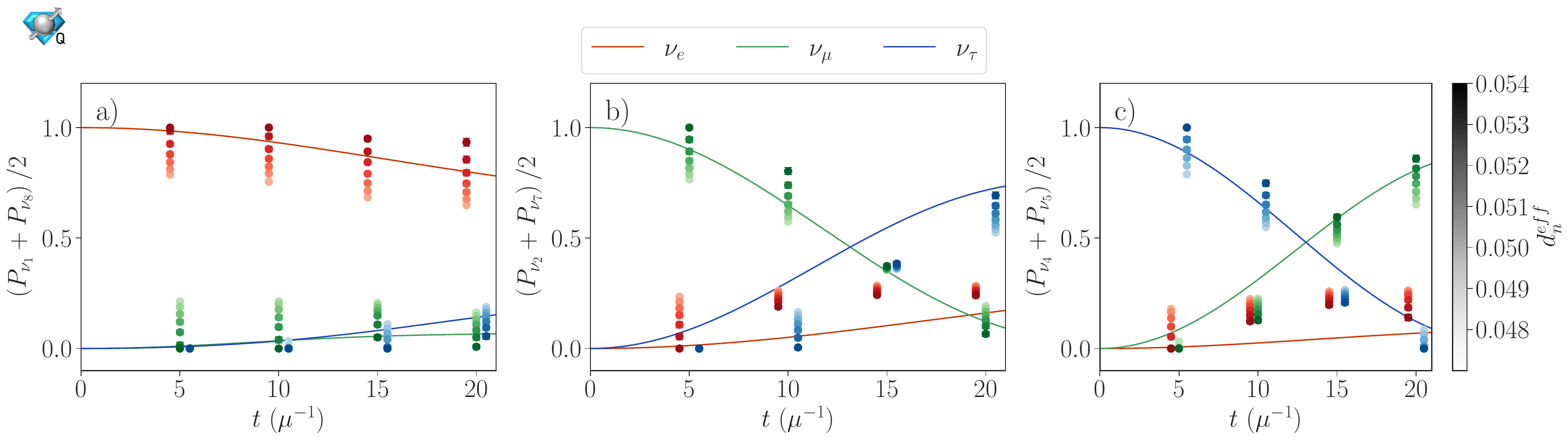}
\caption{Flavor evolution for an eight-neutrino system as a function of time obtained from \texttt{ibm\_torino} device as in Fig.~\ref{fig:ibm_8_neutrinos}, scanning over effective decoherence values $d^{(eff)}_n$ after applying DR and pHS post-selection procedure. Panels (a), (b), and (c) show the flavor evolution of the (symmetrized) first, second and fourth neutrinos, respectively. As shown in the color bar on the far-right, $d^{(eff)}_n$ increases from lighter to darker colors.
}
\label{fig:ibm_8_neutrinos_dn}
\end{figure*}
\begin{table}[ht]
\centering
\renewcommand{\arraystretch}{1.2}
\begin{tabularx}{\columnwidth}{|c|YYY|c|}
\hline
Neutrino  & $P_e$ & $P_\mu$ & $P_\tau$ & $d_n^{(id)}$ \\
\hline
$(P_1+P_8)/2$ & \textbf{0.079(4)} &0.054(3) &0.047(3) &0.033\\
$(P_2+P_7)/2$& 0.055(3) &\textbf{0.081(4)} &0.043(2) &0.033\\
$(P_3+P_6)/2$ & \textbf{0.081(4)} &0.052(3) &0.046(2) &0.033\\
$(P_4+P_5)/2$ & 0.051(3) &0.043(2) &\textbf{0.085(4)} &0.033\\
\hline
\end{tabularx}
\renewcommand{\arraystretch}{1}
\caption{pHS probabilities obtained from implementing the \textit{identity} quantum circuit on \texttt{ibm\_torino} averaging the $i^{\rm th}$ and $(N+1-i)^{\rm th}$ neutrinos. The last column shows the theoretical value for the decoherence line, $d_n^{(id)}=3^7/4^8$. In the noiseless case, the probabilities in bold should be 1.}
\label{tab:identityDR_8neutrinos}
\end{table}

The single-flavor probability results obtained from running the \textit{identity} quantum circuits in the DR method, shown in Table~\ref{tab:identityDR_8neutrinos}, suggest a shift of the decoherence value $d_n$ in Eq.~\eqref{eq:DR}. 
After applying the identity operator, the decoherence line $d_n$ is expected to be the plateau value of the probability that decays due to noise sources, i.e., the initial state probability value goes from 1 to $d_n$, while the other states' probabilities go from $0$ to $d_n$. Therefore, in the ideal case, the state probability results never cross the decoherence line. Instead, Table~\ref{tab:identityDR_8neutrinos} reports that all the obtained probabilities are greater than the theoretical decoherence value $d_n^{(id)}$ (given by $3^7/4^8$).
A possible explanation is the simple depolarizing noise model, assumed when applying DR method, is not enough to describe the noise contributions, and different qubits are subjected to different noise sources. 
Moreover, a non-negligible contribution from relaxation process is observed, increasing the probability of being in the $|0\rangle$ state.

This implies that the decoherence value $d_n$ in Eq.~\eqref{eq:DR} should be empirically changed. Looking at the obtained values, the ``effective'' experimental decoherence value can be estimated to be in the range $d^{(eff)}_n\in[0.048,0.054]$. 
If DR is applied using $d^{(eff)}_n$, the flavor probabilities for the first, second and fourth neutrino are closer to the analytical curves, as shown in Fig.~\ref{fig:ibm_8_neutrinos_dn}.
Also noted was that running a deeper quantum circuit for two Trotter steps causes the empirical decoherence line to move closer to the theoretical decoherence line.

\subsubsection{Entropy and tomography calculations}

\begin{figure*}[th]
    \centering
    \includegraphics[width=\textwidth]{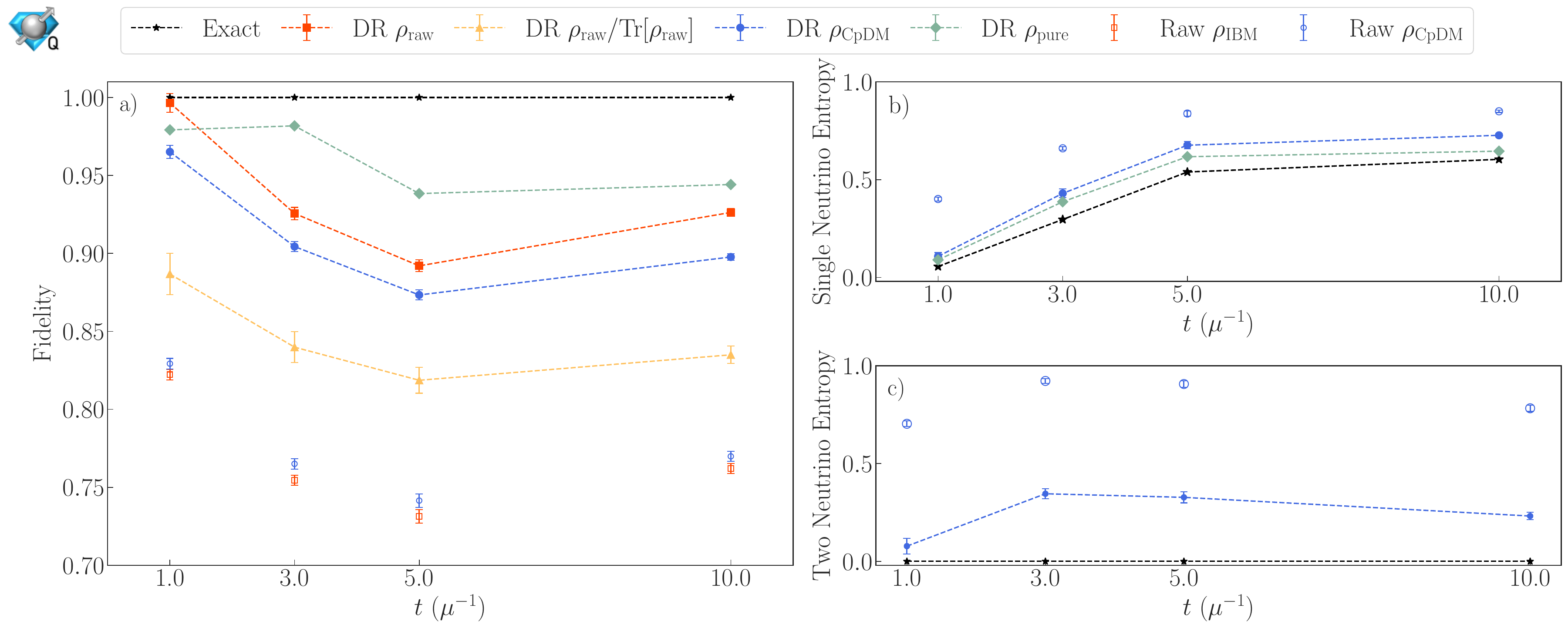}
    \caption{(a) Fidelity and (b) single-neutrino entropy for different $\Delta t$ for the two neutrino system starting from $\ket{e\mu}$ state. The different points correspond to different methods in computing the density matrix (see the main text for details). (c) Two-neutrino entropy computed from $\rho_{\rm CDM}$ density matrix. }
    \label{fig:fidelity}
\end{figure*}
To compute the entanglement entropy and other entanglement witnesses, the density matrix of $n\leq N$ neutrinos is evaluated. 
One approach uses classical shadows~\cite{Huang:2020tih}. 
Here, full state tomography is performed, as in Ref.~\cite{Hall:2021rbv}. 
Since not all $2^{2n}$ states are physical in the qubit mapping, the number of measurements needed to estimate the physical density matrix can be reduced.
For example, for a single neutrino the state-tomography operator pool can be reduced from 15 to 7 (independent) different operators that describe the Gell-Mann decomposition in the qutrit physical space.
The reduced density matrix for one neutrino is given by
\begin{equation}
\rho_\nu= \sum_{i=1}^9 c_i \lambda_i  \ ,
\label{eq:rho_1neut} 
\end{equation}
where $\lambda_9=\mathbb{I}$ and $c_{i}=\Tr(\lambda_i\: \rho)/\mathcal{A}_{i}$, with $\mathcal{A}_{i}=\Tr(\lambda^2_i)$. 
Table~\ref{tab:tomography_pool} in App.~\ref{app:tom} contains the explicit operator pool needed to extract each coefficient $c_i$, where the last column formulates the $c_i$ coefficient from the resulting measurement probabilities.
For a generic system of $N$ neutrinos, all combinations of the 7 operators should be implemented to obtain the corresponding density matrix.
Once these coefficients are fixed, the single-neutrino entanglement von Neumann entropy can be obtained using $S = -\Tr\left[ \rho_\nu \log(\rho_\nu)\right]$.

This three-flavor state tomography procedure is implemented on \texttt{ibm\_torino} for the two-neutrino system using a single Trotter time step. This requires running 49 different quantum circuits (that describe all possible independent combinations of Table~\ref{tab:tomography_pool}). Then, after applying the error mitigation methods, the coefficients $c_{ij}=\Tr(\lambda_i\otimes \lambda_j\: \rho)/\mathcal{A}_{ij}$ are evaluated, with $\mathcal{A}_{ij}=\Tr(\lambda^2_i\otimes \lambda^2_j)$, via the probabilities listed in Table~\ref{tab:tomography_pool}. 
The corresponding density matrix, labeled $\rho_{\rm IBM}$, is obtained in a similar manner to Eq.~\eqref{eq:rho_1neut}, 
\begin{equation}
\rho_{\rm IBM}= \sum_{i,j={1}}^9 c_{ij} \lambda_i \otimes \lambda_j \ .
\end{equation}

Due to hardware noise, $\rho_{\rm IBM}$ is generally not positive semi-definite, therefore it does not represent a physical density matrix. Here, the algorithm from Ref.~\cite{Acharya:2021byb} is applied to find the closest physical density matrix, labeled $\rho_{\rm CpDM}$, via a rescaling of the eigenvalues (more details in App.~\ref{app:tom}). Moreover, because in this case the density matrix of the whole system is computed, the final state is expected to be a pure quantum state. This can be enforced by using the eigenstate with the largest eigenvalue of $\rho_{\rm CpDM}$. This state will correspond to the closest pure quantum state (it has the highest contribution in the Schmidt decomposition), and is labeled as $\rho_{\rm pure}$.

The fidelity between the obtained $\rho_{\rm IBM}$, $\rho_{\rm CpDM}$, $\rho_{\rm pure}$ and the exact one, $\zeta=|\Psi(t)\rangle \langle \Psi(t)|$, is:
\begin{equation}
    F(\rho,\zeta)=\left({\rm Tr}\sqrt{\sqrt{\zeta}\rho\sqrt{\zeta}}\right)^2\, .
\end{equation}
Panel (a) of Fig.~\ref{fig:fidelity} shows the results. It is interesting to note that due to the error mitigation used here (the coefficients $c_{ij}$ are all normalized using the same $P_{id}$ quantity), the raw and DR $\rho_{\rm IBM}/{\rm Tr}(\rho_{\rm IBM})$ completely overlap, since the DR renormalization factor is cancelled out when enforcing the unity of the trace. Similarly, the $\rho_{\rm pure}$ does not depend on whether error mitigation is applied or not, since in this case the largest eigenvalue is the same.

Single neutrino entanglement entropy is also copmuted using the $\rho_{\rm CpDM}$ and $\rho_{\rm pure}$ density matrices,\footnote{The matrix $\rho_{\rm IBM}$ cannot be used since it can have negative eigenvalues.} shown in panel (b) of Fig.~\ref{fig:fidelity}. 
Results from $\rho_{\rm pure}$ (dark blue points) are observed to be closer to the exact entropy behavior than the results from $\rho_{\rm CpDM}$ (green triangles). 
As a further test on the fidelity, the two-neutrino entropy for $\rho_{\rm CpDM}$ is computed to diagnose the effect of noise, since the analytical two-body entropy remains zero. These results are reported in panel (c) of Fig.~\ref{fig:fidelity}, where while the fidelity is seen to be $>90\%$, the density matrix $\rho_{\rm CpDM}$ still exhibits features of a mixed state.

\section{Conclusions}

In this work, we introduce new quantum circuits for simulating the collective dynamics of three-flavor neutrinos on gate-based quantum computers, and also provide an implementation of the two-neutrino flavor-exchange operator on qutrit-based computers using 4 $CX$ gates.
The implementation of the same dynamics on qubit-based platforms is demonstrated, where each neutrino is mapped to two qubits. The corresponding circuits require at least 18 CNOT gates.
Additionally, we introduce a simplification that allows the realization of the required all-to-all connectivity in a linear chain without any additional computational costs (same circuit depth).

We have executed the qubit-based quantum circuits with up to eight neutrinos on the Quantinuum \texttt{H1-1} and IBM \texttt{ibm\_torino} devices, and computed the probabilities of finding each neutrino in a specific flavor state, as well as the initial state persistence. These are key observables used to study the thermalization and equilibration of such systems~\cite{Martin:2023gbo}.
For the smaller systems, with two and four neutrinos, the circuit depth is small enough that using multiple Trotter steps to perform time evolution is feasible.
For eight neutrinos, while only one Trotter step was used, the resulting Trotter errors were smaller than the statistical and systematic uncertainties from the device.
Hardware noise was corrected through various error mitigation techniques and post-selection procedures.
The quality of the results (after mitigation) are higher in the trapped-ion device than the superconducting one, though the different number of shots used in both hinders a direct comparison.

Using the IBM quantum computer, it was also possible to test the proposed partial state tomography, which required implementing 49 operators, allowing us to evaluate the full density matrix of two neutrinos and the entanglement entropy.

The algorithms needed to perform realistic simulations require quantum computers with longer coherence times. That is because one might need to start from a thermal state (and not a pure state)~\cite{Motta:2019yya,sagastizabal2021variational,turro2023quantum,Davoudi:2022uzo}, include the time-dependence in the two-neutrino term in the Hamiltonian~\cite{Rajput:2021khs,Siwach:2023wzy}, or include the effect of anti-neutrinos.

\section{Acknowledgements}

We would like to thank the IQuS group, especially Martin Savage, Xiaojun Yao, Saurabh Vasant Kadam, Niklas Mueller, and Henry Froland for useful discussions, as well as Pooja Siwach. Additionally, we thank Niklas Mueller and Henry Froland for sharing their code to implement the algorithm from Ref.~\cite{Acharya:2021byb}.

This work was supported, in part, by U.S. Department of Energy, Office of Science, Office of Nuclear Physics, InQubator for Quantum Simulation (IQuS)\footnote{\url{https://iqus.uw.edu/}} under DOE (NP) Award No.\ DE-SC0020970 via the program on Quantum Horizons: QIS Research and Innovation for Nuclear Science\footnote{\url{https://science.osti.gov/np/Research/Quantum-Information-Science}} (Turro, Bhaskar, Chernyshev), and the Quantum Science Center (QSC)\footnote{\url{https://qscience.org}} which is a National Quantum Information Science Research Center of the U.S.\ Department of Energy (DOE) (Illa).
This work is also supported, in part, through the Department of Physics\footnote{\url{https://phys.washington.edu}} and the College of Arts and Sciences\footnote{\url{https://www.artsci.washington.edu}} at the University of Washington.
This research used resources of the Oak Ridge Leadership Computing Facility (OLCF), which is a DOE Office of Science User Facility supported under Contract DE-AC05-00OR22725.
We acknowledge the use of IBM Quantum services for this work. The views expressed are those of the authors, and do not reflect the official policy or position of IBM or the IBM Quantum team.

\clearpage
\appendix

\section{Gell-Mann matrices}
\label{app:Gell_Mann}

Our notation for the Gell-Mann matrices is as follows,
\begin{align}
&\lambda_1=\begin{pmatrix}
    0 & 1 & 0\\
    1 & 0 & 0\\
    0 & 0 & 0\\
\end{pmatrix} , \
\lambda_2=\begin{pmatrix}
    0 & i & 0\\
    -i & 0 & 0\\
    0 & 0 & 0\\
\end{pmatrix} , \ 
\lambda_3=\begin{pmatrix}
    1 & 0 & 0\\
    0 & -1 & 0\\
    0 & 0 & 0\\
\end{pmatrix} , \nonumber \\
&\lambda_4=\begin{pmatrix}
    0 & 0 & 1\\
    0 & 0 & 0\\
    1 & 0 & 0\\
\end{pmatrix} ,  \
\lambda_5=\begin{pmatrix}
    0 & 0 & -i\\
    0 & 0 & 0\\
    i & 0 & 0\\
\end{pmatrix} , \nonumber    \\
&\lambda_6=\begin{pmatrix}
    0 & 0 & 0\\
    0 & 0 & 1\\
    0 & 1& 0\\
\end{pmatrix} ,  \  
\lambda_7=\begin{pmatrix}
    0 & 0 & 0\\
    0 & 0 & -i\\
    0 & i & 0\\
\end{pmatrix}  ,  \nonumber    \\
&\lambda_8=\sqrt{\frac{1}{3}}\,\begin{pmatrix}
    1 & 0 & 0\\
    0 & 1 & 0\\
    0 & 0 & -2\\
\end{pmatrix} .
\end{align}

\section{Qutrit gates}
\label{app:qutrit_gate}

This App. writes the explicit matrix representation of the qutrit gates.
The single-qutrit gates used in the circuits described in the main text are~\cite{Goss:2022bqd}
\begin{align}
X^{12}_{\alpha} & = \begin{pmatrix}
     1 & 0 & 0\\
     0 & \cos \tfrac{\alpha}{2} & -i\sin \tfrac{\alpha}{2}\\
     0 & -i\sin \tfrac{\alpha}{2} & \cos \tfrac{\alpha}{2}\\
 \end{pmatrix} \ ,  \\
 R_y^{01}(\alpha) & = \begin{pmatrix}
     \cos\tfrac{\alpha}{2} & -\sin\tfrac{\alpha}{2} & 0\\
     \sin\tfrac{\alpha}{2} & \cos\tfrac{\alpha}{2} & 0\\
     0 & 0 & 1\\
 \end{pmatrix} \ ,  \\
 R_y^{12}(\alpha) & = \begin{pmatrix}
     1 & 0 & 0\\
     0 & \cos \tfrac{\alpha}{2} & -\sin \tfrac{\alpha}{2}\\
     0 & \sin \tfrac{\alpha}{2} & \cos \tfrac{\alpha}{2}\\
 \end{pmatrix} \ ,  \\
  {\rm Ph}(\theta,\phi,\lambda) & = \begin{pmatrix}
    e^{i\theta} & 0 &0\\
    0 & e^{i\phi} &0\\
    0 & 0 &e^{i\lambda}\\
\end{pmatrix} \ ,\\
R_Z^{01}(\theta)&=  {\rm Ph}(-\frac{\theta}{2},\frac{\theta}{2},0)\ ,\\
R_Z^{12}(\phi)&= {\rm Ph}(0,-\frac{\phi}{2},\frac{\phi}{2})\ .
\end{align}
The two-qutrit $CX$ gate, whose action is given by $CX\ket{x,y}=\ket{x,{\rm mod}(x+y,3)}$, implements the following operation,
\begin{equation}
    CX = \begin{pmatrix}
1 & 0 & 0 & 0 & 0 & 0 & 0 & 0 & 0\\0 & 1 & 0 & 0 & 0 & 0 & 0 & 0 & 0\\0 & 0 & 1 & 0 & 0 & 0 & 0 & 0 & 0\\0 & 0 & 0 & 0 & 1 & 0 & 0 & 0 & 0\\0 & 0 & 0 & 0 & 0 & 1 & 0 & 0 & 0\\0 & 0 & 0 & 1 & 0 & 0 & 0 & 0 & 0\\0 & 0 & 0 & 0 & 0 & 0 & 0 & 0 & 1\\0 & 0 & 0 & 0 & 0 & 0 & 1 & 0 & 0\\0 & 0 & 0 & 0 & 0 & 0 & 0 & 1 & 0
    \end{pmatrix} \ .
\end{equation}
While the $CX$ gate might not be a native two-qutrit gate on current qutrit quantum devices, it is straightforward to transform to alternative entangling gates, like the $CZ$ gate~\cite{Goss:2022bqd},
\begin{equation}
    CZ = \sum_{i,j\in\{0,1,2\}}\tilde{\omega}^{ij}|ij\rangle \langle ij | \ , \quad  \tilde{\omega}= e^{i\frac{2\pi}{3}} \ .
\end{equation}
Using the qutrit Hadamard gate~\cite{Morvan:2021qju}, the $CX$ gate can transformed into a $CZ$ gate,
\begin{equation}
    CZ = (1 \otimes H^\dag) \cdot  CX \cdot  (1 \otimes H) \ , \quad H = \frac{1}{\sqrt{3}}\begin{pmatrix}
        1 & 1 & 1 \\
        1 & \tilde{\omega} & \tilde{\omega}^2 \\
        1 & \tilde{\omega}^2 & \tilde{\omega}
    \end{pmatrix} \ ,
\end{equation}
with a similar transformation for $CX^\dag$.

\section{Quantinuum emulator \label{app:4neutrinos}}

\begin{figure*}[ht]
\centering
\includegraphics[width=\textwidth]{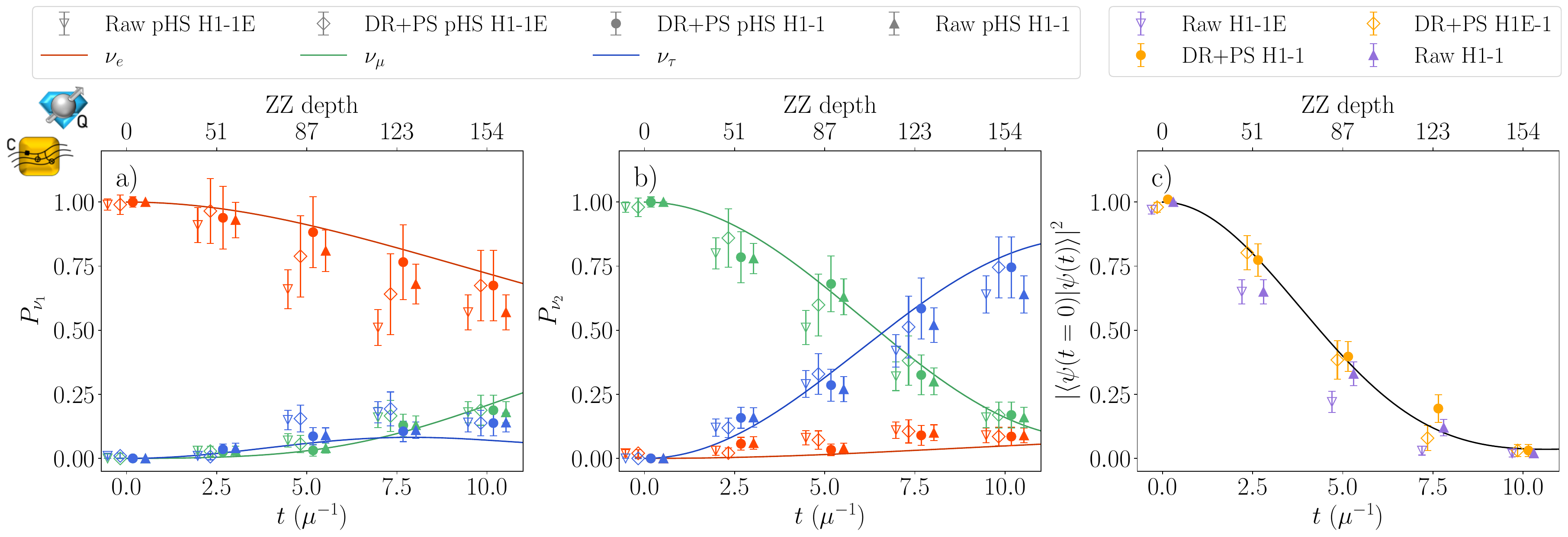}
\caption{Flavor evolution for four neutrinos as a function of time from the \texttt{H1-1E} emulator and \texttt{H1-1} device. Empty triangles and circles represent the emulator's raw and DR+PS results, respectively. Solid symbols show the corresponding results from \texttt{H1-1}. We use the same conventions as in Fig.~\ref{fig:quantinuum_4_neutrinos}.}

\label{fig:emulator_4_neutrinos}
\end{figure*}

Results from \texttt{H1-1} and its emulator for the four neutrino dynamics is reported in Fig.~\ref{fig:emulator_4_neutrinos}. DR and the pHS post-selecting procedure were implemented in both cases. 
The emulator's results (empty symbols) were observed to be compatible with the \texttt{H1-1} device's results (solid symbols).

\section{Simulations of eight neutrinos on IBM quantum computer\label{app:8ibm} }

\begin{figure*}[ht]
    \centering
    \includegraphics[width=\textwidth]{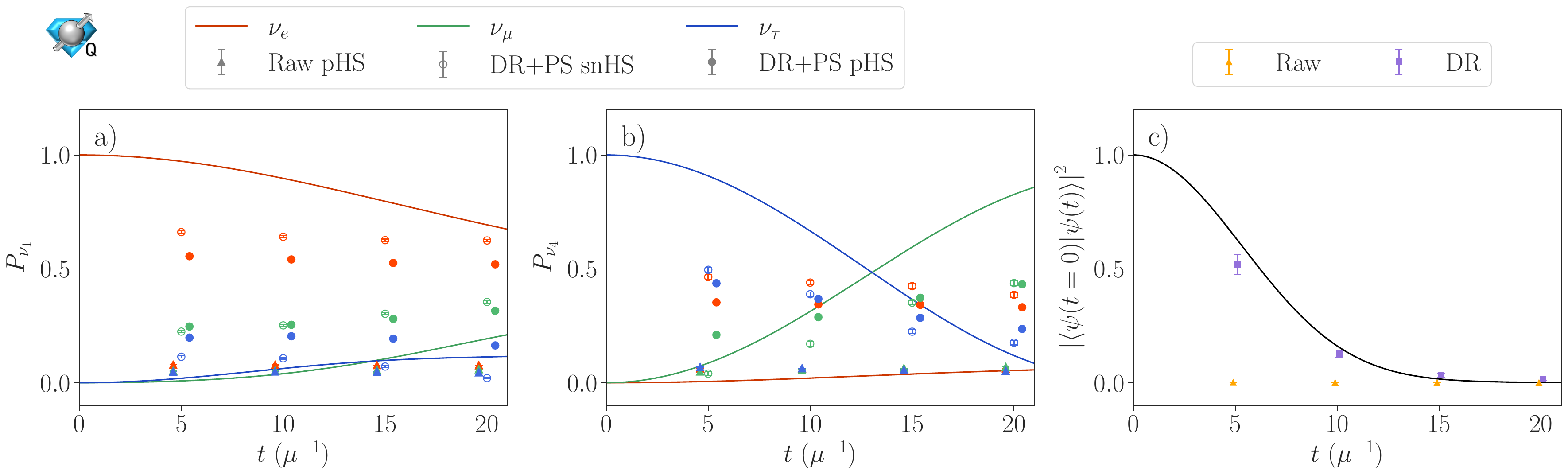}
    \caption{Flavor evolution for eight neutrinos as a function of time obtained from the \texttt{ibm\_torino} device, with $\ket{\nu_e \nu_\mu \nu_e \nu_\tau \nu_e \nu_\mu \nu_e \nu_\tau}$ as the initial state.  Panels (a) and (b) show the flavor evolution of the first and fourth neutrinos, respectively. 
    Panel (c) shows the persistence probability of the initial state.
    We use the same conventions as in Fig.~\ref{fig:quantinuum_4_neutrinos}.}
    \label{fig:enter-label}
\end{figure*}

Figure~\ref{fig:enter-label} shows the results from the evolution of eight neutrinos starting from the non-symmetric state $\ket{\nu_e \nu_\mu \nu_e \nu_\tau \nu_e \nu_\mu \nu_e \nu_\tau}$. 
Compared to the results in Fig.~\ref{fig:ibm_8_neutrinos}, the averaging procedure appears helpful for averaging out device errors.
A clear example is the electron flavor evolution for the fourth neutrino, displayed in panel (b).

\section{Details about the tomography study}
\label{app:tom}

This App. contains details on the operator pool used to reconstruct the density matrix. 
Each row in Table~\ref{tab:tomography_pool} shows how to evaluate the coefficient $c_i$ from Eq.~\eqref{eq:rho_1neut}.
The second column contains the operator needed to change the basis in which to measure the two qubits. By measuring the state probability $P_i$ with the expression in the third column, the value of $c_i$ is recovered. Note that the operators for $\lambda_3$, $\lambda_8$, and $\lambda_9$ are identity operators, thus the operator pool is composed of 7 independent operators (instead of 9). 

The goal of the algorithm from Ref.~\cite{Acharya:2021byb} is to find the closest positive semi-definite density matrix to the one obtained from {\tt ibm\_torino}. The general idea of the algorithm is to find the the density matrix $\rho_{\rm CpDM}$ that minimizes the trace distance with $\rho_{\rm IBM}$ while having all eigenvalues positive. This is done by shifting the eigenvalues of $\rho_{\rm IBM}$ using the algorithm of Ref.~\cite{Wang:2013}, while leaving the eigenvectors of $\rho_{\rm IBM}$ unmodified.

\begin{table}[ht]
    \centering
    \begin{tabularx}{\columnwidth}{|c|Y|Y|}
    \hline
    $\lambda_i$ & Operator & Probability \\
    \hline
    $\lambda_1$ & $\frac{1}{\sqrt{2}} \begin{pmatrix}
            1 & 1 & 0 &0\\
            1 & -1 & 0 &0\\
            0 & 0 & \sqrt{2}&0\\ 
            0 &  0 &0 &\sqrt{2}\\
        \end{pmatrix}$ & $c_1= \frac{1}{2}(P_{00}-P_{01})$\\
    $\lambda_2$ & $\frac{1}{\sqrt{2}} \begin{pmatrix}
            1 & -i & 0 &0\\
            1 & i& 0 &0\\
            0 & 0 & \sqrt{2}&0\\   
            0 &  0 &0 &\sqrt{2}\\
        \end{pmatrix}$ & $c_2= \frac{1}{2}(P_{00}-P_{01})$\\
    $\lambda_3$ & $\begin{pmatrix}
            1 & 0 & 0 &0\\
           0 & 1 & 0 &0\\
            0 & 0 &1&0\\  
            0 &  0 &0 &1\\
        \end{pmatrix}$ & $c_3= \frac{1}{2}(P_{00}-P_{01})$\\            
    $\lambda_4$ & $\frac{1}{\sqrt{2}} \begin{pmatrix}
            1 & 0 & 1 &0\\
            0 & \sqrt{2} & 0 &0\\
            1 & 0 & -1&0\\
            0 &  0 &0 &\sqrt{2}\\
        \end{pmatrix}$ & $c_4=\frac{1}{2}(P_{00}-P_{10})$\\        
    $\lambda_5$ & $\frac{1}{\sqrt{2}} \begin{pmatrix}
            1 & 0 & -i &0\\
            0 & \sqrt{2} & 0 &0\\
            1 & 0 & i&0\\  
            0 &  0 &0 &\sqrt{2}\\
        \end{pmatrix}$ & $c_5=\frac{1}{2}(P_{00}-P_{10})$\\
    $\lambda_6$ & $\frac{1}{\sqrt{2}} \begin{pmatrix}
           \sqrt{2} & 0 & 0 &0\\
            0 & 1 & 1 &0\\
            0 & 1 & -1&0\\ 
            0 &  0 &0 &\sqrt{2}\\
        \end{pmatrix}$ & $c_6=\frac{1}{2}(P_{01}-P_{10})$\\         
    $\lambda_7$ & $\frac{1}{\sqrt{2}} \begin{pmatrix}
           \sqrt{2} & 0 & 0 &0\\
            0 & 1 & -i&0\\
            0 & 1 & i&0\\
            0 & 0 &0 &\sqrt{2}\\
        \end{pmatrix}$ & $c_7=\frac{1}{2}(P_{01}-P_{10})$\\ 
    $\lambda_8$ & $ \begin{pmatrix}
           1 & 0 & 0 &0\\
            0 & 1 &0 &0\\
            0 & 0 & 1&0\\
            0 &  0 &0 &1\\
        \end{pmatrix}$ & $c_8=\frac{1}{2\sqrt{3}}(P_{00}+P_{01}-2\,P_{10})$\\ 
     $\lambda_9$ & $ \begin{pmatrix}
           1 & 0 & 0 &0\\
            0 & 1 &0 &0\\
            0 & 0 & 1&0\\
            0 &  0 &0 &1\\
        \end{pmatrix}$ & $c_9=\frac{1}{3}(P_{00}+P_{01}+P_{10})$\\ \hline
    \end{tabularx}
    \caption{Tomographic pool for computing $c_i=\Tr (\rho \lambda_i)/\mathcal{A}_i$, where $\lambda_i$ represents the Gell-Mann matrix. We perform the change of basis by implementing the operator shown in the second column. The coefficient $c_i$ are then given by the linear combination of the obtained probabilities $P_{ij}=\{ P_{00},P_{01},P_{10},P_{11}\}$ given in the third column.
    }
    \label{tab:tomography_pool}
\end{table}

\section{Device parameters\label{app:device_parameters}}

In this appendix we report the experimental parameters of the quantum computers used in this paper.

In Table~\ref{tab:quantinuum_parameters}, we report the Quantinuum \texttt{H1-1} device parameters, and in Table~\ref{tab:ibm_parameters}, we report the IBM \texttt{ibm\_torino} device parameters.

\begin{table}[ht]
\centering
\begin{tabular}{|c|c|}
\hline
Total number of qubits & 20  \\
Typical Single-qubit gate  infidelity & $2 \cdot 10^{-5}$ \\
Typical Two-qubit gate  infidelity & $1 \cdot 10^{-3}$ \\
SPAM error &  $2 \cdot 10^{-4}$\\
\hline
\end{tabular}
\caption{Quantinuum H1-1 device parameters, as reported in Ref.~\cite{quantinuum}.}
\label{tab:quantinuum_parameters}
\end{table}

\begin{table}[ht]
\centering
\begin{tabular}{|c|c|c|c|}
\hline
Total number of qubits & \multicolumn{3}{c|}{133} \\ \hline
Neutrinos & 2 & 4 & 8\\
Date accessed & 6/12/24 & 6/21/24 & 6/25/14 \\
\hline

Number of qubits used & 4 & 8 & 16 \\
Median T1 coherence time ($\mu$s) & 150 & 133 & 142 \\
Median T2 coherence time ($\mu$s) & 147 & 127 & 151 \\
Median $X$-gate error & $3.2 \cdot 10^{-4}$ & $3.3 \cdot 10^{-4}$ & $2.8 \cdot 10^{-4}$ \\
Median $CZ$-gate error & $9.4 \cdot 10^{-3}$ & $7.8 \cdot 10^{-3}$ & $4.0\cdot 10^{-3}$ \\
Median readout error &  $2.6 \cdot 10^{-2}$ &  $2.9 \cdot 10^{-2}$ &  $2.3 \cdot 10^{-2}$\\
\hline

\end{tabular}
\caption{IBM \texttt{ibm\_torino} device parameters.}
\label{tab:ibm_parameters}
\end{table}

\bibliography{bib}

%apsrev4-2.bst 2019-01-14 (MD) hand-edited version of apsrev4-1.bst
%Control: key (0)
%Control: author (8) initials jnrlst
%Control: editor formatted (1) identically to author
%Control: production of article title (0) allowed
%Control: page (0) single
%Control: year (1) truncated
%Control: production of eprint (0) enabled
\begin{thebibliography}{183}%
\makeatletter
\providecommand \@ifxundefined [1]{%
 \@ifx{#1\undefined}
}%
\providecommand \@ifnum [1]{%
 \ifnum #1\expandafter \@firstoftwo
 \else \expandafter \@secondoftwo
 \fi
}%
\providecommand \@ifx [1]{%
 \ifx #1\expandafter \@firstoftwo
 \else \expandafter \@secondoftwo
 \fi
}%
\providecommand \natexlab [1]{#1}%
\providecommand \enquote  [1]{``#1''}%
\providecommand \bibnamefont  [1]{#1}%
\providecommand \bibfnamefont [1]{#1}%
\providecommand \citenamefont [1]{#1}%
\providecommand \href@noop [0]{\@secondoftwo}%
\providecommand \href [0]{\begingroup \@sanitize@url \@href}%
\providecommand \@href[1]{\@@startlink{#1}\@@href}%
\providecommand \@@href[1]{\endgroup#1\@@endlink}%
\providecommand \@sanitize@url [0]{\catcode `\\12\catcode `\$12\catcode
  `\&12\catcode `\#12\catcode `\^12\catcode `\_12\catcode `\%12\relax}%
\providecommand \@@startlink[1]{}%
\providecommand \@@endlink[0]{}%
\providecommand \url  [0]{\begingroup\@sanitize@url \@url }%
\providecommand \@url [1]{\endgroup\@href {#1}{\urlprefix }}%
\providecommand \urlprefix  [0]{URL }%
\providecommand \Eprint [0]{\href }%
\providecommand \doibase [0]{https://doi.org/}%
\providecommand \selectlanguage [0]{\@gobble}%
\providecommand \bibinfo  [0]{\@secondoftwo}%
\providecommand \bibfield  [0]{\@secondoftwo}%
\providecommand \translation [1]{[#1]}%
\providecommand \BibitemOpen [0]{}%
\providecommand \bibitemStop [0]{}%
\providecommand \bibitemNoStop [0]{.\EOS\space}%
\providecommand \EOS [0]{\spacefactor3000\relax}%
\providecommand \BibitemShut  [1]{\csname bibitem#1\endcsname}%
\let\auto@bib@innerbib\@empty
%</preamble>
\bibitem [{\citenamefont {Gottesman}(1999)}]{Gottesman:1998se}%
  \BibitemOpen
  \bibfield  {author} {\bibinfo {author} {\bibfnamefont {D.}~\bibnamefont
  {Gottesman}},\ }\bibfield  {title} {\bibinfo {title} {{Fault tolerant quantum
  computation with higher dimensional systems}},\ }\href
  {https://doi.org/10.1016/S0960-0779(98)00218-5} {\bibfield  {journal}
  {\bibinfo  {journal} {Chaos Solitons Fractals}\ }\textbf {\bibinfo {volume}
  {10}},\ \bibinfo {pages} {1749} (\bibinfo {year} {1999})},\ \Eprint
  {https://arxiv.org/abs/quant-ph/9802007} {arXiv:quant-ph/9802007}
  \BibitemShut {NoStop}%
\bibitem [{\citenamefont {Ciavarella}\ \emph {et~al.}(2021)\citenamefont
  {Ciavarella}, \citenamefont {Klco},\ and\ \citenamefont
  {Savage}}]{Ciavarella:2021nmj}%
  \BibitemOpen
  \bibfield  {author} {\bibinfo {author} {\bibfnamefont {A.}~\bibnamefont
  {Ciavarella}}, \bibinfo {author} {\bibfnamefont {N.}~\bibnamefont {Klco}},\
  and\ \bibinfo {author} {\bibfnamefont {M.~J.}\ \bibnamefont {Savage}},\
  }\bibfield  {title} {\bibinfo {title} {{Trailhead for quantum simulation of
  SU(3) Yang-Mills lattice gauge theory in the local multiplet basis}},\ }\href
  {https://doi.org/10.1103/PhysRevD.103.094501} {\bibfield  {journal} {\bibinfo
   {journal} {Phys. Rev. D}\ }\textbf {\bibinfo {volume} {103}},\ \bibinfo
  {pages} {094501} (\bibinfo {year} {2021})},\ \Eprint
  {https://arxiv.org/abs/2101.10227} {arXiv:2101.10227 [quant-ph]} \BibitemShut
  {NoStop}%
\bibitem [{\citenamefont {Gustafson}(2021)}]{Gustafson:2021qbt}%
  \BibitemOpen
  \bibfield  {author} {\bibinfo {author} {\bibfnamefont {E.~J.}\ \bibnamefont
  {Gustafson}},\ }\bibfield  {title} {\bibinfo {title} {Prospects for
  simulating a qudit-based model of $(1+1)\mathrm{D}$ scalar qed},\ }\href
  {https://doi.org/10.1103/PhysRevD.103.114505} {\bibfield  {journal} {\bibinfo
   {journal} {Phys. Rev. D}\ }\textbf {\bibinfo {volume} {103}},\ \bibinfo
  {pages} {114505} (\bibinfo {year} {2021})}\BibitemShut {NoStop}%
\bibitem [{\citenamefont {Calixto}\ \emph {et~al.}(2021)\citenamefont
  {Calixto}, \citenamefont {Mayorgas},\ and\ \citenamefont
  {Guerrero}}]{Calixto_2021}%
  \BibitemOpen
  \bibfield  {author} {\bibinfo {author} {\bibfnamefont {M.}~\bibnamefont
  {Calixto}}, \bibinfo {author} {\bibfnamefont {A.}~\bibnamefont {Mayorgas}},\
  and\ \bibinfo {author} {\bibfnamefont {J.}~\bibnamefont {Guerrero}},\
  }\bibfield  {title} {\bibinfo {title} {{Entanglement and {U(D)}-spin
  squeezing in symmetric multi-{quDit} systems and applications to quantum
  phase transitions in Lipkin{\textendash}Meshkov{\textendash}Glick {D}-level
  atom models}},\ }\href {https://doi.org/10.1007/s11128-021-03218-6}
  {\bibfield  {journal} {\bibinfo  {journal} {Quantum Inf. Process.}\ }\textbf
  {\bibinfo {volume} {20}},\ \bibinfo {pages} {304} (\bibinfo {year} {2021})},\
  \Eprint {https://arxiv.org/abs/2104.10581} {arXiv:2104.10581 [quant-ph]}
  \BibitemShut {NoStop}%
\bibitem [{\citenamefont {Gustafson}(2022)}]{Gustafson:2022xlj}%
  \BibitemOpen
  \bibfield  {author} {\bibinfo {author} {\bibfnamefont {E.}~\bibnamefont
  {Gustafson}},\ }\href@noop {} {\bibinfo {title} {{Noise Improvements in
  Quantum Simulations of sQED using Qutrits}}} (\bibinfo {year} {2022}),\
  \Eprint {https://arxiv.org/abs/2201.04546} {arXiv:2201.04546 [quant-ph]}
  \BibitemShut {NoStop}%
\bibitem [{\citenamefont {Gonz\'alez-Cuadra}\ \emph {et~al.}(2022)\citenamefont
  {Gonz\'alez-Cuadra}, \citenamefont {Zache}, \citenamefont {Carrasco},
  \citenamefont {Kraus},\ and\ \citenamefont
  {Zoller}}]{Gonzalez-Cuadra:2022hxt}%
  \BibitemOpen
  \bibfield  {author} {\bibinfo {author} {\bibfnamefont {D.}~\bibnamefont
  {Gonz\'alez-Cuadra}}, \bibinfo {author} {\bibfnamefont {T.~V.}\ \bibnamefont
  {Zache}}, \bibinfo {author} {\bibfnamefont {J.}~\bibnamefont {Carrasco}},
  \bibinfo {author} {\bibfnamefont {B.}~\bibnamefont {Kraus}},\ and\ \bibinfo
  {author} {\bibfnamefont {P.}~\bibnamefont {Zoller}},\ }\bibfield  {title}
  {\bibinfo {title} {{Hardware Efficient Quantum Simulation of Non-Abelian
  Gauge Theories with Qudits on Rydberg Platforms}},\ }\href
  {https://doi.org/10.1103/PhysRevLett.129.160501} {\bibfield  {journal}
  {\bibinfo  {journal} {Phys. Rev. Lett.}\ }\textbf {\bibinfo {volume} {129}},\
  \bibinfo {pages} {160501} (\bibinfo {year} {2022})},\ \Eprint
  {https://arxiv.org/abs/2203.15541} {arXiv:2203.15541 [quant-ph]} \BibitemShut
  {NoStop}%
\bibitem [{\citenamefont {Gustafson}\ \emph {et~al.}(2022)\citenamefont
  {Gustafson}, \citenamefont {Lamm}, \citenamefont {Lovelace},\ and\
  \citenamefont {Musk}}]{Gustafson:2022xdt}%
  \BibitemOpen
  \bibfield  {author} {\bibinfo {author} {\bibfnamefont {E.~J.}\ \bibnamefont
  {Gustafson}}, \bibinfo {author} {\bibfnamefont {H.}~\bibnamefont {Lamm}},
  \bibinfo {author} {\bibfnamefont {F.}~\bibnamefont {Lovelace}},\ and\
  \bibinfo {author} {\bibfnamefont {D.}~\bibnamefont {Musk}},\ }\bibfield
  {title} {\bibinfo {title} {{Primitive quantum gates for an SU(2) discrete
  subgroup: Binary tetrahedral}},\ }\href
  {https://doi.org/10.1103/PhysRevD.106.114501} {\bibfield  {journal} {\bibinfo
   {journal} {Phys. Rev. D}\ }\textbf {\bibinfo {volume} {106}},\ \bibinfo
  {pages} {114501} (\bibinfo {year} {2022})},\ \Eprint
  {https://arxiv.org/abs/2208.12309} {arXiv:2208.12309 [quant-ph]} \BibitemShut
  {NoStop}%
\bibitem [{\citenamefont {Gustafson}\ and\ \citenamefont
  {Lamm}(2023)}]{Gustafson:2023swx}%
  \BibitemOpen
  \bibfield  {author} {\bibinfo {author} {\bibfnamefont {E.~J.}\ \bibnamefont
  {Gustafson}}\ and\ \bibinfo {author} {\bibfnamefont {H.}~\bibnamefont
  {Lamm}},\ }\href@noop {} {\bibinfo {title} {{Robustness of Gauge Digitization
  to Quantum Noise}}} (\bibinfo {year} {2023}),\ \Eprint
  {https://arxiv.org/abs/2301.10207} {arXiv:2301.10207 [hep-lat]} \BibitemShut
  {NoStop}%
\bibitem [{\citenamefont {Zache}\ \emph {et~al.}(2023)\citenamefont {Zache},
  \citenamefont {Gonz\'alez-Cuadra},\ and\ \citenamefont
  {Zoller}}]{Zache:2023cfj}%
  \BibitemOpen
  \bibfield  {author} {\bibinfo {author} {\bibfnamefont {T.~V.}\ \bibnamefont
  {Zache}}, \bibinfo {author} {\bibfnamefont {D.}~\bibnamefont
  {Gonz\'alez-Cuadra}},\ and\ \bibinfo {author} {\bibfnamefont
  {P.}~\bibnamefont {Zoller}},\ }\bibfield  {title} {\bibinfo {title}
  {{Fermion-qudit quantum processors for simulating lattice gauge theories with
  matter}},\ }\href {https://doi.org/10.22331/q-2023-10-16-1140} {\bibfield
  {journal} {\bibinfo  {journal} {Quantum}\ }\textbf {\bibinfo {volume} {7}},\
  \bibinfo {pages} {1140} (\bibinfo {year} {2023})},\ \Eprint
  {https://arxiv.org/abs/2303.08683} {arXiv:2303.08683 [quant-ph]} \BibitemShut
  {NoStop}%
\bibitem [{\citenamefont {Illa}\ \emph {et~al.}(2023)\citenamefont {Illa},
  \citenamefont {Robin},\ and\ \citenamefont {Savage}}]{Illa:2023scc}%
  \BibitemOpen
  \bibfield  {author} {\bibinfo {author} {\bibfnamefont {M.}~\bibnamefont
  {Illa}}, \bibinfo {author} {\bibfnamefont {C.~E.~P.}\ \bibnamefont {Robin}},\
  and\ \bibinfo {author} {\bibfnamefont {M.~J.}\ \bibnamefont {Savage}},\
  }\bibfield  {title} {\bibinfo {title} {{Quantum simulations of SO(5)
  many-fermion systems using qudits}},\ }\href
  {https://doi.org/10.1103/PhysRevC.108.064306} {\bibfield  {journal} {\bibinfo
   {journal} {Phys. Rev. C}\ }\textbf {\bibinfo {volume} {108}},\ \bibinfo
  {pages} {064306} (\bibinfo {year} {2023})},\ \Eprint
  {https://arxiv.org/abs/2305.11941} {arXiv:2305.11941 [quant-ph]} \BibitemShut
  {NoStop}%
\bibitem [{\citenamefont {Popov}\ \emph {et~al.}(2024)\citenamefont {Popov},
  \citenamefont {Meth}, \citenamefont {Lewenstein}, \citenamefont {Hauke},
  \citenamefont {Ringbauer}, \citenamefont {Zohar},\ and\ \citenamefont
  {Kasper}}]{Popov:2023xft}%
  \BibitemOpen
  \bibfield  {author} {\bibinfo {author} {\bibfnamefont {P.~P.}\ \bibnamefont
  {Popov}}, \bibinfo {author} {\bibfnamefont {M.}~\bibnamefont {Meth}},
  \bibinfo {author} {\bibfnamefont {M.}~\bibnamefont {Lewenstein}}, \bibinfo
  {author} {\bibfnamefont {P.}~\bibnamefont {Hauke}}, \bibinfo {author}
  {\bibfnamefont {M.}~\bibnamefont {Ringbauer}}, \bibinfo {author}
  {\bibfnamefont {E.}~\bibnamefont {Zohar}},\ and\ \bibinfo {author}
  {\bibfnamefont {V.}~\bibnamefont {Kasper}},\ }\bibfield  {title} {\bibinfo
  {title} {{Variational quantum simulation of U(1) lattice gauge theories with
  qudit systems}},\ }\href {https://doi.org/10.1103/PhysRevResearch.6.013202}
  {\bibfield  {journal} {\bibinfo  {journal} {Phys. Rev. Res.}\ }\textbf
  {\bibinfo {volume} {6}},\ \bibinfo {pages} {013202} (\bibinfo {year}
  {2024})},\ \Eprint {https://arxiv.org/abs/2307.15173} {arXiv:2307.15173
  [quant-ph]} \BibitemShut {NoStop}%
\bibitem [{\citenamefont {Meth}\ \emph {et~al.}(2023)\citenamefont {Meth} \emph
  {et~al.}}]{Meth:2023wzd}%
  \BibitemOpen
  \bibfield  {author} {\bibinfo {author} {\bibfnamefont {M.}~\bibnamefont
  {Meth}} \emph {et~al.},\ }\href@noop {} {\bibinfo {title} {{Simulating 2D
  lattice gauge theories on a qudit quantum computer}}} (\bibinfo {year}
  {2023}),\ \Eprint {https://arxiv.org/abs/2310.12110} {arXiv:2310.12110
  [quant-ph]} \BibitemShut {NoStop}%
\bibitem [{\citenamefont {Calaj\`o}\ \emph {et~al.}(2024)\citenamefont
  {Calaj\`o}, \citenamefont {Magnifico}, \citenamefont {Edmunds}, \citenamefont
  {Ringbauer}, \citenamefont {Montangero},\ and\ \citenamefont
  {Silvi}}]{Calajo:2024qrc}%
  \BibitemOpen
  \bibfield  {author} {\bibinfo {author} {\bibfnamefont {G.}~\bibnamefont
  {Calaj\`o}}, \bibinfo {author} {\bibfnamefont {G.}~\bibnamefont {Magnifico}},
  \bibinfo {author} {\bibfnamefont {C.}~\bibnamefont {Edmunds}}, \bibinfo
  {author} {\bibfnamefont {M.}~\bibnamefont {Ringbauer}}, \bibinfo {author}
  {\bibfnamefont {S.}~\bibnamefont {Montangero}},\ and\ \bibinfo {author}
  {\bibfnamefont {P.}~\bibnamefont {Silvi}},\ }\href@noop {} {\bibinfo {title}
  {{Digital quantum simulation of a (1+1)D SU(2) lattice gauge theory with ion
  qudits}}} (\bibinfo {year} {2024}),\ \Eprint
  {https://arxiv.org/abs/2402.07987} {arXiv:2402.07987 [quant-ph]} \BibitemShut
  {NoStop}%
\bibitem [{\citenamefont {Carena}\ \emph {et~al.}(2024)\citenamefont {Carena},
  \citenamefont {Lamm}, \citenamefont {Li},\ and\ \citenamefont
  {Liu}}]{Carena:2024dzu}%
  \BibitemOpen
  \bibfield  {author} {\bibinfo {author} {\bibfnamefont {M.}~\bibnamefont
  {Carena}}, \bibinfo {author} {\bibfnamefont {H.}~\bibnamefont {Lamm}},
  \bibinfo {author} {\bibfnamefont {Y.-Y.}\ \bibnamefont {Li}},\ and\ \bibinfo
  {author} {\bibfnamefont {W.}~\bibnamefont {Liu}},\ }\href@noop {} {\bibinfo
  {title} {{Quantum error thresholds for gauge-redundant digitizations of
  lattice field theories}}} (\bibinfo {year} {2024}),\ \Eprint
  {https://arxiv.org/abs/2402.16780} {arXiv:2402.16780 [hep-lat]} \BibitemShut
  {NoStop}%
\bibitem [{\citenamefont {Illa}\ \emph {et~al.}(2024)\citenamefont {Illa},
  \citenamefont {Robin},\ and\ \citenamefont {Savage}}]{Illa:2024kmf}%
  \BibitemOpen
  \bibfield  {author} {\bibinfo {author} {\bibfnamefont {M.}~\bibnamefont
  {Illa}}, \bibinfo {author} {\bibfnamefont {C.~E.~P.}\ \bibnamefont {Robin}},\
  and\ \bibinfo {author} {\bibfnamefont {M.~J.}\ \bibnamefont {Savage}},\
  }\href@noop {} {\bibinfo {title} {{Qu8its for Quantum Simulations of Lattice
  Quantum Chromodynamics}}} (\bibinfo {year} {2024}),\ \Eprint
  {https://arxiv.org/abs/2403.14537} {arXiv:2403.14537 [quant-ph]} \BibitemShut
  {NoStop}%
\bibitem [{\citenamefont {Gustafson}\ \emph {et~al.}(2024)\citenamefont
  {Gustafson}, \citenamefont {Ji}, \citenamefont {Lamm}, \citenamefont
  {Murairi},\ and\ \citenamefont {Zhu}}]{Gustafson:2024kym}%
  \BibitemOpen
  \bibfield  {author} {\bibinfo {author} {\bibfnamefont {E.~J.}\ \bibnamefont
  {Gustafson}}, \bibinfo {author} {\bibfnamefont {Y.}~\bibnamefont {Ji}},
  \bibinfo {author} {\bibfnamefont {H.}~\bibnamefont {Lamm}}, \bibinfo {author}
  {\bibfnamefont {E.~M.}\ \bibnamefont {Murairi}},\ and\ \bibinfo {author}
  {\bibfnamefont {S.}~\bibnamefont {Zhu}},\ }\href@noop {} {\bibinfo {title}
  {{Primitive Quantum Gates for an SU(3) Discrete Subgroup:
  $\Sigma(36\times3)$}}} (\bibinfo {year} {2024}),\ \Eprint
  {https://arxiv.org/abs/2405.05973} {arXiv:2405.05973 [hep-lat]} \BibitemShut
  {NoStop}%
\bibitem [{\citenamefont {Low}\ \emph {et~al.}(2020)\citenamefont {Low},
  \citenamefont {White}, \citenamefont {Cox}, \citenamefont {Day},\ and\
  \citenamefont {Senko}}]{Low:2019}%
  \BibitemOpen
  \bibfield  {author} {\bibinfo {author} {\bibfnamefont {P.~J.}\ \bibnamefont
  {Low}}, \bibinfo {author} {\bibfnamefont {B.~M.}\ \bibnamefont {White}},
  \bibinfo {author} {\bibfnamefont {A.~A.}\ \bibnamefont {Cox}}, \bibinfo
  {author} {\bibfnamefont {M.~L.}\ \bibnamefont {Day}},\ and\ \bibinfo {author}
  {\bibfnamefont {C.}~\bibnamefont {Senko}},\ }\bibfield  {title} {\bibinfo
  {title} {Practical trapped-ion protocols for universal qudit-based quantum
  computing},\ }\href {https://doi.org/10.1103/PhysRevResearch.2.033128}
  {\bibfield  {journal} {\bibinfo  {journal} {Phys. Rev. Res.}\ }\textbf
  {\bibinfo {volume} {2}},\ \bibinfo {pages} {033128} (\bibinfo {year}
  {2020})}\BibitemShut {NoStop}%
\bibitem [{\citenamefont {Ringbauer}\ \emph {et~al.}(2022)\citenamefont
  {Ringbauer}, \citenamefont {Meth}, \citenamefont {Postler}, \citenamefont
  {Stricker}, \citenamefont {Blatt}, \citenamefont {Schindler},\ and\
  \citenamefont {Monz}}]{Ringbauer:2021lhi}%
  \BibitemOpen
  \bibfield  {author} {\bibinfo {author} {\bibfnamefont {M.}~\bibnamefont
  {Ringbauer}}, \bibinfo {author} {\bibfnamefont {M.}~\bibnamefont {Meth}},
  \bibinfo {author} {\bibfnamefont {L.}~\bibnamefont {Postler}}, \bibinfo
  {author} {\bibfnamefont {R.}~\bibnamefont {Stricker}}, \bibinfo {author}
  {\bibfnamefont {R.}~\bibnamefont {Blatt}}, \bibinfo {author} {\bibfnamefont
  {P.}~\bibnamefont {Schindler}},\ and\ \bibinfo {author} {\bibfnamefont
  {T.}~\bibnamefont {Monz}},\ }\bibfield  {title} {\bibinfo {title} {{A
  universal qudit quantum processor with trapped ions}},\ }\href
  {https://doi.org/10.1038/s41567-022-01658-0} {\bibfield  {journal} {\bibinfo
  {journal} {Nature Phys.}\ }\textbf {\bibinfo {volume} {18}},\ \bibinfo
  {pages} {1053} (\bibinfo {year} {2022})},\ \Eprint
  {https://arxiv.org/abs/2109.06903} {arXiv:2109.06903 [quant-ph]} \BibitemShut
  {NoStop}%
\bibitem [{\citenamefont {Low}\ \emph {et~al.}(2023)\citenamefont {Low},
  \citenamefont {White},\ and\ \citenamefont {Senko}}]{Low:2023dlg}%
  \BibitemOpen
  \bibfield  {author} {\bibinfo {author} {\bibfnamefont {P.~J.}\ \bibnamefont
  {Low}}, \bibinfo {author} {\bibfnamefont {B.}~\bibnamefont {White}},\ and\
  \bibinfo {author} {\bibfnamefont {C.}~\bibnamefont {Senko}},\ }\href@noop {}
  {\bibinfo {title} {{Control and Readout of a 13-level Trapped Ion Qudit}}}
  (\bibinfo {year} {2023}),\ \Eprint {https://arxiv.org/abs/2306.03340}
  {arXiv:2306.03340 [quant-ph]} \BibitemShut {NoStop}%
\bibitem [{\citenamefont {Zalivako}\ \emph {et~al.}(2024)\citenamefont
  {Zalivako} \emph {et~al.}}]{Zalivako:2024bjm}%
  \BibitemOpen
  \bibfield  {author} {\bibinfo {author} {\bibfnamefont {I.~V.}\ \bibnamefont
  {Zalivako}} \emph {et~al.},\ }\href@noop {} {\bibinfo {title} {{Towards
  multiqudit quantum processor based on a $^{171}$Yb$^{+}$ ion string:
  Realizing basic quantum algorithms}}} (\bibinfo {year} {2024}),\ \Eprint
  {https://arxiv.org/abs/2402.03121} {arXiv:2402.03121 [quant-ph]} \BibitemShut
  {NoStop}%
\bibitem [{\citenamefont {Nikolaeva}\ \emph {et~al.}(2024)\citenamefont
  {Nikolaeva} \emph {et~al.}}]{Nikolaeva:2024wxl}%
  \BibitemOpen
  \bibfield  {author} {\bibinfo {author} {\bibfnamefont {A.~S.}\ \bibnamefont
  {Nikolaeva}} \emph {et~al.},\ }\href@noop {} {\bibinfo {title} {{Scalable
  improvement of the generalized Toffoli gate realization using
  trapped-ion-based qutrits}}} (\bibinfo {year} {2024}),\ \Eprint
  {https://arxiv.org/abs/2407.07758} {arXiv:2407.07758 [quant-ph]} \BibitemShut
  {NoStop}%
\bibitem [{\citenamefont {Blok}\ \emph {et~al.}(2021)\citenamefont {Blok},
  \citenamefont {Ramasesh}, \citenamefont {Schuster}, \citenamefont {O'Brien},
  \citenamefont {Kreikebaum}, \citenamefont {Dahlen}, \citenamefont {Morvan},
  \citenamefont {Yoshida}, \citenamefont {Yao},\ and\ \citenamefont
  {Siddiqi}}]{Blok:2020may}%
  \BibitemOpen
  \bibfield  {author} {\bibinfo {author} {\bibfnamefont {M.~S.}\ \bibnamefont
  {Blok}}, \bibinfo {author} {\bibfnamefont {V.~V.}\ \bibnamefont {Ramasesh}},
  \bibinfo {author} {\bibfnamefont {T.}~\bibnamefont {Schuster}}, \bibinfo
  {author} {\bibfnamefont {K.}~\bibnamefont {O'Brien}}, \bibinfo {author}
  {\bibfnamefont {J.~M.}\ \bibnamefont {Kreikebaum}}, \bibinfo {author}
  {\bibfnamefont {D.}~\bibnamefont {Dahlen}}, \bibinfo {author} {\bibfnamefont
  {A.}~\bibnamefont {Morvan}}, \bibinfo {author} {\bibfnamefont
  {B.}~\bibnamefont {Yoshida}}, \bibinfo {author} {\bibfnamefont {N.~Y.}\
  \bibnamefont {Yao}},\ and\ \bibinfo {author} {\bibfnamefont {I.}~\bibnamefont
  {Siddiqi}},\ }\bibfield  {title} {\bibinfo {title} {Quantum information
  scrambling on a superconducting qutrit processor},\ }\href
  {https://doi.org/10.1103/PhysRevX.11.021010} {\bibfield  {journal} {\bibinfo
  {journal} {Phys. Rev. X}\ }\textbf {\bibinfo {volume} {11}},\ \bibinfo
  {pages} {021010} (\bibinfo {year} {2021})},\ \Eprint
  {https://arxiv.org/abs/2003.03307} {arXiv:2003.03307 [quant-ph]} \BibitemShut
  {NoStop}%
\bibitem [{\citenamefont {Seifert}\ \emph {et~al.}(2023)\citenamefont
  {Seifert}, \citenamefont {Li}, \citenamefont {Roy}, \citenamefont {Schuster},
  \citenamefont {Chong},\ and\ \citenamefont {Baker}}]{Seifert:2023ous}%
  \BibitemOpen
  \bibfield  {author} {\bibinfo {author} {\bibfnamefont {L.~M.}\ \bibnamefont
  {Seifert}}, \bibinfo {author} {\bibfnamefont {Z.}~\bibnamefont {Li}},
  \bibinfo {author} {\bibfnamefont {T.}~\bibnamefont {Roy}}, \bibinfo {author}
  {\bibfnamefont {D.~I.}\ \bibnamefont {Schuster}}, \bibinfo {author}
  {\bibfnamefont {F.~T.}\ \bibnamefont {Chong}},\ and\ \bibinfo {author}
  {\bibfnamefont {J.~M.}\ \bibnamefont {Baker}},\ }\bibfield  {title} {\bibinfo
  {title} {{Exploring ququart computation on a transmon using optimal
  control}},\ }\href {https://doi.org/10.1103/PhysRevA.108.062609} {\bibfield
  {journal} {\bibinfo  {journal} {Phys. Rev. A}\ }\textbf {\bibinfo {volume}
  {108}},\ \bibinfo {pages} {062609} (\bibinfo {year} {2023})},\ \Eprint
  {https://arxiv.org/abs/2304.11159} {arXiv:2304.11159 [quant-ph]} \BibitemShut
  {NoStop}%
\bibitem [{\citenamefont {Nguyen}\ \emph
  {et~al.}(2023{\natexlab{a}})\citenamefont {Nguyen}, \citenamefont {Goss},
  \citenamefont {Siva}, \citenamefont {Kim}, \citenamefont {Younis},
  \citenamefont {Qing}, \citenamefont {Hashim}, \citenamefont {Santiago},\ and\
  \citenamefont {Siddiqi}}]{Nguyen:2023svc}%
  \BibitemOpen
  \bibfield  {author} {\bibinfo {author} {\bibfnamefont {L.~B.}\ \bibnamefont
  {Nguyen}}, \bibinfo {author} {\bibfnamefont {N.}~\bibnamefont {Goss}},
  \bibinfo {author} {\bibfnamefont {K.}~\bibnamefont {Siva}}, \bibinfo {author}
  {\bibfnamefont {Y.}~\bibnamefont {Kim}}, \bibinfo {author} {\bibnamefont
  {Younis}}, \bibinfo {author} {\bibfnamefont {B.}~\bibnamefont {Qing}},
  \bibinfo {author} {\bibfnamefont {A.}~\bibnamefont {Hashim}}, \bibinfo
  {author} {\bibfnamefont {D.~I.}\ \bibnamefont {Santiago}},\ and\ \bibinfo
  {author} {\bibfnamefont {I.}~\bibnamefont {Siddiqi}},\ }\href@noop {}
  {\bibinfo {title} {{Empowering high-dimensional quantum computing by
  traversing the dual bosonic ladder}}} (\bibinfo {year}
  {2023}{\natexlab{a}}),\ \Eprint {https://arxiv.org/abs/2312.17741}
  {arXiv:2312.17741 [quant-ph]} \BibitemShut {NoStop}%
\bibitem [{\citenamefont {Champion}\ \emph {et~al.}(2024)\citenamefont
  {Champion}, \citenamefont {Wang}, \citenamefont {Parker},\ and\ \citenamefont
  {Blok}}]{Champion:2024ufp}%
  \BibitemOpen
  \bibfield  {author} {\bibinfo {author} {\bibfnamefont {E.}~\bibnamefont
  {Champion}}, \bibinfo {author} {\bibfnamefont {Z.}~\bibnamefont {Wang}},
  \bibinfo {author} {\bibfnamefont {R.}~\bibnamefont {Parker}},\ and\ \bibinfo
  {author} {\bibfnamefont {M.}~\bibnamefont {Blok}},\ }\href@noop {} {\bibinfo
  {title} {{Multi-frequency control and measurement of a spin-7/2 system
  encoded in a transmon qudit}}} (\bibinfo {year} {2024}),\ \Eprint
  {https://arxiv.org/abs/2405.15857} {arXiv:2405.15857 [quant-ph]} \BibitemShut
  {NoStop}%
\bibitem [{\citenamefont {Roy}\ \emph {et~al.}(2024)\citenamefont {Roy},
  \citenamefont {Kim}, \citenamefont {Romanenko},\ and\ \citenamefont
  {Grassellino}}]{Roy:2024uro}%
  \BibitemOpen
  \bibfield  {author} {\bibinfo {author} {\bibfnamefont {T.}~\bibnamefont
  {Roy}}, \bibinfo {author} {\bibfnamefont {T.}~\bibnamefont {Kim}}, \bibinfo
  {author} {\bibfnamefont {A.}~\bibnamefont {Romanenko}},\ and\ \bibinfo
  {author} {\bibfnamefont {A.}~\bibnamefont {Grassellino}},\ }\bibfield
  {title} {\bibinfo {title} {{Qudit-based quantum computing with SRF cavities
  at Fermilab}},\ }\href {https://doi.org/10.22323/1.453.0127} {\bibfield
  {journal} {\bibinfo  {journal} {PoS}\ }\textbf {\bibinfo {volume}
  {LATTICE2023}},\ \bibinfo {pages} {127} (\bibinfo {year} {2024})}\BibitemShut
  {NoStop}%
\bibitem [{\citenamefont {{Chi}}\ \emph {et~al.}(2022)\citenamefont {{Chi}}
  \emph {et~al.}}]{Chi:2022}%
  \BibitemOpen
  \bibfield  {author} {\bibinfo {author} {\bibfnamefont {Y.}~\bibnamefont
  {{Chi}}} \emph {et~al.},\ }\bibfield  {title} {\bibinfo {title} {{A
  programmable qudit-based quantum processor}},\ }\href
  {https://doi.org/10.1038/s41467-022-28767-x} {\bibfield  {journal} {\bibinfo
  {journal} {Nat. Commun.}\ }\textbf {\bibinfo {volume} {13}},\ \bibinfo {eid}
  {1166} (\bibinfo {year} {2022})}\BibitemShut {NoStop}%
\bibitem [{\citenamefont {{Cerf}}\ \emph {et~al.}(2002)\citenamefont {{Cerf}},
  \citenamefont {{Bourennane}}, \citenamefont {{Karlsson}},\ and\ \citenamefont
  {{Gisin}}}]{Cerf_2002}%
  \BibitemOpen
  \bibfield  {author} {\bibinfo {author} {\bibfnamefont {N.~J.}\ \bibnamefont
  {{Cerf}}}, \bibinfo {author} {\bibfnamefont {M.}~\bibnamefont
  {{Bourennane}}}, \bibinfo {author} {\bibfnamefont {A.}~\bibnamefont
  {{Karlsson}}},\ and\ \bibinfo {author} {\bibfnamefont {N.}~\bibnamefont
  {{Gisin}}},\ }\bibfield  {title} {\bibinfo {title} {{Security of Quantum Key
  Distribution Using d-Level Systems}},\ }\href
  {https://doi.org/10.1103/PhysRevLett.88.127902} {\bibfield  {journal}
  {\bibinfo  {journal} {Phys. Rev. Lett.}\ }\textbf {\bibinfo {volume} {88}},\
  \bibinfo {eid} {127902} (\bibinfo {year} {2002})},\ \Eprint
  {https://arxiv.org/abs/quant-ph/0107130} {arXiv:quant-ph/0107130 [quant-ph]}
  \BibitemShut {NoStop}%
\bibitem [{\citenamefont {{Gedik}}\ \emph {et~al.}(2015)\citenamefont
  {{Gedik}}, \citenamefont {{Silva}}, \citenamefont {{{\c{C}}akmak}},
  \citenamefont {{Karpat}}, \citenamefont {{Vidoto}}, \citenamefont
  {{Soares-Pinto}}, \citenamefont {{Deazevedo}},\ and\ \citenamefont
  {{Fanchini}}}]{Gedik:2015}%
  \BibitemOpen
  \bibfield  {author} {\bibinfo {author} {\bibfnamefont {Z.}~\bibnamefont
  {{Gedik}}}, \bibinfo {author} {\bibfnamefont {I.~A.}\ \bibnamefont
  {{Silva}}}, \bibinfo {author} {\bibfnamefont {B.}~\bibnamefont
  {{{\c{C}}akmak}}}, \bibinfo {author} {\bibfnamefont {G.}~\bibnamefont
  {{Karpat}}}, \bibinfo {author} {\bibfnamefont {E.~L.~G.}\ \bibnamefont
  {{Vidoto}}}, \bibinfo {author} {\bibfnamefont {D.~O.}\ \bibnamefont
  {{Soares-Pinto}}}, \bibinfo {author} {\bibfnamefont {E.~R.}\ \bibnamefont
  {{Deazevedo}}},\ and\ \bibinfo {author} {\bibfnamefont {F.~F.}\ \bibnamefont
  {{Fanchini}}},\ }\bibfield  {title} {\bibinfo {title} {{Computational
  speed-up with a single qudit}},\ }\href {https://doi.org/10.1038/srep14671}
  {\bibfield  {journal} {\bibinfo  {journal} {Sci. Rep.}\ }\textbf {\bibinfo
  {volume} {5}},\ \bibinfo {eid} {14671} (\bibinfo {year} {2015})},\ \Eprint
  {https://arxiv.org/abs/1403.5861} {arXiv:1403.5861 [quant-ph]} \BibitemShut
  {NoStop}%
\bibitem [{\citenamefont {Gokhale}\ \emph {et~al.}(2019)\citenamefont
  {Gokhale}, \citenamefont {Baker}, \citenamefont {Duckering}, \citenamefont
  {Brown}, \citenamefont {Brown},\ and\ \citenamefont {Chong}}]{3307650}%
  \BibitemOpen
  \bibfield  {author} {\bibinfo {author} {\bibfnamefont {P.}~\bibnamefont
  {Gokhale}}, \bibinfo {author} {\bibfnamefont {J.~M.}\ \bibnamefont {Baker}},
  \bibinfo {author} {\bibfnamefont {C.}~\bibnamefont {Duckering}}, \bibinfo
  {author} {\bibfnamefont {N.~C.}\ \bibnamefont {Brown}}, \bibinfo {author}
  {\bibfnamefont {K.~R.}\ \bibnamefont {Brown}},\ and\ \bibinfo {author}
  {\bibfnamefont {F.~T.}\ \bibnamefont {Chong}},\ }\bibfield  {title} {\bibinfo
  {title} {Asymptotic improvements to quantum circuits via qutrits},\ }in\
  \href {https://doi.org/https://doi.org/10.1145/3307650.33222} {\emph
  {\bibinfo {booktitle} {Proceedings of the 46th International Symposium on
  Computer Architecture}}}\ (\bibinfo {year} {2019})\ pp.\ \bibinfo {pages}
  {554--566}\BibitemShut {NoStop}%
\bibitem [{\citenamefont {Baker}\ \emph {et~al.}(2020)\citenamefont {Baker},
  \citenamefont {Duckering},\ and\ \citenamefont {Chong}}]{Baker:2020}%
  \BibitemOpen
  \bibfield  {author} {\bibinfo {author} {\bibfnamefont {J.~M.}\ \bibnamefont
  {Baker}}, \bibinfo {author} {\bibfnamefont {C.}~\bibnamefont {Duckering}},\
  and\ \bibinfo {author} {\bibfnamefont {F.~T.}\ \bibnamefont {Chong}},\
  }\bibfield  {title} {\bibinfo {title} {Efficient quantum circuit
  decompositions via intermediate qudits},\ }in\ \href
  {https://doi.org/10.1109/ISMVL49045.2020.9345604} {\emph {\bibinfo
  {booktitle} {2020 IEEE 50th International Symposium on Multiple-Valued Logic
  (ISMVL)}}}\ (\bibinfo {year} {2020})\ pp.\ \bibinfo {pages} {303--308},\
  \Eprint {https://arxiv.org/abs/2002.10592} {arXiv:2002.10592 [quant-ph]}
  \BibitemShut {NoStop}%
\bibitem [{\citenamefont {Lim}\ \emph {et~al.}(2023)\citenamefont {Lim},
  \citenamefont {Liu},\ and\ \citenamefont {Ardavan}}]{Lim:2023hkb}%
  \BibitemOpen
  \bibfield  {author} {\bibinfo {author} {\bibfnamefont {S.}~\bibnamefont
  {Lim}}, \bibinfo {author} {\bibfnamefont {J.}~\bibnamefont {Liu}},\ and\
  \bibinfo {author} {\bibfnamefont {A.}~\bibnamefont {Ardavan}},\ }\bibfield
  {title} {\bibinfo {title} {{Fault-tolerant qubit encoding using a spin-7/2
  qudit}},\ }\href {https://doi.org/10.1103/PhysRevA.108.062403} {\bibfield
  {journal} {\bibinfo  {journal} {Phys. Rev. A}\ }\textbf {\bibinfo {volume}
  {108}},\ \bibinfo {pages} {062403} (\bibinfo {year} {2023})},\ \Eprint
  {https://arxiv.org/abs/2303.02084} {arXiv:2303.02084 [quant-ph]} \BibitemShut
  {NoStop}%
\bibitem [{\citenamefont {{Wang}}\ \emph {et~al.}(2020)\citenamefont {{Wang}},
  \citenamefont {{Hu}}, \citenamefont {{Sanders}},\ and\ \citenamefont
  {{Kais}}}]{10.3389}%
  \BibitemOpen
  \bibfield  {author} {\bibinfo {author} {\bibfnamefont {Y.}~\bibnamefont
  {{Wang}}}, \bibinfo {author} {\bibfnamefont {Z.}~\bibnamefont {{Hu}}},
  \bibinfo {author} {\bibfnamefont {B.~C.}\ \bibnamefont {{Sanders}}},\ and\
  \bibinfo {author} {\bibfnamefont {S.}~\bibnamefont {{Kais}}},\ }\bibfield
  {title} {\bibinfo {title} {{Qudits and high-dimensional quantum computing}},\
  }\href {https://doi.org/10.3389/fphy.2020.589504} {\bibfield  {journal}
  {\bibinfo  {journal} {Front. Phys.}\ }\textbf {\bibinfo {volume} {8}},\
  \bibinfo {eid} {479} (\bibinfo {year} {2020})},\ \Eprint
  {https://arxiv.org/abs/2008.00959} {arXiv:2008.00959 [quant-ph]} \BibitemShut
  {NoStop}%
\bibitem [{\citenamefont {{Yurtalan}}\ \emph {et~al.}(2020)\citenamefont
  {{Yurtalan}}, \citenamefont {{Shi}}, \citenamefont {{Kononenko}},
  \citenamefont {{Lupascu}},\ and\ \citenamefont {{Ashhab}}}]{Yurtalan:2020}%
  \BibitemOpen
  \bibfield  {author} {\bibinfo {author} {\bibfnamefont {M.~A.}\ \bibnamefont
  {{Yurtalan}}}, \bibinfo {author} {\bibfnamefont {J.}~\bibnamefont {{Shi}}},
  \bibinfo {author} {\bibfnamefont {M.}~\bibnamefont {{Kononenko}}}, \bibinfo
  {author} {\bibfnamefont {A.}~\bibnamefont {{Lupascu}}},\ and\ \bibinfo
  {author} {\bibfnamefont {S.}~\bibnamefont {{Ashhab}}},\ }\bibfield  {title}
  {\bibinfo {title} {{Implementation of a Walsh-Hadamard Gate in a
  Superconducting Qutrit}},\ }\href
  {https://doi.org/10.1103/PhysRevLett.125.180504} {\bibfield  {journal}
  {\bibinfo  {journal} {Phys. Rev. Lett.}\ }\textbf {\bibinfo {volume} {125}},\
  \bibinfo {eid} {180504} (\bibinfo {year} {2020})},\ \Eprint
  {https://arxiv.org/abs/2003.04879} {arXiv:2003.04879 [quant-ph]} \BibitemShut
  {NoStop}%
\bibitem [{\citenamefont {Morvan}\ \emph {et~al.}(2021)\citenamefont {Morvan},
  \citenamefont {Ramasesh}, \citenamefont {Blok}, \citenamefont {Kreikebaum},
  \citenamefont {O'Brien}, \citenamefont {Chen}, \citenamefont {Mitchell},
  \citenamefont {Naik}, \citenamefont {Santiago},\ and\ \citenamefont
  {Siddiqi}}]{Morvan:2021qju}%
  \BibitemOpen
  \bibfield  {author} {\bibinfo {author} {\bibfnamefont {A.}~\bibnamefont
  {Morvan}}, \bibinfo {author} {\bibfnamefont {V.~V.}\ \bibnamefont
  {Ramasesh}}, \bibinfo {author} {\bibfnamefont {M.~S.}\ \bibnamefont {Blok}},
  \bibinfo {author} {\bibfnamefont {J.~M.}\ \bibnamefont {Kreikebaum}},
  \bibinfo {author} {\bibfnamefont {K.}~\bibnamefont {O'Brien}}, \bibinfo
  {author} {\bibfnamefont {L.}~\bibnamefont {Chen}}, \bibinfo {author}
  {\bibfnamefont {B.~K.}\ \bibnamefont {Mitchell}}, \bibinfo {author}
  {\bibfnamefont {R.~K.}\ \bibnamefont {Naik}}, \bibinfo {author}
  {\bibfnamefont {D.~I.}\ \bibnamefont {Santiago}},\ and\ \bibinfo {author}
  {\bibfnamefont {I.}~\bibnamefont {Siddiqi}},\ }\bibfield  {title} {\bibinfo
  {title} {Qutrit randomized benchmarking},\ }\href
  {https://doi.org/10.1103/PhysRevLett.126.210504} {\bibfield  {journal}
  {\bibinfo  {journal} {Phys. Rev. Lett.}\ }\textbf {\bibinfo {volume} {126}},\
  \bibinfo {pages} {210504} (\bibinfo {year} {2021})},\ \Eprint
  {https://arxiv.org/abs/2008.09134} {arXiv:2008.09134 [quant-ph]} \BibitemShut
  {NoStop}%
\bibitem [{\citenamefont {Cervera-Lierta}\ \emph {et~al.}(2022)\citenamefont
  {Cervera-Lierta}, \citenamefont {Krenn}, \citenamefont {Aspuru-Guzik},\ and\
  \citenamefont {Galda}}]{Cervera-Lierta:2021nhp}%
  \BibitemOpen
  \bibfield  {author} {\bibinfo {author} {\bibfnamefont {A.}~\bibnamefont
  {Cervera-Lierta}}, \bibinfo {author} {\bibfnamefont {M.}~\bibnamefont
  {Krenn}}, \bibinfo {author} {\bibfnamefont {A.}~\bibnamefont
  {Aspuru-Guzik}},\ and\ \bibinfo {author} {\bibfnamefont {A.}~\bibnamefont
  {Galda}},\ }\bibfield  {title} {\bibinfo {title} {{Experimental
  high-dimensional Greenberger-Horne-Zeilinger entanglement with
  superconducting transmon qutrits}},\ }\href
  {https://doi.org/10.1103/PhysRevApplied.17.024062} {\bibfield  {journal}
  {\bibinfo  {journal} {Phys. Rev. Applied}\ }\textbf {\bibinfo {volume}
  {17}},\ \bibinfo {pages} {024062} (\bibinfo {year} {2022})},\ \Eprint
  {https://arxiv.org/abs/2104.05627} {arXiv:2104.05627 [quant-ph]} \BibitemShut
  {NoStop}%
\bibitem [{\citenamefont {Hrmo}\ \emph {et~al.}(2023)\citenamefont {Hrmo},
  \citenamefont {Wilhelm}, \citenamefont {Gerster}, \citenamefont {van Mourik},
  \citenamefont {Huber}, \citenamefont {Blatt}, \citenamefont {Schindler},
  \citenamefont {Monz},\ and\ \citenamefont {Ringbauer}}]{Hrmo:2022bvo}%
  \BibitemOpen
  \bibfield  {author} {\bibinfo {author} {\bibfnamefont {P.}~\bibnamefont
  {Hrmo}}, \bibinfo {author} {\bibfnamefont {B.}~\bibnamefont {Wilhelm}},
  \bibinfo {author} {\bibfnamefont {L.}~\bibnamefont {Gerster}}, \bibinfo
  {author} {\bibfnamefont {M.~W.}\ \bibnamefont {van Mourik}}, \bibinfo
  {author} {\bibfnamefont {M.}~\bibnamefont {Huber}}, \bibinfo {author}
  {\bibfnamefont {R.}~\bibnamefont {Blatt}}, \bibinfo {author} {\bibfnamefont
  {P.}~\bibnamefont {Schindler}}, \bibinfo {author} {\bibfnamefont
  {T.}~\bibnamefont {Monz}},\ and\ \bibinfo {author} {\bibfnamefont
  {M.}~\bibnamefont {Ringbauer}},\ }\bibfield  {title} {\bibinfo {title}
  {{Native qudit entanglement in a trapped ion quantum processor}},\ }\href
  {https://doi.org/10.1038/s41467-023-37375-2} {\bibfield  {journal} {\bibinfo
  {journal} {Nature Commun.}\ }\textbf {\bibinfo {volume} {14}},\ \bibinfo
  {pages} {2242} (\bibinfo {year} {2023})},\ \Eprint
  {https://arxiv.org/abs/2206.04104} {arXiv:2206.04104 [quant-ph]} \BibitemShut
  {NoStop}%
\bibitem [{\citenamefont {Goss}\ \emph {et~al.}(2022)\citenamefont {Goss} \emph
  {et~al.}}]{Goss:2022bqd}%
  \BibitemOpen
  \bibfield  {author} {\bibinfo {author} {\bibfnamefont {N.}~\bibnamefont
  {Goss}} \emph {et~al.},\ }\bibfield  {title} {\bibinfo {title}
  {{High-fidelity qutrit entangling gates for superconducting circuits}},\
  }\href {https://doi.org/10.1038/s41467-022-34851-z} {\bibfield  {journal}
  {\bibinfo  {journal} {Nature Commun.}\ }\textbf {\bibinfo {volume} {13}},\
  \bibinfo {pages} {7481} (\bibinfo {year} {2022})},\ \bibinfo {note}
  {[Erratum: Nature Commun. 14, 4256 (2023)]},\ \Eprint
  {https://arxiv.org/abs/2206.07216} {arXiv:2206.07216 [quant-ph]} \BibitemShut
  {NoStop}%
\bibitem [{\citenamefont {Subramanian}\ and\ \citenamefont
  {Lupascu}(2023)}]{Subramanian:2023xzi}%
  \BibitemOpen
  \bibfield  {author} {\bibinfo {author} {\bibfnamefont {M.}~\bibnamefont
  {Subramanian}}\ and\ \bibinfo {author} {\bibfnamefont {A.}~\bibnamefont
  {Lupascu}},\ }\bibfield  {title} {\bibinfo {title} {Efficient two-qutrit
  gates in superconducting circuits using parametric coupling},\ }\href
  {https://doi.org/10.1103/PhysRevA.108.062616} {\bibfield  {journal} {\bibinfo
   {journal} {Phys. Rev. A}\ }\textbf {\bibinfo {volume} {108}},\ \bibinfo
  {pages} {062616} (\bibinfo {year} {2023})},\ \Eprint
  {https://arxiv.org/abs/2309.05766} {arXiv:2309.05766 [quant-ph]} \BibitemShut
  {NoStop}%
\bibitem [{\citenamefont {Pantaleone}(1992{\natexlab{a}})}]{Pantaleone:1992xh}%
  \BibitemOpen
  \bibfield  {author} {\bibinfo {author} {\bibfnamefont {J.}~\bibnamefont
  {Pantaleone}},\ }\bibfield  {title} {\bibinfo {title} {Dirac neutrinos in
  dense matter},\ }\href {https://doi.org/10.1103/PhysRevD.46.510} {\bibfield
  {journal} {\bibinfo  {journal} {Phys. Rev. D}\ }\textbf {\bibinfo {volume}
  {46}},\ \bibinfo {pages} {510} (\bibinfo {year}
  {1992}{\natexlab{a}})}\BibitemShut {NoStop}%
\bibitem [{\citenamefont {Pantaleone}(1992{\natexlab{b}})}]{Pantaleone:1992eq}%
  \BibitemOpen
  \bibfield  {author} {\bibinfo {author} {\bibfnamefont {J.~T.}\ \bibnamefont
  {Pantaleone}},\ }\bibfield  {title} {\bibinfo {title} {{Neutrino oscillations
  at high densities}},\ }\href {https://doi.org/10.1016/0370-2693(92)91887-F}
  {\bibfield  {journal} {\bibinfo  {journal} {Phys. Lett. B}\ }\textbf
  {\bibinfo {volume} {287}},\ \bibinfo {pages} {128} (\bibinfo {year}
  {1992}{\natexlab{b}})}\BibitemShut {NoStop}%
\bibitem [{\citenamefont {Qian}\ and\ \citenamefont
  {Fuller}(1995)}]{Qian:1994wh}%
  \BibitemOpen
  \bibfield  {author} {\bibinfo {author} {\bibfnamefont {Y.~Z.}\ \bibnamefont
  {Qian}}\ and\ \bibinfo {author} {\bibfnamefont {G.~M.}\ \bibnamefont
  {Fuller}},\ }\bibfield  {title} {\bibinfo {title} {{Neutrino-neutrino
  scattering and matter enhanced neutrino flavor transformation in
  Supernovae}},\ }\href {https://doi.org/10.1103/PhysRevD.51.1479} {\bibfield
  {journal} {\bibinfo  {journal} {Phys. Rev. D}\ }\textbf {\bibinfo {volume}
  {51}},\ \bibinfo {pages} {1479} (\bibinfo {year} {1995})},\ \Eprint
  {https://arxiv.org/abs/astro-ph/9406073} {arXiv:astro-ph/9406073}
  \BibitemShut {NoStop}%
\bibitem [{\citenamefont {Pastor}\ and\ \citenamefont
  {Raffelt}(2002)}]{Pastor:2002we}%
  \BibitemOpen
  \bibfield  {author} {\bibinfo {author} {\bibfnamefont {S.}~\bibnamefont
  {Pastor}}\ and\ \bibinfo {author} {\bibfnamefont {G.}~\bibnamefont
  {Raffelt}},\ }\bibfield  {title} {\bibinfo {title} {{Flavor oscillations in
  the supernova hot bubble region: Nonlinear effects of neutrino background}},\
  }\href {https://doi.org/10.1103/PhysRevLett.89.191101} {\bibfield  {journal}
  {\bibinfo  {journal} {Phys. Rev. Lett.}\ }\textbf {\bibinfo {volume} {89}},\
  \bibinfo {pages} {191101} (\bibinfo {year} {2002})},\ \Eprint
  {https://arxiv.org/abs/astro-ph/0207281} {arXiv:astro-ph/0207281}
  \BibitemShut {NoStop}%
\bibitem [{\citenamefont {Mirizzi}\ \emph {et~al.}(2016)\citenamefont
  {Mirizzi}, \citenamefont {Tamborra}, \citenamefont {Janka}, \citenamefont
  {Saviano}, \citenamefont {Scholberg}, \citenamefont {Bollig}, \citenamefont
  {Hudepohl},\ and\ \citenamefont {Chakraborty}}]{Mirizzi:2015eza}%
  \BibitemOpen
  \bibfield  {author} {\bibinfo {author} {\bibfnamefont {A.}~\bibnamefont
  {Mirizzi}}, \bibinfo {author} {\bibfnamefont {I.}~\bibnamefont {Tamborra}},
  \bibinfo {author} {\bibfnamefont {H.-T.}\ \bibnamefont {Janka}}, \bibinfo
  {author} {\bibfnamefont {N.}~\bibnamefont {Saviano}}, \bibinfo {author}
  {\bibfnamefont {K.}~\bibnamefont {Scholberg}}, \bibinfo {author}
  {\bibfnamefont {R.}~\bibnamefont {Bollig}}, \bibinfo {author} {\bibfnamefont
  {L.}~\bibnamefont {Hudepohl}},\ and\ \bibinfo {author} {\bibfnamefont
  {S.}~\bibnamefont {Chakraborty}},\ }\bibfield  {title} {\bibinfo {title}
  {{Supernova Neutrinos: Production, Oscillations and Detection}},\ }\href
  {https://doi.org/10.1393/ncr/i2016-10120-8} {\bibfield  {journal} {\bibinfo
  {journal} {Riv. Nuovo Cim.}\ }\textbf {\bibinfo {volume} {39}},\ \bibinfo
  {pages} {1} (\bibinfo {year} {2016})},\ \Eprint
  {https://arxiv.org/abs/1508.00785} {arXiv:1508.00785 [astro-ph.HE]}
  \BibitemShut {NoStop}%
\bibitem [{\citenamefont {Malkus}\ \emph {et~al.}(2012)\citenamefont {Malkus},
  \citenamefont {Kneller}, \citenamefont {McLaughlin},\ and\ \citenamefont
  {Surman}}]{Malkus:2012ts}%
  \BibitemOpen
  \bibfield  {author} {\bibinfo {author} {\bibfnamefont {A.}~\bibnamefont
  {Malkus}}, \bibinfo {author} {\bibfnamefont {J.~P.}\ \bibnamefont {Kneller}},
  \bibinfo {author} {\bibfnamefont {G.~C.}\ \bibnamefont {McLaughlin}},\ and\
  \bibinfo {author} {\bibfnamefont {R.}~\bibnamefont {Surman}},\ }\bibfield
  {title} {\bibinfo {title} {{Neutrino oscillations above black hole accretion
  disks: disks with electron-flavor emission}},\ }\href
  {https://doi.org/10.1103/PhysRevD.86.085015} {\bibfield  {journal} {\bibinfo
  {journal} {Phys. Rev. D}\ }\textbf {\bibinfo {volume} {86}},\ \bibinfo
  {pages} {085015} (\bibinfo {year} {2012})},\ \Eprint
  {https://arxiv.org/abs/1207.6648} {arXiv:1207.6648 [hep-ph]} \BibitemShut
  {NoStop}%
\bibitem [{\citenamefont {Malkus}\ \emph {et~al.}(2016)\citenamefont {Malkus},
  \citenamefont {McLaughlin},\ and\ \citenamefont {Surman}}]{Malkus:2015mda}%
  \BibitemOpen
  \bibfield  {author} {\bibinfo {author} {\bibfnamefont {A.}~\bibnamefont
  {Malkus}}, \bibinfo {author} {\bibfnamefont {G.~C.}\ \bibnamefont
  {McLaughlin}},\ and\ \bibinfo {author} {\bibfnamefont {R.}~\bibnamefont
  {Surman}},\ }\bibfield  {title} {\bibinfo {title} {{Symmetric and Standard
  Matter-Neutrino Resonances Above Merging Compact Objects}},\ }\href
  {https://doi.org/10.1103/PhysRevD.93.045021} {\bibfield  {journal} {\bibinfo
  {journal} {Phys. Rev. D}\ }\textbf {\bibinfo {volume} {93}},\ \bibinfo
  {pages} {045021} (\bibinfo {year} {2016})},\ \Eprint
  {https://arxiv.org/abs/1507.00946} {arXiv:1507.00946 [hep-ph]} \BibitemShut
  {NoStop}%
\bibitem [{\citenamefont {Zhu}\ \emph {et~al.}(2016)\citenamefont {Zhu},
  \citenamefont {Perego},\ and\ \citenamefont {McLaughlin}}]{Zhu:2016mwa}%
  \BibitemOpen
  \bibfield  {author} {\bibinfo {author} {\bibfnamefont {Y.-L.}\ \bibnamefont
  {Zhu}}, \bibinfo {author} {\bibfnamefont {A.}~\bibnamefont {Perego}},\ and\
  \bibinfo {author} {\bibfnamefont {G.~C.}\ \bibnamefont {McLaughlin}},\
  }\bibfield  {title} {\bibinfo {title} {{Matter Neutrino Resonance Transitions
  above a Neutron Star Merger Remnant}},\ }\href
  {https://doi.org/10.1103/PhysRevD.94.105006} {\bibfield  {journal} {\bibinfo
  {journal} {Phys. Rev. D}\ }\textbf {\bibinfo {volume} {94}},\ \bibinfo
  {pages} {105006} (\bibinfo {year} {2016})},\ \Eprint
  {https://arxiv.org/abs/1607.04671} {arXiv:1607.04671 [hep-ph]} \BibitemShut
  {NoStop}%
\bibitem [{\citenamefont {Frensel}\ \emph {et~al.}(2017)\citenamefont
  {Frensel}, \citenamefont {Wu}, \citenamefont {Volpe},\ and\ \citenamefont
  {Perego}}]{Frensel:2016fge}%
  \BibitemOpen
  \bibfield  {author} {\bibinfo {author} {\bibfnamefont {M.}~\bibnamefont
  {Frensel}}, \bibinfo {author} {\bibfnamefont {M.-R.}\ \bibnamefont {Wu}},
  \bibinfo {author} {\bibfnamefont {C.}~\bibnamefont {Volpe}},\ and\ \bibinfo
  {author} {\bibfnamefont {A.}~\bibnamefont {Perego}},\ }\bibfield  {title}
  {\bibinfo {title} {{Neutrino Flavor Evolution in Binary Neutron Star Merger
  Remnants}},\ }\href {https://doi.org/10.1103/PhysRevD.95.023011} {\bibfield
  {journal} {\bibinfo  {journal} {Phys. Rev. D}\ }\textbf {\bibinfo {volume}
  {95}},\ \bibinfo {pages} {023011} (\bibinfo {year} {2017})},\ \Eprint
  {https://arxiv.org/abs/1607.05938} {arXiv:1607.05938 [astro-ph.HE]}
  \BibitemShut {NoStop}%
\bibitem [{\citenamefont {Chatelain}\ and\ \citenamefont
  {Volpe}(2017)}]{Chatelain:2016xva}%
  \BibitemOpen
  \bibfield  {author} {\bibinfo {author} {\bibfnamefont {A.}~\bibnamefont
  {Chatelain}}\ and\ \bibinfo {author} {\bibfnamefont {C.}~\bibnamefont
  {Volpe}},\ }\bibfield  {title} {\bibinfo {title} {{Helicity coherence in
  binary neutron star mergers and non-linear feedback}},\ }\href
  {https://doi.org/10.1103/PhysRevD.95.043005} {\bibfield  {journal} {\bibinfo
  {journal} {Phys. Rev. D}\ }\textbf {\bibinfo {volume} {95}},\ \bibinfo
  {pages} {043005} (\bibinfo {year} {2017})},\ \Eprint
  {https://arxiv.org/abs/1611.01862} {arXiv:1611.01862 [hep-ph]} \BibitemShut
  {NoStop}%
\bibitem [{\citenamefont {Wu}\ and\ \citenamefont
  {Tamborra}(2017)}]{Wu:2017qpc}%
  \BibitemOpen
  \bibfield  {author} {\bibinfo {author} {\bibfnamefont {M.-R.}\ \bibnamefont
  {Wu}}\ and\ \bibinfo {author} {\bibfnamefont {I.}~\bibnamefont {Tamborra}},\
  }\bibfield  {title} {\bibinfo {title} {{Fast neutrino conversions: Ubiquitous
  in compact binary merger remnants}},\ }\href
  {https://doi.org/10.1103/PhysRevD.95.103007} {\bibfield  {journal} {\bibinfo
  {journal} {Phys. Rev. D}\ }\textbf {\bibinfo {volume} {95}},\ \bibinfo
  {pages} {103007} (\bibinfo {year} {2017})},\ \Eprint
  {https://arxiv.org/abs/1701.06580} {arXiv:1701.06580 [astro-ph.HE]}
  \BibitemShut {NoStop}%
\bibitem [{\citenamefont {Tian}\ \emph {et~al.}(2017)\citenamefont {Tian},
  \citenamefont {Patwardhan},\ and\ \citenamefont {Fuller}}]{Tian:2017xbr}%
  \BibitemOpen
  \bibfield  {author} {\bibinfo {author} {\bibfnamefont {J.~Y.}\ \bibnamefont
  {Tian}}, \bibinfo {author} {\bibfnamefont {A.~V.}\ \bibnamefont
  {Patwardhan}},\ and\ \bibinfo {author} {\bibfnamefont {G.~M.}\ \bibnamefont
  {Fuller}},\ }\bibfield  {title} {\bibinfo {title} {{Neutrino Flavor Evolution
  in Neutron Star Mergers}},\ }\href
  {https://doi.org/10.1103/PhysRevD.96.043001} {\bibfield  {journal} {\bibinfo
  {journal} {Phys. Rev. D}\ }\textbf {\bibinfo {volume} {96}},\ \bibinfo
  {pages} {043001} (\bibinfo {year} {2017})},\ \Eprint
  {https://arxiv.org/abs/1703.03039} {arXiv:1703.03039 [astro-ph.HE]}
  \BibitemShut {NoStop}%
\bibitem [{\citenamefont {Purcell}\ \emph {et~al.}(2024)\citenamefont
  {Purcell}, \citenamefont {Richers}, \citenamefont {Patwardhan},\ and\
  \citenamefont {Foucart}}]{Purcell:2024bim}%
  \BibitemOpen
  \bibfield  {author} {\bibinfo {author} {\bibfnamefont {H.}~\bibnamefont
  {Purcell}}, \bibinfo {author} {\bibfnamefont {S.}~\bibnamefont {Richers}},
  \bibinfo {author} {\bibfnamefont {A.~V.}\ \bibnamefont {Patwardhan}},\ and\
  \bibinfo {author} {\bibfnamefont {F.}~\bibnamefont {Foucart}},\ }\href@noop
  {} {\bibinfo {title} {{Three-flavor, Full Momentum Space Neutrino Spin
  Oscillations in Neutron Star Mergers}}} (\bibinfo {year} {2024}),\ \Eprint
  {https://arxiv.org/abs/2404.08159} {arXiv:2404.08159 [astro-ph.HE]}
  \BibitemShut {NoStop}%
\bibitem [{\citenamefont {Fuller}\ and\ \citenamefont
  {Meyer}(1995)}]{Fuller:1995ih}%
  \BibitemOpen
  \bibfield  {author} {\bibinfo {author} {\bibfnamefont {G.~M.}\ \bibnamefont
  {Fuller}}\ and\ \bibinfo {author} {\bibfnamefont {B.~S.}\ \bibnamefont
  {Meyer}},\ }\bibfield  {title} {\bibinfo {title} {{Neutrino capture and
  supernova nucleosynthesis}},\ }\href {https://doi.org/10.1086/176442}
  {\bibfield  {journal} {\bibinfo  {journal} {Astrophys. J.}\ }\textbf
  {\bibinfo {volume} {453}},\ \bibinfo {pages} {792} (\bibinfo {year}
  {1995})}\BibitemShut {NoStop}%
\bibitem [{\citenamefont {Balantekin}\ and\ \citenamefont
  {Yuksel}(2005)}]{Balantekin:2004ug}%
  \BibitemOpen
  \bibfield  {author} {\bibinfo {author} {\bibfnamefont {A.~B.}\ \bibnamefont
  {Balantekin}}\ and\ \bibinfo {author} {\bibfnamefont {H.}~\bibnamefont
  {Yuksel}},\ }\bibfield  {title} {\bibinfo {title} {{Neutrino mixing and
  nucleosynthesis in core-collapse supernovae}},\ }\href
  {https://doi.org/10.1088/1367-2630/7/1/051} {\bibfield  {journal} {\bibinfo
  {journal} {New J. Phys.}\ }\textbf {\bibinfo {volume} {7}},\ \bibinfo {pages}
  {51} (\bibinfo {year} {2005})},\ \Eprint
  {https://arxiv.org/abs/astro-ph/0411159} {arXiv:astro-ph/0411159}
  \BibitemShut {NoStop}%
\bibitem [{\citenamefont {Duan}\ \emph {et~al.}(2011)\citenamefont {Duan},
  \citenamefont {Friedland}, \citenamefont {McLaughlin},\ and\ \citenamefont
  {Surman}}]{Duan:2010af}%
  \BibitemOpen
  \bibfield  {author} {\bibinfo {author} {\bibfnamefont {H.}~\bibnamefont
  {Duan}}, \bibinfo {author} {\bibfnamefont {A.}~\bibnamefont {Friedland}},
  \bibinfo {author} {\bibfnamefont {G.}~\bibnamefont {McLaughlin}},\ and\
  \bibinfo {author} {\bibfnamefont {R.}~\bibnamefont {Surman}},\ }\bibfield
  {title} {\bibinfo {title} {{The influence of collective neutrino oscillations
  on a supernova r-process}},\ }\href
  {https://doi.org/10.1088/0954-3899/38/3/035201} {\bibfield  {journal}
  {\bibinfo  {journal} {J. Phys. G}\ }\textbf {\bibinfo {volume} {38}},\
  \bibinfo {pages} {035201} (\bibinfo {year} {2011})},\ \Eprint
  {https://arxiv.org/abs/1012.0532} {arXiv:1012.0532 [astro-ph.SR]}
  \BibitemShut {NoStop}%
\bibitem [{\citenamefont {Xiong}\ \emph {et~al.}(2020)\citenamefont {Xiong},
  \citenamefont {Sieverding}, \citenamefont {Sen},\ and\ \citenamefont
  {Qian}}]{Xiong:2020ntn}%
  \BibitemOpen
  \bibfield  {author} {\bibinfo {author} {\bibfnamefont {Z.}~\bibnamefont
  {Xiong}}, \bibinfo {author} {\bibfnamefont {A.}~\bibnamefont {Sieverding}},
  \bibinfo {author} {\bibfnamefont {M.}~\bibnamefont {Sen}},\ and\ \bibinfo
  {author} {\bibfnamefont {Y.-Z.}\ \bibnamefont {Qian}},\ }\bibfield  {title}
  {\bibinfo {title} {{Potential Impact of Fast Flavor Oscillations on
  Neutrino-driven Winds and Their Nucleosynthesis}},\ }\href
  {https://doi.org/10.3847/1538-4357/abac5e} {\bibfield  {journal} {\bibinfo
  {journal} {Astrophys. J.}\ }\textbf {\bibinfo {volume} {900}},\ \bibinfo
  {pages} {144} (\bibinfo {year} {2020})},\ \Eprint
  {https://arxiv.org/abs/2006.11414} {arXiv:2006.11414 [astro-ph.HE]}
  \BibitemShut {NoStop}%
\bibitem [{\citenamefont {Balantekin}\ \emph {et~al.}(2024)\citenamefont
  {Balantekin}, \citenamefont {Cervia}, \citenamefont {Patwardhan},
  \citenamefont {Surman},\ and\ \citenamefont {Wang}}]{Balantekin:2023ayx}%
  \BibitemOpen
  \bibfield  {author} {\bibinfo {author} {\bibfnamefont {A.~B.}\ \bibnamefont
  {Balantekin}}, \bibinfo {author} {\bibfnamefont {M.~J.}\ \bibnamefont
  {Cervia}}, \bibinfo {author} {\bibfnamefont {A.~V.}\ \bibnamefont
  {Patwardhan}}, \bibinfo {author} {\bibfnamefont {R.}~\bibnamefont {Surman}},\
  and\ \bibinfo {author} {\bibfnamefont {X.}~\bibnamefont {Wang}},\ }\bibfield
  {title} {\bibinfo {title} {{Collective Neutrino Oscillations and
  Heavy-element Nucleosynthesis in Supernovae: Exploring Potential Effects of
  Many-body Neutrino Correlations}},\ }\href
  {https://doi.org/10.3847/1538-4357/ad393d} {\bibfield  {journal} {\bibinfo
  {journal} {Astrophys. J.}\ }\textbf {\bibinfo {volume} {967}},\ \bibinfo
  {pages} {146} (\bibinfo {year} {2024})},\ \Eprint
  {https://arxiv.org/abs/2311.02562} {arXiv:2311.02562 [astro-ph.HE]}
  \BibitemShut {NoStop}%
\bibitem [{\citenamefont {Duan}\ and\ \citenamefont
  {Kneller}(2009)}]{Duan:2009cd}%
  \BibitemOpen
  \bibfield  {author} {\bibinfo {author} {\bibfnamefont {H.}~\bibnamefont
  {Duan}}\ and\ \bibinfo {author} {\bibfnamefont {J.~P.}\ \bibnamefont
  {Kneller}},\ }\bibfield  {title} {\bibinfo {title} {{Neutrino flavour
  transformation in supernovae}},\ }\href
  {https://doi.org/10.1088/0954-3899/36/11/113201} {\bibfield  {journal}
  {\bibinfo  {journal} {J. Phys. G}\ }\textbf {\bibinfo {volume} {36}},\
  \bibinfo {pages} {113201} (\bibinfo {year} {2009})},\ \Eprint
  {https://arxiv.org/abs/0904.0974} {arXiv:0904.0974 [astro-ph.HE]}
  \BibitemShut {NoStop}%
\bibitem [{\citenamefont {Duan}\ \emph {et~al.}(2010)\citenamefont {Duan},
  \citenamefont {Fuller},\ and\ \citenamefont {Qian}}]{Duan:2010bg}%
  \BibitemOpen
  \bibfield  {author} {\bibinfo {author} {\bibfnamefont {H.}~\bibnamefont
  {Duan}}, \bibinfo {author} {\bibfnamefont {G.~M.}\ \bibnamefont {Fuller}},\
  and\ \bibinfo {author} {\bibfnamefont {Y.-Z.}\ \bibnamefont {Qian}},\
  }\bibfield  {title} {\bibinfo {title} {{Collective Neutrino Oscillations}},\
  }\href {https://doi.org/10.1146/annurev.nucl.012809.104524} {\bibfield
  {journal} {\bibinfo  {journal} {Ann. Rev. Nucl. Part. Sci.}\ }\textbf
  {\bibinfo {volume} {60}},\ \bibinfo {pages} {569} (\bibinfo {year} {2010})},\
  \Eprint {https://arxiv.org/abs/1001.2799} {arXiv:1001.2799 [hep-ph]}
  \BibitemShut {NoStop}%
\bibitem [{\citenamefont {Chakraborty}\ \emph {et~al.}(2016)\citenamefont
  {Chakraborty}, \citenamefont {Hansen}, \citenamefont {Izaguirre},\ and\
  \citenamefont {Raffelt}}]{Chakraborty:2016yeg}%
  \BibitemOpen
  \bibfield  {author} {\bibinfo {author} {\bibfnamefont {S.}~\bibnamefont
  {Chakraborty}}, \bibinfo {author} {\bibfnamefont {R.}~\bibnamefont {Hansen}},
  \bibinfo {author} {\bibfnamefont {I.}~\bibnamefont {Izaguirre}},\ and\
  \bibinfo {author} {\bibfnamefont {G.}~\bibnamefont {Raffelt}},\ }\bibfield
  {title} {\bibinfo {title} {{Collective neutrino flavor conversion: Recent
  developments}},\ }\href {https://doi.org/10.1016/j.nuclphysb.2016.02.012}
  {\bibfield  {journal} {\bibinfo  {journal} {Nucl. Phys. B}\ }\textbf
  {\bibinfo {volume} {908}},\ \bibinfo {pages} {366} (\bibinfo {year}
  {2016})},\ \Eprint {https://arxiv.org/abs/1602.02766} {arXiv:1602.02766
  [hep-ph]} \BibitemShut {NoStop}%
\bibitem [{\citenamefont {Tamborra}\ and\ \citenamefont
  {Shalgar}(2021)}]{Tamborra:2020cul}%
  \BibitemOpen
  \bibfield  {author} {\bibinfo {author} {\bibfnamefont {I.}~\bibnamefont
  {Tamborra}}\ and\ \bibinfo {author} {\bibfnamefont {S.}~\bibnamefont
  {Shalgar}},\ }\bibfield  {title} {\bibinfo {title} {{New Developments in
  Flavor Evolution of a Dense Neutrino Gas}},\ }\href
  {https://doi.org/10.1146/annurev-nucl-102920-050505} {\bibfield  {journal}
  {\bibinfo  {journal} {Ann. Rev. Nucl. Part. Sci.}\ }\textbf {\bibinfo
  {volume} {71}},\ \bibinfo {pages} {165} (\bibinfo {year} {2021})},\ \Eprint
  {https://arxiv.org/abs/2011.01948} {arXiv:2011.01948 [astro-ph.HE]}
  \BibitemShut {NoStop}%
\bibitem [{\citenamefont {Capozzi}\ and\ \citenamefont
  {Saviano}(2022)}]{Capozzi:2022slf}%
  \BibitemOpen
  \bibfield  {author} {\bibinfo {author} {\bibfnamefont {F.}~\bibnamefont
  {Capozzi}}\ and\ \bibinfo {author} {\bibfnamefont {N.}~\bibnamefont
  {Saviano}},\ }\bibfield  {title} {\bibinfo {title} {{Neutrino Flavor
  Conversions in High-Density Astrophysical and Cosmological Environments}},\
  }\href {https://doi.org/10.3390/universe8020094} {\bibfield  {journal}
  {\bibinfo  {journal} {Universe}\ }\textbf {\bibinfo {volume} {8}},\ \bibinfo
  {pages} {94} (\bibinfo {year} {2022})},\ \Eprint
  {https://arxiv.org/abs/2202.02494} {arXiv:2202.02494 [hep-ph]} \BibitemShut
  {NoStop}%
\bibitem [{\citenamefont {Richers}\ and\ \citenamefont
  {Sen}(2022)}]{Richers:2022zug}%
  \BibitemOpen
  \bibfield  {author} {\bibinfo {author} {\bibfnamefont {S.}~\bibnamefont
  {Richers}}\ and\ \bibinfo {author} {\bibfnamefont {M.}~\bibnamefont {Sen}},\
  }\bibinfo {title} {{Fast Flavor Transformations}},\ in\ \href
  {https://doi.org/10.1007/978-981-15-8818-1_125-1} {\emph {\bibinfo
  {booktitle} {{Handbook of Nuclear Physics}}}},\ \bibinfo {editor} {edited by\
  \bibinfo {editor} {\bibfnamefont {I.}~\bibnamefont {Tanihata}}, \bibinfo
  {editor} {\bibfnamefont {H.}~\bibnamefont {Toki}},\ and\ \bibinfo {editor}
  {\bibfnamefont {T.}~\bibnamefont {Kajino}}}\ (\bibinfo  {publisher} {Springer
  Nature Singapore},\ \bibinfo {address} {Singapore},\ \bibinfo {year} {2022})\
  pp.\ \bibinfo {pages} {1--17},\ \Eprint {https://arxiv.org/abs/2207.03561}
  {arXiv:2207.03561 [astro-ph.HE]} \BibitemShut {NoStop}%
\bibitem [{\citenamefont {Patwardhan}\ \emph {et~al.}(2023)\citenamefont
  {Patwardhan}, \citenamefont {Cervia}, \citenamefont {Rrapaj}, \citenamefont
  {Siwach},\ and\ \citenamefont {Balantekin}}]{Patwardhan:2022mxg}%
  \BibitemOpen
  \bibfield  {author} {\bibinfo {author} {\bibfnamefont {A.~V.}\ \bibnamefont
  {Patwardhan}}, \bibinfo {author} {\bibfnamefont {M.~J.}\ \bibnamefont
  {Cervia}}, \bibinfo {author} {\bibfnamefont {E.}~\bibnamefont {Rrapaj}},
  \bibinfo {author} {\bibfnamefont {P.}~\bibnamefont {Siwach}},\ and\ \bibinfo
  {author} {\bibfnamefont {A.~B.}\ \bibnamefont {Balantekin}},\ }\bibinfo
  {title} {{Many-Body Collective Neutrino Oscillations: Recent Developments}},\
  in\ \href {https://doi.org/10.1007/978-981-15-8818-1_126-1} {\emph {\bibinfo
  {booktitle} {{Handbook of Nuclear Physics}}}},\ \bibinfo {editor} {edited by\
  \bibinfo {editor} {\bibfnamefont {I.}~\bibnamefont {Tanihata}}, \bibinfo
  {editor} {\bibfnamefont {H.}~\bibnamefont {Toki}},\ and\ \bibinfo {editor}
  {\bibfnamefont {T.}~\bibnamefont {Kajino}}}\ (\bibinfo  {publisher} {Springer
  Nature Singapore},\ \bibinfo {address} {Singapore},\ \bibinfo {year} {2023})\
  pp.\ \bibinfo {pages} {1--16},\ \Eprint {https://arxiv.org/abs/2301.00342}
  {arXiv:2301.00342 [hep-ph]} \BibitemShut {NoStop}%
\bibitem [{\citenamefont {Volpe}(2024)}]{Volpe:2023met}%
  \BibitemOpen
  \bibfield  {author} {\bibinfo {author} {\bibfnamefont {M.~C.}\ \bibnamefont
  {Volpe}},\ }\bibfield  {title} {\bibinfo {title} {{Neutrinos from dense
  environments: Flavor mechanisms, theoretical approaches, observations, and
  new directions}},\ }\href {https://doi.org/10.1103/RevModPhys.96.025004}
  {\bibfield  {journal} {\bibinfo  {journal} {Rev. Mod. Phys.}\ }\textbf
  {\bibinfo {volume} {96}},\ \bibinfo {pages} {025004} (\bibinfo {year}
  {2024})},\ \Eprint {https://arxiv.org/abs/2301.11814} {arXiv:2301.11814
  [hep-ph]} \BibitemShut {NoStop}%
\bibitem [{\citenamefont {Balantekin}\ \emph {et~al.}(2023)\citenamefont
  {Balantekin}, \citenamefont {Cervia}, \citenamefont {Patwardhan},
  \citenamefont {Rrapaj},\ and\ \citenamefont {Siwach}}]{Balantekin:2023qvm}%
  \BibitemOpen
  \bibfield  {author} {\bibinfo {author} {\bibfnamefont {A.~B.}\ \bibnamefont
  {Balantekin}}, \bibinfo {author} {\bibfnamefont {M.~J.}\ \bibnamefont
  {Cervia}}, \bibinfo {author} {\bibfnamefont {A.~V.}\ \bibnamefont
  {Patwardhan}}, \bibinfo {author} {\bibfnamefont {E.}~\bibnamefont {Rrapaj}},\
  and\ \bibinfo {author} {\bibfnamefont {P.}~\bibnamefont {Siwach}},\
  }\bibfield  {title} {\bibinfo {title} {{Quantum information and quantum
  simulation of neutrino physics}},\ }\href
  {https://doi.org/10.1140/epja/s10050-023-01092-7} {\bibfield  {journal}
  {\bibinfo  {journal} {Eur. Phys. J. A}\ }\textbf {\bibinfo {volume} {59}},\
  \bibinfo {pages} {186} (\bibinfo {year} {2023})},\ \Eprint
  {https://arxiv.org/abs/2305.01150} {arXiv:2305.01150 [nucl-th]} \BibitemShut
  {NoStop}%
\bibitem [{\citenamefont {Samuel}(1993)}]{Samuel:1993uw}%
  \BibitemOpen
  \bibfield  {author} {\bibinfo {author} {\bibfnamefont {S.}~\bibnamefont
  {Samuel}},\ }\bibfield  {title} {\bibinfo {title} {{Neutrino oscillations in
  dense neutrino gases}},\ }\href {https://doi.org/10.1103/PhysRevD.48.1462}
  {\bibfield  {journal} {\bibinfo  {journal} {Phys. Rev. D}\ }\textbf {\bibinfo
  {volume} {48}},\ \bibinfo {pages} {1462} (\bibinfo {year}
  {1993})}\BibitemShut {NoStop}%
\bibitem [{\citenamefont {Kostelecky}\ \emph {et~al.}(1993)\citenamefont
  {Kostelecky}, \citenamefont {Pantaleone},\ and\ \citenamefont
  {Samuel}}]{Kostelecky:1993yt}%
  \BibitemOpen
  \bibfield  {author} {\bibinfo {author} {\bibfnamefont {V.~A.}\ \bibnamefont
  {Kostelecky}}, \bibinfo {author} {\bibfnamefont {J.~T.}\ \bibnamefont
  {Pantaleone}},\ and\ \bibinfo {author} {\bibfnamefont {S.}~\bibnamefont
  {Samuel}},\ }\bibfield  {title} {\bibinfo {title} {{Neutrino oscillation in
  the early universe}},\ }\href {https://doi.org/10.1016/0370-2693(93)90156-C}
  {\bibfield  {journal} {\bibinfo  {journal} {Phys. Lett. B}\ }\textbf
  {\bibinfo {volume} {315}},\ \bibinfo {pages} {46} (\bibinfo {year}
  {1993})}\BibitemShut {NoStop}%
\bibitem [{\citenamefont {Kostelecky}\ and\ \citenamefont
  {Samuel}(1993)}]{Kostelecky:1993dm}%
  \BibitemOpen
  \bibfield  {author} {\bibinfo {author} {\bibfnamefont {V.~A.}\ \bibnamefont
  {Kostelecky}}\ and\ \bibinfo {author} {\bibfnamefont {S.}~\bibnamefont
  {Samuel}},\ }\bibfield  {title} {\bibinfo {title} {{Neutrino oscillations in
  the early universe with an inverted neutrino mass hierarchy}},\ }\href
  {https://doi.org/10.1016/0370-2693(93)91795-O} {\bibfield  {journal}
  {\bibinfo  {journal} {Phys. Lett. B}\ }\textbf {\bibinfo {volume} {318}},\
  \bibinfo {pages} {127} (\bibinfo {year} {1993})}\BibitemShut {NoStop}%
\bibitem [{\citenamefont {Kostelecky}\ and\ \citenamefont
  {Samuel}(1994)}]{Kostelecky:1993ys}%
  \BibitemOpen
  \bibfield  {author} {\bibinfo {author} {\bibfnamefont {V.~A.}\ \bibnamefont
  {Kostelecky}}\ and\ \bibinfo {author} {\bibfnamefont {S.}~\bibnamefont
  {Samuel}},\ }\bibfield  {title} {\bibinfo {title} {{Nonlinear neutrino
  oscillations in the expanding universe}},\ }\href
  {https://doi.org/10.1103/PhysRevD.49.1740} {\bibfield  {journal} {\bibinfo
  {journal} {Phys. Rev. D}\ }\textbf {\bibinfo {volume} {49}},\ \bibinfo
  {pages} {1740} (\bibinfo {year} {1994})}\BibitemShut {NoStop}%
\bibitem [{\citenamefont {Kostelecky}\ and\ \citenamefont
  {Samuel}(1995)}]{Kostelecky:1995xc}%
  \BibitemOpen
  \bibfield  {author} {\bibinfo {author} {\bibfnamefont {V.~A.}\ \bibnamefont
  {Kostelecky}}\ and\ \bibinfo {author} {\bibfnamefont {S.}~\bibnamefont
  {Samuel}},\ }\bibfield  {title} {\bibinfo {title} {{Neutrino oscillations in
  the early universe with nonequilibrium neutrino distributions}},\ }\href
  {https://doi.org/10.1103/PhysRevD.52.3184} {\bibfield  {journal} {\bibinfo
  {journal} {Phys. Rev. D}\ }\textbf {\bibinfo {volume} {52}},\ \bibinfo
  {pages} {3184} (\bibinfo {year} {1995})},\ \Eprint
  {https://arxiv.org/abs/hep-ph/9507427} {arXiv:hep-ph/9507427} \BibitemShut
  {NoStop}%
\bibitem [{\citenamefont {Wolfenstein}(1978)}]{Wolfenstein:1977ue}%
  \BibitemOpen
  \bibfield  {author} {\bibinfo {author} {\bibfnamefont {L.}~\bibnamefont
  {Wolfenstein}},\ }\bibfield  {title} {\bibinfo {title} {Neutrino oscillations
  in matter},\ }\href {https://doi.org/10.1103/PhysRevD.17.2369} {\bibfield
  {journal} {\bibinfo  {journal} {Phys. Rev. D}\ }\textbf {\bibinfo {volume}
  {17}},\ \bibinfo {pages} {2369} (\bibinfo {year} {1978})}\BibitemShut
  {NoStop}%
\bibitem [{\citenamefont {Wolfenstein}(1979)}]{Wolfenstein:1979ni}%
  \BibitemOpen
  \bibfield  {author} {\bibinfo {author} {\bibfnamefont {L.}~\bibnamefont
  {Wolfenstein}},\ }\bibfield  {title} {\bibinfo {title} {Neutrino oscillations
  and stellar collapse},\ }\href {https://doi.org/10.1103/PhysRevD.20.2634}
  {\bibfield  {journal} {\bibinfo  {journal} {Phys. Rev. D}\ }\textbf {\bibinfo
  {volume} {20}},\ \bibinfo {pages} {2634} (\bibinfo {year}
  {1979})}\BibitemShut {NoStop}%
\bibitem [{\citenamefont {Mikheyev}\ and\ \citenamefont
  {Smirnov}(1986)}]{Mikheyev:1985zog}%
  \BibitemOpen
  \bibfield  {author} {\bibinfo {author} {\bibfnamefont {S.~P.}\ \bibnamefont
  {Mikheyev}}\ and\ \bibinfo {author} {\bibfnamefont {A.~Y.}\ \bibnamefont
  {Smirnov}},\ }\bibfield  {title} {\bibinfo {title} {{Resonant amplification
  of neutrino oscillations in matter and solar neutrino spectroscopy}},\ }\href
  {https://doi.org/10.1007/BF02508049} {\bibfield  {journal} {\bibinfo
  {journal} {Nuovo Cim. C}\ }\textbf {\bibinfo {volume} {9}},\ \bibinfo {pages}
  {17} (\bibinfo {year} {1986})}\BibitemShut {NoStop}%
\bibitem [{\citenamefont {Mikheyev}\ and\ \citenamefont
  {Smirnov}(1989)}]{Mikheyev:1989dy}%
  \BibitemOpen
  \bibfield  {author} {\bibinfo {author} {\bibfnamefont {S.~P.}\ \bibnamefont
  {Mikheyev}}\ and\ \bibinfo {author} {\bibfnamefont {A.~Y.}\ \bibnamefont
  {Smirnov}},\ }\bibfield  {title} {\bibinfo {title} {{Resonant neutrino
  oscillations in matter}},\ }\href
  {https://doi.org/10.1016/0146-6410(89)90008-2} {\bibfield  {journal}
  {\bibinfo  {journal} {Prog. Part. Nucl. Phys.}\ }\textbf {\bibinfo {volume}
  {23}},\ \bibinfo {pages} {41} (\bibinfo {year} {1989})}\BibitemShut {NoStop}%
\bibitem [{\citenamefont {{Fuller}}\ \emph {et~al.}(1987)\citenamefont
  {{Fuller}}, \citenamefont {{Mayle}}, \citenamefont {{Wilson}},\ and\
  \citenamefont {{Schramm}}}]{1987ApJ322795F}%
  \BibitemOpen
  \bibfield  {author} {\bibinfo {author} {\bibfnamefont {G.~M.}\ \bibnamefont
  {{Fuller}}}, \bibinfo {author} {\bibfnamefont {R.~W.}\ \bibnamefont
  {{Mayle}}}, \bibinfo {author} {\bibfnamefont {J.~R.}\ \bibnamefont
  {{Wilson}}},\ and\ \bibinfo {author} {\bibfnamefont {D.~N.}\ \bibnamefont
  {{Schramm}}},\ }\bibfield  {title} {\bibinfo {title} {{Resonant Neutrino
  Oscillations and Stellar Collapse}},\ }\href {https://doi.org/10.1086/165772}
  {\bibfield  {journal} {\bibinfo  {journal} {\apj}\ }\textbf {\bibinfo
  {volume} {322}},\ \bibinfo {pages} {795} (\bibinfo {year}
  {1987})}\BibitemShut {NoStop}%
\bibitem [{\citenamefont {Notzold}\ and\ \citenamefont
  {Raffelt}(1988)}]{Notzold:1987ik}%
  \BibitemOpen
  \bibfield  {author} {\bibinfo {author} {\bibfnamefont {D.}~\bibnamefont
  {Notzold}}\ and\ \bibinfo {author} {\bibfnamefont {G.}~\bibnamefont
  {Raffelt}},\ }\bibfield  {title} {\bibinfo {title} {{Neutrino dispersion at
  finite temperature and density}},\ }\href
  {https://doi.org/10.1016/0550-3213(88)90113-7} {\bibfield  {journal}
  {\bibinfo  {journal} {Nucl. Phys. B}\ }\textbf {\bibinfo {volume} {307}},\
  \bibinfo {pages} {924} (\bibinfo {year} {1988})}\BibitemShut {NoStop}%
\bibitem [{\citenamefont {Savage}\ \emph {et~al.}(1991)\citenamefont {Savage},
  \citenamefont {Malaney},\ and\ \citenamefont {Fuller}}]{Savage:1990by}%
  \BibitemOpen
  \bibfield  {author} {\bibinfo {author} {\bibfnamefont {M.~J.}\ \bibnamefont
  {Savage}}, \bibinfo {author} {\bibfnamefont {R.~A.}\ \bibnamefont
  {Malaney}},\ and\ \bibinfo {author} {\bibfnamefont {G.~M.}\ \bibnamefont
  {Fuller}},\ }\bibfield  {title} {\bibinfo {title} {{Neutrino Oscillations and
  the Leptonic Charge of the Universe}},\ }\href
  {https://doi.org/10.1086/169665} {\bibfield  {journal} {\bibinfo  {journal}
  {Astrophys. J.}\ }\textbf {\bibinfo {volume} {368}},\ \bibinfo {pages} {1}
  (\bibinfo {year} {1991})}\BibitemShut {NoStop}%
\bibitem [{\citenamefont {{Sigl}}\ and\ \citenamefont
  {{Raffelt}}(1993)}]{Sigl:1993ctk}%
  \BibitemOpen
  \bibfield  {author} {\bibinfo {author} {\bibfnamefont {G.}~\bibnamefont
  {{Sigl}}}\ and\ \bibinfo {author} {\bibfnamefont {G.}~\bibnamefont
  {{Raffelt}}},\ }\bibfield  {title} {\bibinfo {title} {{General kinetic
  description of relativistic mixed neutrinos}},\ }\href
  {https://doi.org/10.1016/0550-3213(93)90175-O} {\bibfield  {journal}
  {\bibinfo  {journal} {Nuclear Physics B}\ }\textbf {\bibinfo {volume}
  {406}},\ \bibinfo {pages} {423} (\bibinfo {year} {1993})}\BibitemShut
  {NoStop}%
\bibitem [{\citenamefont {Samuel}(1996)}]{Samuel:1995ri}%
  \BibitemOpen
  \bibfield  {author} {\bibinfo {author} {\bibfnamefont {S.}~\bibnamefont
  {Samuel}},\ }\bibfield  {title} {\bibinfo {title} {{Bimodal coherence in
  dense selfinteracting neutrino gases}},\ }\href
  {https://doi.org/10.1103/PhysRevD.53.5382} {\bibfield  {journal} {\bibinfo
  {journal} {Phys. Rev. D}\ }\textbf {\bibinfo {volume} {53}},\ \bibinfo
  {pages} {5382} (\bibinfo {year} {1996})},\ \Eprint
  {https://arxiv.org/abs/hep-ph/9604341} {arXiv:hep-ph/9604341} \BibitemShut
  {NoStop}%
\bibitem [{\citenamefont {Hoffman}\ \emph {et~al.}(1997)\citenamefont
  {Hoffman}, \citenamefont {Woosley},\ and\ \citenamefont
  {Qian}}]{Hoffman:1996aj}%
  \BibitemOpen
  \bibfield  {author} {\bibinfo {author} {\bibfnamefont {R.~D.}\ \bibnamefont
  {Hoffman}}, \bibinfo {author} {\bibfnamefont {S.~E.}\ \bibnamefont
  {Woosley}},\ and\ \bibinfo {author} {\bibfnamefont {Y.~Z.}\ \bibnamefont
  {Qian}},\ }\bibfield  {title} {\bibinfo {title} {{Nucleosynthesis in neutrino
  driven winds: 2. Implications for heavy element synthesis}},\ }\href
  {https://doi.org/10.1086/304181} {\bibfield  {journal} {\bibinfo  {journal}
  {Astrophys. J.}\ }\textbf {\bibinfo {volume} {482}},\ \bibinfo {pages} {951}
  (\bibinfo {year} {1997})},\ \Eprint {https://arxiv.org/abs/astro-ph/9611097}
  {arXiv:astro-ph/9611097} \BibitemShut {NoStop}%
\bibitem [{\citenamefont {Balantekin}\ and\ \citenamefont
  {Pehlivan}(2007)}]{Balantekin:2006tg}%
  \BibitemOpen
  \bibfield  {author} {\bibinfo {author} {\bibfnamefont {A.~B.}\ \bibnamefont
  {Balantekin}}\ and\ \bibinfo {author} {\bibfnamefont {Y.}~\bibnamefont
  {Pehlivan}},\ }\bibfield  {title} {\bibinfo {title} {{Neutrino-Neutrino
  Interactions and Flavor Mixing in Dense Matter}},\ }\href
  {https://doi.org/10.1088/0954-3899/34/1/004} {\bibfield  {journal} {\bibinfo
  {journal} {J. Phys. G}\ }\textbf {\bibinfo {volume} {34}},\ \bibinfo {pages}
  {47} (\bibinfo {year} {2007})},\ \Eprint
  {https://arxiv.org/abs/astro-ph/0607527} {arXiv:astro-ph/0607527}
  \BibitemShut {NoStop}%
\bibitem [{\citenamefont {Dasgupta}\ and\ \citenamefont
  {Sen}(2018)}]{Dasgupta:2017oko}%
  \BibitemOpen
  \bibfield  {author} {\bibinfo {author} {\bibfnamefont {B.}~\bibnamefont
  {Dasgupta}}\ and\ \bibinfo {author} {\bibfnamefont {M.}~\bibnamefont {Sen}},\
  }\bibfield  {title} {\bibinfo {title} {{Fast Neutrino Flavor Conversion as
  Oscillations in a Quartic Potential}},\ }\href
  {https://doi.org/10.1103/PhysRevD.97.023017} {\bibfield  {journal} {\bibinfo
  {journal} {Phys. Rev. D}\ }\textbf {\bibinfo {volume} {97}},\ \bibinfo
  {pages} {023017} (\bibinfo {year} {2018})},\ \Eprint
  {https://arxiv.org/abs/1709.08671} {arXiv:1709.08671 [hep-ph]} \BibitemShut
  {NoStop}%
\bibitem [{\citenamefont {Fiorillo}\ \emph
  {et~al.}(2024{\natexlab{a}})\citenamefont {Fiorillo}, \citenamefont
  {Raffelt},\ and\ \citenamefont {Sigl}}]{Fiorillo:2024wej}%
  \BibitemOpen
  \bibfield  {author} {\bibinfo {author} {\bibfnamefont {D.~F.~G.}\
  \bibnamefont {Fiorillo}}, \bibinfo {author} {\bibfnamefont {G.~G.}\
  \bibnamefont {Raffelt}},\ and\ \bibinfo {author} {\bibfnamefont
  {G.}~\bibnamefont {Sigl}},\ }\bibfield  {title} {\bibinfo {title}
  {{Collective neutrino-antineutrino oscillations in dense neutrino
  environments?}},\ }\href {https://doi.org/10.1103/PhysRevD.109.043031}
  {\bibfield  {journal} {\bibinfo  {journal} {Phys. Rev. D}\ }\textbf {\bibinfo
  {volume} {109}},\ \bibinfo {pages} {043031} (\bibinfo {year}
  {2024}{\natexlab{a}})},\ \Eprint {https://arxiv.org/abs/2401.02478}
  {arXiv:2401.02478 [hep-ph]} \BibitemShut {NoStop}%
\bibitem [{\citenamefont {Cirigliano}\ \emph {et~al.}(2024)\citenamefont
  {Cirigliano}, \citenamefont {Sen},\ and\ \citenamefont
  {Yamauchi}}]{Cirigliano:2024pnm}%
  \BibitemOpen
  \bibfield  {author} {\bibinfo {author} {\bibfnamefont {V.}~\bibnamefont
  {Cirigliano}}, \bibinfo {author} {\bibfnamefont {S.}~\bibnamefont {Sen}},\
  and\ \bibinfo {author} {\bibfnamefont {Y.}~\bibnamefont {Yamauchi}},\
  }\href@noop {} {\bibinfo {title} {{Neutrino many-body flavor evolution: the
  full Hamiltonian}}} (\bibinfo {year} {2024}),\ \Eprint
  {https://arxiv.org/abs/2404.16690} {arXiv:2404.16690 [hep-ph]} \BibitemShut
  {NoStop}%
\bibitem [{\citenamefont {Duan}\ \emph
  {et~al.}(2006{\natexlab{a}})\citenamefont {Duan}, \citenamefont {Fuller},
  \citenamefont {Carlson},\ and\ \citenamefont {Qian}}]{Duan:2006jv}%
  \BibitemOpen
  \bibfield  {author} {\bibinfo {author} {\bibfnamefont {H.}~\bibnamefont
  {Duan}}, \bibinfo {author} {\bibfnamefont {G.~M.}\ \bibnamefont {Fuller}},
  \bibinfo {author} {\bibfnamefont {J.}~\bibnamefont {Carlson}},\ and\ \bibinfo
  {author} {\bibfnamefont {Y.-Z.}\ \bibnamefont {Qian}},\ }\bibfield  {title}
  {\bibinfo {title} {{Coherent Development of Neutrino Flavor in the Supernova
  Environment}},\ }\href {https://doi.org/10.1103/PhysRevLett.97.241101}
  {\bibfield  {journal} {\bibinfo  {journal} {Phys. Rev. Lett.}\ }\textbf
  {\bibinfo {volume} {97}},\ \bibinfo {pages} {241101} (\bibinfo {year}
  {2006}{\natexlab{a}})},\ \Eprint {https://arxiv.org/abs/astro-ph/0608050}
  {arXiv:astro-ph/0608050} \BibitemShut {NoStop}%
\bibitem [{\citenamefont {Duan}\ \emph
  {et~al.}(2006{\natexlab{b}})\citenamefont {Duan}, \citenamefont {Fuller},
  \citenamefont {Carlson},\ and\ \citenamefont {Qian}}]{Duan:2006an}%
  \BibitemOpen
  \bibfield  {author} {\bibinfo {author} {\bibfnamefont {H.}~\bibnamefont
  {Duan}}, \bibinfo {author} {\bibfnamefont {G.~M.}\ \bibnamefont {Fuller}},
  \bibinfo {author} {\bibfnamefont {J.}~\bibnamefont {Carlson}},\ and\ \bibinfo
  {author} {\bibfnamefont {Y.-Z.}\ \bibnamefont {Qian}},\ }\bibfield  {title}
  {\bibinfo {title} {{Simulation of Coherent Non-Linear Neutrino Flavor
  Transformation in the Supernova Environment. 1. Correlated Neutrino
  Trajectories}},\ }\href {https://doi.org/10.1103/PhysRevD.74.105014}
  {\bibfield  {journal} {\bibinfo  {journal} {Phys. Rev. D}\ }\textbf {\bibinfo
  {volume} {74}},\ \bibinfo {pages} {105014} (\bibinfo {year}
  {2006}{\natexlab{b}})},\ \Eprint {https://arxiv.org/abs/astro-ph/0606616}
  {arXiv:astro-ph/0606616} \BibitemShut {NoStop}%
\bibitem [{\citenamefont {Izaguirre}\ \emph {et~al.}(2017)\citenamefont
  {Izaguirre}, \citenamefont {Raffelt},\ and\ \citenamefont
  {Tamborra}}]{Izaguirre:2016gsx}%
  \BibitemOpen
  \bibfield  {author} {\bibinfo {author} {\bibfnamefont {I.}~\bibnamefont
  {Izaguirre}}, \bibinfo {author} {\bibfnamefont {G.}~\bibnamefont {Raffelt}},\
  and\ \bibinfo {author} {\bibfnamefont {I.}~\bibnamefont {Tamborra}},\
  }\bibfield  {title} {\bibinfo {title} {{Fast Pairwise Conversion of Supernova
  Neutrinos: A Dispersion-Relation Approach}},\ }\href
  {https://doi.org/10.1103/PhysRevLett.118.021101} {\bibfield  {journal}
  {\bibinfo  {journal} {Phys. Rev. Lett.}\ }\textbf {\bibinfo {volume} {118}},\
  \bibinfo {pages} {021101} (\bibinfo {year} {2017})},\ \Eprint
  {https://arxiv.org/abs/1610.01612} {arXiv:1610.01612 [hep-ph]} \BibitemShut
  {NoStop}%
\bibitem [{\citenamefont {Capozzi}\ \emph {et~al.}(2020)\citenamefont
  {Capozzi}, \citenamefont {Chakraborty}, \citenamefont {Chakraborty},\ and\
  \citenamefont {Sen}}]{Capozzi:2020kge}%
  \BibitemOpen
  \bibfield  {author} {\bibinfo {author} {\bibfnamefont {F.}~\bibnamefont
  {Capozzi}}, \bibinfo {author} {\bibfnamefont {M.}~\bibnamefont
  {Chakraborty}}, \bibinfo {author} {\bibfnamefont {S.}~\bibnamefont
  {Chakraborty}},\ and\ \bibinfo {author} {\bibfnamefont {M.}~\bibnamefont
  {Sen}},\ }\bibfield  {title} {\bibinfo {title} {{Fast flavor conversions in
  supernovae: the rise of mu-tau neutrinos}},\ }\href
  {https://doi.org/10.1103/PhysRevLett.125.251801} {\bibfield  {journal}
  {\bibinfo  {journal} {Phys. Rev. Lett.}\ }\textbf {\bibinfo {volume} {125}},\
  \bibinfo {pages} {251801} (\bibinfo {year} {2020})},\ \Eprint
  {https://arxiv.org/abs/2005.14204} {arXiv:2005.14204 [hep-ph]} \BibitemShut
  {NoStop}%
\bibitem [{\citenamefont {Fiorillo}\ and\ \citenamefont
  {Raffelt}(2023{\natexlab{a}})}]{Fiorillo:2023mze}%
  \BibitemOpen
  \bibfield  {author} {\bibinfo {author} {\bibfnamefont {D.~F.~G.}\
  \bibnamefont {Fiorillo}}\ and\ \bibinfo {author} {\bibfnamefont {G.~G.}\
  \bibnamefont {Raffelt}},\ }\bibfield  {title} {\bibinfo {title} {{Slow and
  fast collective neutrino oscillations: Invariants and reciprocity}},\ }\href
  {https://doi.org/10.1103/PhysRevD.107.043024} {\bibfield  {journal} {\bibinfo
   {journal} {Phys. Rev. D}\ }\textbf {\bibinfo {volume} {107}},\ \bibinfo
  {pages} {043024} (\bibinfo {year} {2023}{\natexlab{a}})},\ \Eprint
  {https://arxiv.org/abs/2301.09650} {arXiv:2301.09650 [hep-ph]} \BibitemShut
  {NoStop}%
\bibitem [{\citenamefont {Fiorillo}\ and\ \citenamefont
  {Raffelt}(2023{\natexlab{b}})}]{Fiorillo:2023hlk}%
  \BibitemOpen
  \bibfield  {author} {\bibinfo {author} {\bibfnamefont {D.~F.~G.}\
  \bibnamefont {Fiorillo}}\ and\ \bibinfo {author} {\bibfnamefont {G.~G.}\
  \bibnamefont {Raffelt}},\ }\bibfield  {title} {\bibinfo {title} {{Flavor
  solitons in dense neutrino gases}},\ }\href
  {https://doi.org/10.1103/PhysRevD.107.123024} {\bibfield  {journal} {\bibinfo
   {journal} {Phys. Rev. D}\ }\textbf {\bibinfo {volume} {107}},\ \bibinfo
  {pages} {123024} (\bibinfo {year} {2023}{\natexlab{b}})},\ \Eprint
  {https://arxiv.org/abs/2303.12143} {arXiv:2303.12143 [hep-ph]} \BibitemShut
  {NoStop}%
\bibitem [{\citenamefont {Fiorillo}\ \emph
  {et~al.}(2024{\natexlab{b}})\citenamefont {Fiorillo}, \citenamefont
  {Raffelt},\ and\ \citenamefont {Sigl}}]{Fiorillo:2024fnl}%
  \BibitemOpen
  \bibfield  {author} {\bibinfo {author} {\bibfnamefont {D.~F.~G.}\
  \bibnamefont {Fiorillo}}, \bibinfo {author} {\bibfnamefont {G.~G.}\
  \bibnamefont {Raffelt}},\ and\ \bibinfo {author} {\bibfnamefont
  {G.}~\bibnamefont {Sigl}},\ }\bibfield  {title} {\bibinfo {title}
  {{Inhomogeneous Kinetic Equation for Mixed Neutrinos: Tracing the Missing
  Energy}},\ }\href {https://doi.org/10.1103/PhysRevLett.133.021002} {\bibfield
   {journal} {\bibinfo  {journal} {Phys. Rev. Lett.}\ }\textbf {\bibinfo
  {volume} {133}},\ \bibinfo {pages} {021002} (\bibinfo {year}
  {2024}{\natexlab{b}})},\ \Eprint {https://arxiv.org/abs/2401.05278}
  {arXiv:2401.05278 [hep-ph]} \BibitemShut {NoStop}%
\bibitem [{\citenamefont {Roggero}(2021{\natexlab{a}})}]{Roggero:2021asb}%
  \BibitemOpen
  \bibfield  {author} {\bibinfo {author} {\bibfnamefont {A.}~\bibnamefont
  {Roggero}},\ }\bibfield  {title} {\bibinfo {title} {Entanglement and
  many-body effects in collective neutrino oscillations},\ }\href
  {https://doi.org/10.1103/PhysRevD.104.103016} {\bibfield  {journal} {\bibinfo
   {journal} {Phys. Rev. D}\ }\textbf {\bibinfo {volume} {104}},\ \bibinfo
  {pages} {103016} (\bibinfo {year} {2021}{\natexlab{a}})},\ \Eprint
  {https://arxiv.org/abs/2102.10188} {arXiv:2102.10188 [hep-ph]} \BibitemShut
  {NoStop}%
\bibitem [{\citenamefont {Xiong}(2022)}]{Xiong:2021evk}%
  \BibitemOpen
  \bibfield  {author} {\bibinfo {author} {\bibfnamefont {Z.}~\bibnamefont
  {Xiong}},\ }\bibfield  {title} {\bibinfo {title} {{Many-body effects of
  collective neutrino oscillations}},\ }\href
  {https://doi.org/10.1103/PhysRevD.105.103002} {\bibfield  {journal} {\bibinfo
   {journal} {Phys. Rev. D}\ }\textbf {\bibinfo {volume} {105}},\ \bibinfo
  {pages} {103002} (\bibinfo {year} {2022})},\ \Eprint
  {https://arxiv.org/abs/2111.00437} {arXiv:2111.00437 [astro-ph.HE]}
  \BibitemShut {NoStop}%
\bibitem [{\citenamefont {Roggero}\ \emph {et~al.}(2022)\citenamefont
  {Roggero}, \citenamefont {Rrapaj},\ and\ \citenamefont
  {Xiong}}]{Roggero:2022hpy}%
  \BibitemOpen
  \bibfield  {author} {\bibinfo {author} {\bibfnamefont {A.}~\bibnamefont
  {Roggero}}, \bibinfo {author} {\bibfnamefont {E.}~\bibnamefont {Rrapaj}},\
  and\ \bibinfo {author} {\bibfnamefont {Z.}~\bibnamefont {Xiong}},\ }\bibfield
   {title} {\bibinfo {title} {{Entanglement and correlations in fast collective
  neutrino flavor oscillations}},\ }\href
  {https://doi.org/10.1103/PhysRevD.106.043022} {\bibfield  {journal} {\bibinfo
   {journal} {Phys. Rev. D}\ }\textbf {\bibinfo {volume} {106}},\ \bibinfo
  {pages} {043022} (\bibinfo {year} {2022})},\ \Eprint
  {https://arxiv.org/abs/2203.02783} {arXiv:2203.02783 [astro-ph.HE]}
  \BibitemShut {NoStop}%
\bibitem [{\citenamefont {Bhaskar}\ \emph {et~al.}(2023)\citenamefont
  {Bhaskar}, \citenamefont {Roggero},\ and\ \citenamefont
  {Savage}}]{Bhaskar:2023sta}%
  \BibitemOpen
  \bibfield  {author} {\bibinfo {author} {\bibfnamefont {R.}~\bibnamefont
  {Bhaskar}}, \bibinfo {author} {\bibfnamefont {A.}~\bibnamefont {Roggero}},\
  and\ \bibinfo {author} {\bibfnamefont {M.~J.}\ \bibnamefont {Savage}},\
  }\href@noop {} {\bibinfo {title} {{Time Scales in Many-Body Fast Neutrino
  Flavor Conversion}}} (\bibinfo {year} {2023}),\ \Eprint
  {https://arxiv.org/abs/2312.16212} {arXiv:2312.16212 [nucl-th]} \BibitemShut
  {NoStop}%
\bibitem [{\citenamefont {Kost}\ \emph {et~al.}(2024)\citenamefont {Kost},
  \citenamefont {Johns},\ and\ \citenamefont {Duan}}]{Kost:2024esc}%
  \BibitemOpen
  \bibfield  {author} {\bibinfo {author} {\bibfnamefont {A.}~\bibnamefont
  {Kost}}, \bibinfo {author} {\bibfnamefont {L.}~\bibnamefont {Johns}},\ and\
  \bibinfo {author} {\bibfnamefont {H.}~\bibnamefont {Duan}},\ }\bibfield
  {title} {\bibinfo {title} {{Once-in-a-lifetime encounter models for neutrino
  media: From coherent oscillations to flavor equilibration}},\ }\href
  {https://doi.org/10.1103/PhysRevD.109.103037} {\bibfield  {journal} {\bibinfo
   {journal} {Phys. Rev. D}\ }\textbf {\bibinfo {volume} {109}},\ \bibinfo
  {pages} {103037} (\bibinfo {year} {2024})},\ \Eprint
  {https://arxiv.org/abs/2402.05022} {arXiv:2402.05022 [hep-ph]} \BibitemShut
  {NoStop}%
\bibitem [{\citenamefont {Borrelli}\ \emph {et~al.}(1990)\citenamefont
  {Borrelli}, \citenamefont {Maiani}, \citenamefont {Rossi}, \citenamefont
  {Sisto},\ and\ \citenamefont {Testa}}]{BORRELLI1990335}%
  \BibitemOpen
  \bibfield  {author} {\bibinfo {author} {\bibfnamefont {A.}~\bibnamefont
  {Borrelli}}, \bibinfo {author} {\bibfnamefont {L.}~\bibnamefont {Maiani}},
  \bibinfo {author} {\bibfnamefont {G.}~\bibnamefont {Rossi}}, \bibinfo
  {author} {\bibfnamefont {R.}~\bibnamefont {Sisto}},\ and\ \bibinfo {author}
  {\bibfnamefont {M.}~\bibnamefont {Testa}},\ }\bibfield  {title} {\bibinfo
  {title} {Neutrinos on the lattice. the regularization of a chiral gauge
  theory},\ }\href
  {https://doi.org/https://doi.org/10.1016/0550-3213(90)90041-B} {\bibfield
  {journal} {\bibinfo  {journal} {Nuclear Physics B}\ }\textbf {\bibinfo
  {volume} {333}},\ \bibinfo {pages} {335} (\bibinfo {year}
  {1990})}\BibitemShut {NoStop}%
\bibitem [{\citenamefont {Kaplan}(1992)}]{KAPLAN1992342}%
  \BibitemOpen
  \bibfield  {author} {\bibinfo {author} {\bibfnamefont {D.~B.}\ \bibnamefont
  {Kaplan}},\ }\bibfield  {title} {\bibinfo {title} {A method for simulating
  chiral fermions on the lattice},\ }\href
  {https://doi.org/https://doi.org/10.1016/0370-2693(92)91112-M} {\bibfield
  {journal} {\bibinfo  {journal} {Physics Letters B}\ }\textbf {\bibinfo
  {volume} {288}},\ \bibinfo {pages} {342} (\bibinfo {year}
  {1992})}\BibitemShut {NoStop}%
\bibitem [{\citenamefont {Kaplan}(2024)}]{PhysRevLett.132.141603}%
  \BibitemOpen
  \bibfield  {author} {\bibinfo {author} {\bibfnamefont {D.~B.}\ \bibnamefont
  {Kaplan}},\ }\bibfield  {title} {\bibinfo {title} {Chiral gauge theory at the
  boundary between topological phases},\ }\href
  {https://doi.org/10.1103/PhysRevLett.132.141603} {\bibfield  {journal}
  {\bibinfo  {journal} {Phys. Rev. Lett.}\ }\textbf {\bibinfo {volume} {132}},\
  \bibinfo {pages} {141603} (\bibinfo {year} {2024})}\BibitemShut {NoStop}%
\bibitem [{\citenamefont {Kaplan}\ and\ \citenamefont
  {Sen}(2024)}]{PhysRevLett.132.141604}%
  \BibitemOpen
  \bibfield  {author} {\bibinfo {author} {\bibfnamefont {D.~B.}\ \bibnamefont
  {Kaplan}}\ and\ \bibinfo {author} {\bibfnamefont {S.}~\bibnamefont {Sen}},\
  }\bibfield  {title} {\bibinfo {title} {Weyl fermions on a finite lattice},\
  }\href {https://doi.org/10.1103/PhysRevLett.132.141604} {\bibfield  {journal}
  {\bibinfo  {journal} {Phys. Rev. Lett.}\ }\textbf {\bibinfo {volume} {132}},\
  \bibinfo {pages} {141604} (\bibinfo {year} {2024})}\BibitemShut {NoStop}%
\bibitem [{\citenamefont {Shalgar}\ and\ \citenamefont
  {Tamborra}(2023)}]{Shalgar:2023ooi}%
  \BibitemOpen
  \bibfield  {author} {\bibinfo {author} {\bibfnamefont {S.}~\bibnamefont
  {Shalgar}}\ and\ \bibinfo {author} {\bibfnamefont {I.}~\bibnamefont
  {Tamborra}},\ }\bibfield  {title} {\bibinfo {title} {{Do we have enough
  evidence to invalidate the mean-field approximation adopted to model
  collective neutrino oscillations?}},\ }\href
  {https://doi.org/10.1103/PhysRevD.107.123004} {\bibfield  {journal} {\bibinfo
   {journal} {Phys. Rev. D}\ }\textbf {\bibinfo {volume} {107}},\ \bibinfo
  {pages} {123004} (\bibinfo {year} {2023})},\ \Eprint
  {https://arxiv.org/abs/2304.13050} {arXiv:2304.13050 [astro-ph.HE]}
  \BibitemShut {NoStop}%
\bibitem [{\citenamefont {Johns}(2023)}]{Johns:2023ewj}%
  \BibitemOpen
  \bibfield  {author} {\bibinfo {author} {\bibfnamefont {L.}~\bibnamefont
  {Johns}},\ }\href@noop {} {\bibinfo {title} {{Neutrino many-body
  correlations}}} (\bibinfo {year} {2023}),\ \Eprint
  {https://arxiv.org/abs/2305.04916} {arXiv:2305.04916 [hep-ph]} \BibitemShut
  {NoStop}%
\bibitem [{\citenamefont {Hall}\ \emph {et~al.}(2021)\citenamefont {Hall},
  \citenamefont {Roggero}, \citenamefont {Baroni},\ and\ \citenamefont
  {Carlson}}]{Hall:2021rbv}%
  \BibitemOpen
  \bibfield  {author} {\bibinfo {author} {\bibfnamefont {B.}~\bibnamefont
  {Hall}}, \bibinfo {author} {\bibfnamefont {A.}~\bibnamefont {Roggero}},
  \bibinfo {author} {\bibfnamefont {A.}~\bibnamefont {Baroni}},\ and\ \bibinfo
  {author} {\bibfnamefont {J.}~\bibnamefont {Carlson}},\ }\bibfield  {title}
  {\bibinfo {title} {{Simulation of collective neutrino oscillations on a
  quantum computer}},\ }\href {https://doi.org/10.1103/PhysRevD.104.063009}
  {\bibfield  {journal} {\bibinfo  {journal} {Phys. Rev. D}\ }\textbf {\bibinfo
  {volume} {104}},\ \bibinfo {pages} {063009} (\bibinfo {year} {2021})},\
  \Eprint {https://arxiv.org/abs/2102.12556} {arXiv:2102.12556 [quant-ph]}
  \BibitemShut {NoStop}%
\bibitem [{\citenamefont {Yeter-Aydeniz}\ \emph {et~al.}(2022)\citenamefont
  {Yeter-Aydeniz}, \citenamefont {Bangar}, \citenamefont {Siopsis},\ and\
  \citenamefont {Pooser}}]{Yeter-Aydeniz:2021olz}%
  \BibitemOpen
  \bibfield  {author} {\bibinfo {author} {\bibfnamefont {K.}~\bibnamefont
  {Yeter-Aydeniz}}, \bibinfo {author} {\bibfnamefont {S.}~\bibnamefont
  {Bangar}}, \bibinfo {author} {\bibfnamefont {G.}~\bibnamefont {Siopsis}},\
  and\ \bibinfo {author} {\bibfnamefont {R.~C.}\ \bibnamefont {Pooser}},\
  }\bibfield  {title} {\bibinfo {title} {{Collective neutrino oscillations on a
  quantum computer}},\ }\href {https://doi.org/10.1007/s11128-021-03348-x}
  {\bibfield  {journal} {\bibinfo  {journal} {Quantum Inf. Process}\ }\textbf
  {\bibinfo {volume} {21}},\ \bibinfo {pages} {84} (\bibinfo {year} {2022})},\
  \Eprint {https://arxiv.org/abs/2104.03273} {arXiv:2104.03273 [quant-ph]}
  \BibitemShut {NoStop}%
\bibitem [{\citenamefont {Illa}\ and\ \citenamefont
  {Savage}(2022)}]{Illa:2022jqb}%
  \BibitemOpen
  \bibfield  {author} {\bibinfo {author} {\bibfnamefont {M.}~\bibnamefont
  {Illa}}\ and\ \bibinfo {author} {\bibfnamefont {M.~J.}\ \bibnamefont
  {Savage}},\ }\bibfield  {title} {\bibinfo {title} {{Basic elements for
  simulations of standard-model physics with quantum annealers: Multigrid and
  clock states}},\ }\href {https://doi.org/10.1103/PhysRevA.106.052605}
  {\bibfield  {journal} {\bibinfo  {journal} {Phys. Rev. A}\ }\textbf {\bibinfo
  {volume} {106}},\ \bibinfo {pages} {052605} (\bibinfo {year} {2022})},\
  \Eprint {https://arxiv.org/abs/2202.12340} {arXiv:2202.12340 [quant-ph]}
  \BibitemShut {NoStop}%
\bibitem [{\citenamefont {Amitrano}\ \emph {et~al.}(2023)\citenamefont
  {Amitrano}, \citenamefont {Roggero}, \citenamefont {Luchi}, \citenamefont
  {Turro}, \citenamefont {Vespucci},\ and\ \citenamefont
  {Pederiva}}]{Amitrano:2022yyn}%
  \BibitemOpen
  \bibfield  {author} {\bibinfo {author} {\bibfnamefont {V.}~\bibnamefont
  {Amitrano}}, \bibinfo {author} {\bibfnamefont {A.}~\bibnamefont {Roggero}},
  \bibinfo {author} {\bibfnamefont {P.}~\bibnamefont {Luchi}}, \bibinfo
  {author} {\bibfnamefont {F.}~\bibnamefont {Turro}}, \bibinfo {author}
  {\bibfnamefont {L.}~\bibnamefont {Vespucci}},\ and\ \bibinfo {author}
  {\bibfnamefont {F.}~\bibnamefont {Pederiva}},\ }\bibfield  {title} {\bibinfo
  {title} {{Trapped-ion quantum simulation of collective neutrino
  oscillations}},\ }\href {https://doi.org/10.1103/PhysRevD.107.023007}
  {\bibfield  {journal} {\bibinfo  {journal} {Phys. Rev. D}\ }\textbf {\bibinfo
  {volume} {107}},\ \bibinfo {pages} {023007} (\bibinfo {year} {2023})},\
  \Eprint {https://arxiv.org/abs/2207.03189} {arXiv:2207.03189 [quant-ph]}
  \BibitemShut {NoStop}%
\bibitem [{\citenamefont {Illa}\ and\ \citenamefont
  {Savage}(2023)}]{Illa:2022zgu}%
  \BibitemOpen
  \bibfield  {author} {\bibinfo {author} {\bibfnamefont {M.}~\bibnamefont
  {Illa}}\ and\ \bibinfo {author} {\bibfnamefont {M.~J.}\ \bibnamefont
  {Savage}},\ }\bibfield  {title} {\bibinfo {title} {{Multi-Neutrino
  Entanglement and Correlations in Dense Neutrino Systems}},\ }\href
  {https://doi.org/10.1103/PhysRevLett.130.221003} {\bibfield  {journal}
  {\bibinfo  {journal} {Phys. Rev. Lett.}\ }\textbf {\bibinfo {volume} {130}},\
  \bibinfo {pages} {221003} (\bibinfo {year} {2023})},\ \Eprint
  {https://arxiv.org/abs/2210.08656} {arXiv:2210.08656 [nucl-th]} \BibitemShut
  {NoStop}%
\bibitem [{\citenamefont {Siwach}\ \emph
  {et~al.}(2023{\natexlab{a}})\citenamefont {Siwach}, \citenamefont
  {Harrison},\ and\ \citenamefont {Balantekin}}]{Siwach:2023wzy}%
  \BibitemOpen
  \bibfield  {author} {\bibinfo {author} {\bibfnamefont {P.}~\bibnamefont
  {Siwach}}, \bibinfo {author} {\bibfnamefont {K.}~\bibnamefont {Harrison}},\
  and\ \bibinfo {author} {\bibfnamefont {A.~B.}\ \bibnamefont {Balantekin}},\
  }\bibfield  {title} {\bibinfo {title} {{Collective neutrino oscillations on a
  quantum computer with hybrid quantum-classical algorithm}},\ }\href
  {https://doi.org/10.1103/PhysRevD.108.083039} {\bibfield  {journal} {\bibinfo
   {journal} {Phys. Rev. D}\ }\textbf {\bibinfo {volume} {108}},\ \bibinfo
  {pages} {083039} (\bibinfo {year} {2023}{\natexlab{a}})},\ \Eprint
  {https://arxiv.org/abs/2308.09123} {arXiv:2308.09123 [quant-ph]} \BibitemShut
  {NoStop}%
\bibitem [{\citenamefont {Feynman}(1982)}]{feynman}%
  \BibitemOpen
  \bibfield  {author} {\bibinfo {author} {\bibfnamefont {R.~P.}\ \bibnamefont
  {Feynman}},\ }\bibfield  {title} {\bibinfo {title} {{Simulating physics with
  computers}},\ }\href {https://doi.org/10.1007/BF02650179} {\bibfield
  {journal} {\bibinfo  {journal} {Int. J. Theor. Phys.}\ }\textbf {\bibinfo
  {volume} {21}},\ \bibinfo {pages} {467} (\bibinfo {year} {1982})}\BibitemShut
  {NoStop}%
\bibitem [{\citenamefont {Rrapaj}(2020)}]{Rrapaj:2019pxz}%
  \BibitemOpen
  \bibfield  {author} {\bibinfo {author} {\bibfnamefont {E.}~\bibnamefont
  {Rrapaj}},\ }\bibfield  {title} {\bibinfo {title} {{Exact solution of
  multiangle quantum many-body collective neutrino-flavor oscillations}},\
  }\href {https://doi.org/10.1103/PhysRevC.101.065805} {\bibfield  {journal}
  {\bibinfo  {journal} {Phys. Rev. C}\ }\textbf {\bibinfo {volume} {101}},\
  \bibinfo {pages} {065805} (\bibinfo {year} {2020})},\ \Eprint
  {https://arxiv.org/abs/1905.13335} {arXiv:1905.13335 [hep-ph]} \BibitemShut
  {NoStop}%
\bibitem [{\citenamefont {Patwardhan}\ \emph {et~al.}(2019)\citenamefont
  {Patwardhan}, \citenamefont {Cervia},\ and\ \citenamefont
  {Balantekin}}]{Patwardhan:2019zta}%
  \BibitemOpen
  \bibfield  {author} {\bibinfo {author} {\bibfnamefont {A.~V.}\ \bibnamefont
  {Patwardhan}}, \bibinfo {author} {\bibfnamefont {M.~J.}\ \bibnamefont
  {Cervia}},\ and\ \bibinfo {author} {\bibfnamefont {A.~B.}\ \bibnamefont
  {Balantekin}},\ }\bibfield  {title} {\bibinfo {title} {Eigenvalues and
  eigenstates of the many-body collective neutrino oscillation problem},\
  }\href {https://doi.org/10.1103/PhysRevD.99.123013} {\bibfield  {journal}
  {\bibinfo  {journal} {Phys. Rev. D}\ }\textbf {\bibinfo {volume} {99}},\
  \bibinfo {pages} {123013} (\bibinfo {year} {2019})}\BibitemShut {NoStop}%
\bibitem [{\citenamefont {Patwardhan}\ \emph {et~al.}(2021)\citenamefont
  {Patwardhan}, \citenamefont {Cervia},\ and\ \citenamefont
  {Balantekin}}]{Patwardhan:2021rej}%
  \BibitemOpen
  \bibfield  {author} {\bibinfo {author} {\bibfnamefont {A.~V.}\ \bibnamefont
  {Patwardhan}}, \bibinfo {author} {\bibfnamefont {M.~J.}\ \bibnamefont
  {Cervia}},\ and\ \bibinfo {author} {\bibfnamefont {A.~B.}\ \bibnamefont
  {Balantekin}},\ }\bibfield  {title} {\bibinfo {title} {{Spectral splits and
  entanglement entropy in collective neutrino oscillations}},\ }\href
  {https://doi.org/10.1103/PhysRevD.104.123035} {\bibfield  {journal} {\bibinfo
   {journal} {Phys. Rev. D}\ }\textbf {\bibinfo {volume} {104}},\ \bibinfo
  {pages} {123035} (\bibinfo {year} {2021})},\ \Eprint
  {https://arxiv.org/abs/2109.08995} {arXiv:2109.08995 [hep-ph]} \BibitemShut
  {NoStop}%
\bibitem [{\citenamefont {Martin}\ \emph
  {et~al.}(2023{\natexlab{a}})\citenamefont {Martin}, \citenamefont {Neill},
  \citenamefont {Roggero}, \citenamefont {Duan},\ and\ \citenamefont
  {Carlson}}]{Martin:2023gbo}%
  \BibitemOpen
  \bibfield  {author} {\bibinfo {author} {\bibfnamefont {J.~D.}\ \bibnamefont
  {Martin}}, \bibinfo {author} {\bibfnamefont {D.}~\bibnamefont {Neill}},
  \bibinfo {author} {\bibfnamefont {A.}~\bibnamefont {Roggero}}, \bibinfo
  {author} {\bibfnamefont {H.}~\bibnamefont {Duan}},\ and\ \bibinfo {author}
  {\bibfnamefont {J.}~\bibnamefont {Carlson}},\ }\bibfield  {title} {\bibinfo
  {title} {{Equilibration of quantum many-body fast neutrino flavor
  oscillations}},\ }\href {https://doi.org/10.1103/PhysRevD.108.123010}
  {\bibfield  {journal} {\bibinfo  {journal} {Phys. Rev. D}\ }\textbf {\bibinfo
  {volume} {108}},\ \bibinfo {pages} {123010} (\bibinfo {year}
  {2023}{\natexlab{a}})},\ \Eprint {https://arxiv.org/abs/2307.16793}
  {arXiv:2307.16793 [hep-ph]} \BibitemShut {NoStop}%
\bibitem [{\citenamefont {Neill}\ \emph {et~al.}(2024)\citenamefont {Neill},
  \citenamefont {Liu}, \citenamefont {Martin},\ and\ \citenamefont
  {Roggero}}]{Neill:2024klc}%
  \BibitemOpen
  \bibfield  {author} {\bibinfo {author} {\bibfnamefont {D.}~\bibnamefont
  {Neill}}, \bibinfo {author} {\bibfnamefont {H.}~\bibnamefont {Liu}}, \bibinfo
  {author} {\bibfnamefont {J.}~\bibnamefont {Martin}},\ and\ \bibinfo {author}
  {\bibfnamefont {A.}~\bibnamefont {Roggero}},\ }\href@noop {} {\bibinfo
  {title} {{Scattering Neutrinos, Spin Models, and Permutations}}} (\bibinfo
  {year} {2024}),\ \Eprint {https://arxiv.org/abs/2406.18677} {arXiv:2406.18677
  [hep-ph]} \BibitemShut {NoStop}%
\bibitem [{\citenamefont {Siwach}\ \emph
  {et~al.}(2023{\natexlab{b}})\citenamefont {Siwach}, \citenamefont {Suliga},\
  and\ \citenamefont {Balantekin}}]{Siwach:2022xhx}%
  \BibitemOpen
  \bibfield  {author} {\bibinfo {author} {\bibfnamefont {P.}~\bibnamefont
  {Siwach}}, \bibinfo {author} {\bibfnamefont {A.~M.}\ \bibnamefont {Suliga}},\
  and\ \bibinfo {author} {\bibfnamefont {A.~B.}\ \bibnamefont {Balantekin}},\
  }\bibfield  {title} {\bibinfo {title} {{Entanglement in three-flavor
  collective neutrino oscillations}},\ }\href
  {https://doi.org/10.1103/PhysRevD.107.023019} {\bibfield  {journal} {\bibinfo
   {journal} {Phys. Rev. D}\ }\textbf {\bibinfo {volume} {107}},\ \bibinfo
  {pages} {023019} (\bibinfo {year} {2023}{\natexlab{b}})},\ \Eprint
  {https://arxiv.org/abs/2211.07678} {arXiv:2211.07678 [hep-ph]} \BibitemShut
  {NoStop}%
\bibitem [{\citenamefont {Chernyshev}(2024)}]{Chernyshev:2024kpu}%
  \BibitemOpen
  \bibfield  {author} {\bibinfo {author} {\bibfnamefont {I.~A.}\ \bibnamefont
  {Chernyshev}},\ }\href@noop {} {\bibinfo {title} {{Three-flavor Collective
  Neutrino Oscillations on D-Wave's Advantage Quantum Annealer}}} (\bibinfo
  {year} {2024}),\ \Eprint {https://arxiv.org/abs/2405.20436} {arXiv:2405.20436
  [quant-ph]} \BibitemShut {NoStop}%
\bibitem [{ibm(2021)}]{ibm}%
  \BibitemOpen
  \href {https://quantum.ibm.com/} {\bibinfo {title} {{IBM Quantum}}} (\bibinfo
  {year} {2021})\BibitemShut {NoStop}%
\bibitem [{qua(2024)}]{quantinuum}%
  \BibitemOpen
  \href {https://www.quantinuum.com/} {\bibinfo {title} {Quantinuum}} (\bibinfo
  {year} {2024})\BibitemShut {NoStop}%
\bibitem [{\citenamefont {Pehlivan}\ \emph {et~al.}(2011)\citenamefont
  {Pehlivan}, \citenamefont {Balantekin}, \citenamefont {Kajino},\ and\
  \citenamefont {Yoshida}}]{Pehlivan:2011hp}%
  \BibitemOpen
  \bibfield  {author} {\bibinfo {author} {\bibfnamefont {Y.}~\bibnamefont
  {Pehlivan}}, \bibinfo {author} {\bibfnamefont {A.~B.}\ \bibnamefont
  {Balantekin}}, \bibinfo {author} {\bibfnamefont {T.}~\bibnamefont {Kajino}},\
  and\ \bibinfo {author} {\bibfnamefont {T.}~\bibnamefont {Yoshida}},\
  }\bibfield  {title} {\bibinfo {title} {{Invariants of Collective Neutrino
  Oscillations}},\ }\href {https://doi.org/10.1103/PhysRevD.84.065008}
  {\bibfield  {journal} {\bibinfo  {journal} {Phys. Rev. D}\ }\textbf {\bibinfo
  {volume} {84}},\ \bibinfo {pages} {065008} (\bibinfo {year} {2011})},\
  \Eprint {https://arxiv.org/abs/1105.1182} {arXiv:1105.1182 [astro-ph.CO]}
  \BibitemShut {NoStop}%
\bibitem [{\citenamefont {Cervia}\ \emph {et~al.}(2022)\citenamefont {Cervia},
  \citenamefont {Siwach}, \citenamefont {Patwardhan}, \citenamefont
  {Balantekin}, \citenamefont {Coppersmith},\ and\ \citenamefont
  {Johnson}}]{Cervia:2022pro}%
  \BibitemOpen
  \bibfield  {author} {\bibinfo {author} {\bibfnamefont {M.~J.}\ \bibnamefont
  {Cervia}}, \bibinfo {author} {\bibfnamefont {P.}~\bibnamefont {Siwach}},
  \bibinfo {author} {\bibfnamefont {A.~V.}\ \bibnamefont {Patwardhan}},
  \bibinfo {author} {\bibfnamefont {A.~B.}\ \bibnamefont {Balantekin}},
  \bibinfo {author} {\bibfnamefont {S.~N.}\ \bibnamefont {Coppersmith}},\ and\
  \bibinfo {author} {\bibfnamefont {C.~W.}\ \bibnamefont {Johnson}},\
  }\bibfield  {title} {\bibinfo {title} {{Collective neutrino oscillations with
  tensor networks using a time-dependent variational principle}},\ }\href
  {https://doi.org/10.1103/PhysRevD.105.123025} {\bibfield  {journal} {\bibinfo
   {journal} {Phys. Rev. D}\ }\textbf {\bibinfo {volume} {105}},\ \bibinfo
  {pages} {123025} (\bibinfo {year} {2022})},\ \Eprint
  {https://arxiv.org/abs/2202.01865} {arXiv:2202.01865 [hep-ph]} \BibitemShut
  {NoStop}%
\bibitem [{\citenamefont {Pantaleone}(1992{\natexlab{c}})}]{PANTALEONE1992128}%
  \BibitemOpen
  \bibfield  {author} {\bibinfo {author} {\bibfnamefont {J.}~\bibnamefont
  {Pantaleone}},\ }\bibfield  {title} {\bibinfo {title} {Neutrino oscillations
  at high densities},\ }\href
  {https://doi.org/https://doi.org/10.1016/0370-2693(92)91887-F} {\bibfield
  {journal} {\bibinfo  {journal} {Physics Letters B}\ }\textbf {\bibinfo
  {volume} {287}},\ \bibinfo {pages} {128} (\bibinfo {year}
  {1992}{\natexlab{c}})}\BibitemShut {NoStop}%
\bibitem [{\citenamefont {Friedland}\ and\ \citenamefont
  {Lunardini}(2003{\natexlab{a}})}]{friedland2003manyparticle}%
  \BibitemOpen
  \bibfield  {author} {\bibinfo {author} {\bibfnamefont {A.}~\bibnamefont
  {Friedland}}\ and\ \bibinfo {author} {\bibfnamefont {C.}~\bibnamefont
  {Lunardini}},\ }\bibfield  {title} {\bibinfo {title} {{Do many particle
  neutrino interactions cause a novel coherent effect?}},\ }\href
  {https://doi.org/10.1088/1126-6708/2003/10/043} {\bibfield  {journal}
  {\bibinfo  {journal} {JHEP}\ }\textbf {\bibinfo {volume} {10}},\ \bibinfo
  {pages} {043}},\ \Eprint {https://arxiv.org/abs/hep-ph/0307140}
  {arXiv:hep-ph/0307140} \BibitemShut {NoStop}%
\bibitem [{\citenamefont {Friedland}\ and\ \citenamefont
  {Lunardini}(2003{\natexlab{b}})}]{Friedland2003Neutrino}%
  \BibitemOpen
  \bibfield  {author} {\bibinfo {author} {\bibfnamefont {A.}~\bibnamefont
  {Friedland}}\ and\ \bibinfo {author} {\bibfnamefont {C.}~\bibnamefont
  {Lunardini}},\ }\bibfield  {title} {\bibinfo {title} {{Neutrino flavor
  conversion in a neutrino background: Single particle versus multiparticle
  description}},\ }\href {https://doi.org/10.1103/PhysRevD.68.013007}
  {\bibfield  {journal} {\bibinfo  {journal} {Phys. Rev. D}\ }\textbf {\bibinfo
  {volume} {68}},\ \bibinfo {pages} {013007} (\bibinfo {year}
  {2003}{\natexlab{b}})},\ \Eprint {https://arxiv.org/abs/hep-ph/0304055}
  {arXiv:hep-ph/0304055} \BibitemShut {NoStop}%
\bibitem [{\citenamefont {Bell}\ \emph {et~al.}(2003)\citenamefont {Bell},
  \citenamefont {Rawlinson},\ and\ \citenamefont {Sawyer}}]{Bell2003Speedup}%
  \BibitemOpen
  \bibfield  {author} {\bibinfo {author} {\bibfnamefont {N.~F.}\ \bibnamefont
  {Bell}}, \bibinfo {author} {\bibfnamefont {A.~A.}\ \bibnamefont
  {Rawlinson}},\ and\ \bibinfo {author} {\bibfnamefont {R.~F.}\ \bibnamefont
  {Sawyer}},\ }\bibfield  {title} {\bibinfo {title} {Speedup through
  entanglement: {M}any-body effects in neutrino processes},\ }\href
  {https://doi.org/10.1016/j.physletb.2003.08.086} {\bibfield  {journal}
  {\bibinfo  {journal} {Phys. Lett. B}\ }\textbf {\bibinfo {volume} {573}},\
  \bibinfo {pages} {86} (\bibinfo {year} {2003})}\BibitemShut {NoStop}%
\bibitem [{\citenamefont {Sawyer}(2004)}]{Sawyer2004Instabilities}%
  \BibitemOpen
  \bibfield  {author} {\bibinfo {author} {\bibfnamefont {R.~F.}\ \bibnamefont
  {Sawyer}},\ }\href@noop {} {\bibinfo {title} {{'Classical' instabilities and
  'quantum' speed-up in the evolution of neutrino clouds}}} (\bibinfo {year}
  {2004}),\ \Eprint {https://arxiv.org/abs/hep-ph/0408265}
  {arXiv:hep-ph/0408265} \BibitemShut {NoStop}%
\bibitem [{\citenamefont {Friedland}\ \emph {et~al.}(2006)\citenamefont
  {Friedland}, \citenamefont {McKellar},\ and\ \citenamefont
  {Okuniewicz}}]{Friedland2006ManyBody}%
  \BibitemOpen
  \bibfield  {author} {\bibinfo {author} {\bibfnamefont {A.}~\bibnamefont
  {Friedland}}, \bibinfo {author} {\bibfnamefont {B.~H.~J.}\ \bibnamefont
  {McKellar}},\ and\ \bibinfo {author} {\bibfnamefont {I.}~\bibnamefont
  {Okuniewicz}},\ }\bibfield  {title} {\bibinfo {title} {Construction and
  analysis of a simplified many-body neutrino model},\ }\href
  {https://doi.org/10.1103/PhysRevD.73.093002} {\bibfield  {journal} {\bibinfo
  {journal} {Phys. Rev. D}\ }\textbf {\bibinfo {volume} {73}},\ \bibinfo
  {pages} {093002} (\bibinfo {year} {2006})}\BibitemShut {NoStop}%
\bibitem [{\citenamefont {Birol}\ \emph {et~al.}(2018)\citenamefont {Birol},
  \citenamefont {Pehlivan}, \citenamefont {Balantekin},\ and\ \citenamefont
  {Kajino}}]{birol2018neutrino}%
  \BibitemOpen
  \bibfield  {author} {\bibinfo {author} {\bibfnamefont {S.}~\bibnamefont
  {Birol}}, \bibinfo {author} {\bibfnamefont {Y.}~\bibnamefont {Pehlivan}},
  \bibinfo {author} {\bibfnamefont {A.~B.}\ \bibnamefont {Balantekin}},\ and\
  \bibinfo {author} {\bibfnamefont {T.}~\bibnamefont {Kajino}},\ }\bibfield
  {title} {\bibinfo {title} {Neutrino spectral split in the exact many-body
  formalism},\ }\href {https://doi.org/10.1103/PhysRevD.98.083002} {\bibfield
  {journal} {\bibinfo  {journal} {Phys. Rev. D}\ }\textbf {\bibinfo {volume}
  {98}},\ \bibinfo {pages} {083002} (\bibinfo {year} {2018})}\BibitemShut
  {NoStop}%
\bibitem [{\citenamefont {Cervia}\ \emph {et~al.}(2019)\citenamefont {Cervia},
  \citenamefont {Patwardhan}, \citenamefont {Balantekin}, \citenamefont
  {Coppersmith},\ and\ \citenamefont {Johnson}}]{Cervia2019Entanglement}%
  \BibitemOpen
  \bibfield  {author} {\bibinfo {author} {\bibfnamefont {M.~J.}\ \bibnamefont
  {Cervia}}, \bibinfo {author} {\bibfnamefont {A.~V.}\ \bibnamefont
  {Patwardhan}}, \bibinfo {author} {\bibfnamefont {A.~B.}\ \bibnamefont
  {Balantekin}}, \bibinfo {author} {\bibfnamefont {S.~N.}\ \bibnamefont
  {Coppersmith}},\ and\ \bibinfo {author} {\bibfnamefont {C.~W.}\ \bibnamefont
  {Johnson}},\ }\bibfield  {title} {\bibinfo {title} {Entanglement and
  collective flavor oscillations in a dense neutrino gas},\ }\href
  {https://doi.org/10.1103/PhysRevD.100.083001} {\bibfield  {journal} {\bibinfo
   {journal} {Phys. Rev. D}\ }\textbf {\bibinfo {volume} {100}},\ \bibinfo
  {pages} {083001} (\bibinfo {year} {2019})}\BibitemShut {NoStop}%
\bibitem [{\citenamefont {Roggero}(2021{\natexlab{b}})}]{roggero2021dynamical}%
  \BibitemOpen
  \bibfield  {author} {\bibinfo {author} {\bibfnamefont {A.}~\bibnamefont
  {Roggero}},\ }\bibfield  {title} {\bibinfo {title} {{Dynamical phase
  transitions in models of collective neutrino oscillations}},\ }\href
  {https://doi.org/10.1103/PhysRevD.104.123023} {\bibfield  {journal} {\bibinfo
   {journal} {Phys. Rev. D}\ }\textbf {\bibinfo {volume} {104}},\ \bibinfo
  {pages} {123023} (\bibinfo {year} {2021}{\natexlab{b}})},\ \Eprint
  {https://arxiv.org/abs/2103.11497} {arXiv:2103.11497 [hep-ph]} \BibitemShut
  {NoStop}%
\bibitem [{\citenamefont {Martin}\ \emph {et~al.}(2022)\citenamefont {Martin},
  \citenamefont {Roggero}, \citenamefont {Duan}, \citenamefont {Carlson},\ and\
  \citenamefont {Cirigliano}}]{Martin2022Classical}%
  \BibitemOpen
  \bibfield  {author} {\bibinfo {author} {\bibfnamefont {J.~D.}\ \bibnamefont
  {Martin}}, \bibinfo {author} {\bibfnamefont {A.}~\bibnamefont {Roggero}},
  \bibinfo {author} {\bibfnamefont {H.}~\bibnamefont {Duan}}, \bibinfo {author}
  {\bibfnamefont {J.}~\bibnamefont {Carlson}},\ and\ \bibinfo {author}
  {\bibfnamefont {V.}~\bibnamefont {Cirigliano}},\ }\bibfield  {title}
  {\bibinfo {title} {Classical and quantum evolution in a simple coherent
  neutrino problem},\ }\href {https://doi.org/10.1103/PhysRevD.105.083020}
  {\bibfield  {journal} {\bibinfo  {journal} {Phys. Rev. D}\ }\textbf {\bibinfo
  {volume} {105}},\ \bibinfo {pages} {083020} (\bibinfo {year}
  {2022})}\BibitemShut {NoStop}%
\bibitem [{\citenamefont {Lacroix}\ \emph {et~al.}(2022)\citenamefont
  {Lacroix}, \citenamefont {Balantekin}, \citenamefont {Cervia}, \citenamefont
  {Patwardhan},\ and\ \citenamefont {Siwach}}]{lacroix2022role}%
  \BibitemOpen
  \bibfield  {author} {\bibinfo {author} {\bibfnamefont {D.}~\bibnamefont
  {Lacroix}}, \bibinfo {author} {\bibfnamefont {A.~B.}\ \bibnamefont
  {Balantekin}}, \bibinfo {author} {\bibfnamefont {M.~J.}\ \bibnamefont
  {Cervia}}, \bibinfo {author} {\bibfnamefont {A.~V.}\ \bibnamefont
  {Patwardhan}},\ and\ \bibinfo {author} {\bibfnamefont {P.}~\bibnamefont
  {Siwach}},\ }\bibfield  {title} {\bibinfo {title} {Role of non-{Gaussian}
  quantum fluctuations in neutrino entanglement},\ }\href
  {https://doi.org/10.1103/PhysRevD.106.123006} {\bibfield  {journal} {\bibinfo
   {journal} {Phys. Rev. D}\ }\textbf {\bibinfo {volume} {106}},\ \bibinfo
  {pages} {123006} (\bibinfo {year} {2022})}\BibitemShut {NoStop}%
\bibitem [{\citenamefont {Martin}\ \emph
  {et~al.}(2023{\natexlab{b}})\citenamefont {Martin}, \citenamefont {Roggero},
  \citenamefont {Duan},\ and\ \citenamefont {Carlson}}]{Martin2023ManyBody}%
  \BibitemOpen
  \bibfield  {author} {\bibinfo {author} {\bibfnamefont {J.~D.}\ \bibnamefont
  {Martin}}, \bibinfo {author} {\bibfnamefont {A.}~\bibnamefont {Roggero}},
  \bibinfo {author} {\bibfnamefont {H.}~\bibnamefont {Duan}},\ and\ \bibinfo
  {author} {\bibfnamefont {J.}~\bibnamefont {Carlson}},\ }\bibfield  {title}
  {\bibinfo {title} {Many-body neutrino flavor entanglement in a simple dynamic
  model},\ }\href@noop {} {\  (\bibinfo {year} {2023}{\natexlab{b}})},\ \Eprint
  {https://arxiv.org/abs/2301.07049} {arXiv:2301.07049 [hep-ph]} \BibitemShut
  {NoStop}%
\bibitem [{\citenamefont {Esteban}\ \emph {et~al.}(2020)\citenamefont
  {Esteban}, \citenamefont {Gonz{\'a}lez-Garc{\'\i}a}, \citenamefont {Maltoni},
  \citenamefont {Schwetz},\ and\ \citenamefont {Zhou}}]{Esteban:2020cvm}%
  \BibitemOpen
  \bibfield  {author} {\bibinfo {author} {\bibfnamefont {I.}~\bibnamefont
  {Esteban}}, \bibinfo {author} {\bibfnamefont {M.~C.}\ \bibnamefont
  {Gonz{\'a}lez-Garc{\'\i}a}}, \bibinfo {author} {\bibfnamefont
  {M.}~\bibnamefont {Maltoni}}, \bibinfo {author} {\bibfnamefont
  {T.}~\bibnamefont {Schwetz}},\ and\ \bibinfo {author} {\bibfnamefont
  {A.}~\bibnamefont {Zhou}},\ }\bibfield  {title} {\bibinfo {title} {The fate
  of hints: updated global analysis of three-flavor neutrino oscillations},\
  }\href {https://doi.org/10.1007/JHEP09(2020)178} {\bibfield  {journal}
  {\bibinfo  {journal} {JHEP}\ }\textbf {\bibinfo {volume} {09}},\ \bibinfo
  {pages} {178}},\ \Eprint {https://arxiv.org/abs/2007.14792} {arXiv:2007.14792
  [hep-ph]} \BibitemShut {NoStop}%
\bibitem [{nuf(2024)}]{nufit}%
  \BibitemOpen
  \href@noop {} {\bibinfo {title} {Nufit 5.3, \url{www.nu-fit.org}}} (\bibinfo
  {year} {2024})\BibitemShut {NoStop}%
\bibitem [{\citenamefont {de~Salas}\ \emph {et~al.}(2021)\citenamefont
  {de~Salas}, \citenamefont {Forero}, \citenamefont {Gariazzo}, \citenamefont
  {Mart\'\i{}nez-Mirav\'e}, \citenamefont {Mena}, \citenamefont {Ternes},
  \citenamefont {T\'ortola},\ and\ \citenamefont {Valle}}]{deSalas:2020pgw}%
  \BibitemOpen
  \bibfield  {author} {\bibinfo {author} {\bibfnamefont {P.~F.}\ \bibnamefont
  {de~Salas}}, \bibinfo {author} {\bibfnamefont {D.~V.}\ \bibnamefont
  {Forero}}, \bibinfo {author} {\bibfnamefont {S.}~\bibnamefont {Gariazzo}},
  \bibinfo {author} {\bibfnamefont {P.}~\bibnamefont {Mart\'\i{}nez-Mirav\'e}},
  \bibinfo {author} {\bibfnamefont {O.}~\bibnamefont {Mena}}, \bibinfo {author}
  {\bibfnamefont {C.~A.}\ \bibnamefont {Ternes}}, \bibinfo {author}
  {\bibfnamefont {M.}~\bibnamefont {T\'ortola}},\ and\ \bibinfo {author}
  {\bibfnamefont {J.~W.~F.}\ \bibnamefont {Valle}},\ }\bibfield  {title}
  {\bibinfo {title} {{2020 global reassessment of the neutrino oscillation
  picture}},\ }\href {https://doi.org/10.1007/JHEP02(2021)071} {\bibfield
  {journal} {\bibinfo  {journal} {JHEP}\ }\textbf {\bibinfo {volume} {02}},\
  \bibinfo {pages} {071}},\ \Eprint {https://arxiv.org/abs/2006.11237}
  {arXiv:2006.11237 [hep-ph]} \BibitemShut {NoStop}%
\bibitem [{\citenamefont {Capozzi}\ \emph {et~al.}(2021)\citenamefont
  {Capozzi}, \citenamefont {Di~Valentino}, \citenamefont {Lisi}, \citenamefont
  {Marrone}, \citenamefont {Melchiorri},\ and\ \citenamefont
  {Palazzo}}]{Capozzi:2021fjo}%
  \BibitemOpen
  \bibfield  {author} {\bibinfo {author} {\bibfnamefont {F.}~\bibnamefont
  {Capozzi}}, \bibinfo {author} {\bibfnamefont {E.}~\bibnamefont
  {Di~Valentino}}, \bibinfo {author} {\bibfnamefont {E.}~\bibnamefont {Lisi}},
  \bibinfo {author} {\bibfnamefont {A.}~\bibnamefont {Marrone}}, \bibinfo
  {author} {\bibfnamefont {A.}~\bibnamefont {Melchiorri}},\ and\ \bibinfo
  {author} {\bibfnamefont {A.}~\bibnamefont {Palazzo}},\ }\bibfield  {title}
  {\bibinfo {title} {{Unfinished fabric of the three neutrino paradigm}},\
  }\href {https://doi.org/10.1103/PhysRevD.104.083031} {\bibfield  {journal}
  {\bibinfo  {journal} {Phys. Rev. D}\ }\textbf {\bibinfo {volume} {104}},\
  \bibinfo {pages} {083031} (\bibinfo {year} {2021})},\ \Eprint
  {https://arxiv.org/abs/2107.00532} {arXiv:2107.00532 [hep-ph]} \BibitemShut
  {NoStop}%
\bibitem [{\citenamefont {Nguyen}\ \emph
  {et~al.}(2023{\natexlab{b}})\citenamefont {Nguyen}, \citenamefont {Bach},
  \citenamefont {Nguyen}, \citenamefont {Tran}, \citenamefont {Nguyen},\ and\
  \citenamefont {Nguyen}}]{Nguyen:2022snr}%
  \BibitemOpen
  \bibfield  {author} {\bibinfo {author} {\bibfnamefont {H.~C.}\ \bibnamefont
  {Nguyen}}, \bibinfo {author} {\bibfnamefont {B.~G.}\ \bibnamefont {Bach}},
  \bibinfo {author} {\bibfnamefont {T.~D.}\ \bibnamefont {Nguyen}}, \bibinfo
  {author} {\bibfnamefont {D.~M.}\ \bibnamefont {Tran}}, \bibinfo {author}
  {\bibfnamefont {D.~V.}\ \bibnamefont {Nguyen}},\ and\ \bibinfo {author}
  {\bibfnamefont {H.~Q.}\ \bibnamefont {Nguyen}},\ }\bibfield  {title}
  {\bibinfo {title} {{Simulating neutrino oscillations on a superconducting
  qutrit}},\ }\href {https://doi.org/10.1103/PhysRevD.108.023013} {\bibfield
  {journal} {\bibinfo  {journal} {Phys. Rev. D}\ }\textbf {\bibinfo {volume}
  {108}},\ \bibinfo {pages} {023013} (\bibinfo {year} {2023}{\natexlab{b}})},\
  \Eprint {https://arxiv.org/abs/2212.14170} {arXiv:2212.14170 [quant-ph]}
  \BibitemShut {NoStop}%
\bibitem [{\citenamefont {Kivlichan}\ \emph {et~al.}(2018)\citenamefont
  {Kivlichan}, \citenamefont {McClean}, \citenamefont {Wiebe}, \citenamefont
  {Gidney}, \citenamefont {Aspuru-Guzik}, \citenamefont {Chan},\ and\
  \citenamefont {Babbush}}]{Kivlichan2018prl}%
  \BibitemOpen
  \bibfield  {author} {\bibinfo {author} {\bibfnamefont {I.~D.}\ \bibnamefont
  {Kivlichan}}, \bibinfo {author} {\bibfnamefont {J.}~\bibnamefont {McClean}},
  \bibinfo {author} {\bibfnamefont {N.}~\bibnamefont {Wiebe}}, \bibinfo
  {author} {\bibfnamefont {C.}~\bibnamefont {Gidney}}, \bibinfo {author}
  {\bibfnamefont {A.}~\bibnamefont {Aspuru-Guzik}}, \bibinfo {author}
  {\bibfnamefont {G.~K.-L.}\ \bibnamefont {Chan}},\ and\ \bibinfo {author}
  {\bibfnamefont {R.}~\bibnamefont {Babbush}},\ }\bibfield  {title} {\bibinfo
  {title} {Quantum simulation of electronic structure with linear depth and
  connectivity},\ }\href {https://doi.org/10.1103/PhysRevLett.120.110501}
  {\bibfield  {journal} {\bibinfo  {journal} {Phys. Rev. Lett.}\ }\textbf
  {\bibinfo {volume} {120}},\ \bibinfo {pages} {110501} (\bibinfo {year}
  {2018})}\BibitemShut {NoStop}%
\bibitem [{\citenamefont {O'Gorman}\ \emph {et~al.}(2019)\citenamefont
  {O'Gorman}, \citenamefont {Huggins}, \citenamefont {Rieffel},\ and\
  \citenamefont {Whaley}}]{OGorman:2019mll}%
  \BibitemOpen
  \bibfield  {author} {\bibinfo {author} {\bibfnamefont {B.}~\bibnamefont
  {O'Gorman}}, \bibinfo {author} {\bibfnamefont {W.~J.}\ \bibnamefont
  {Huggins}}, \bibinfo {author} {\bibfnamefont {E.~G.}\ \bibnamefont
  {Rieffel}},\ and\ \bibinfo {author} {\bibfnamefont {K.~B.}\ \bibnamefont
  {Whaley}},\ }\href@noop {} {\bibinfo {title} {{Generalized swap networks for
  near-term quantum computing}}} (\bibinfo {year} {2019}),\ \Eprint
  {https://arxiv.org/abs/1905.05118} {arXiv:1905.05118 [quant-ph]} \BibitemShut
  {NoStop}%
\bibitem [{\citenamefont {{Suzuki}}(1990)}]{Suzuki1990}%
  \BibitemOpen
  \bibfield  {author} {\bibinfo {author} {\bibfnamefont {M.}~\bibnamefont
  {{Suzuki}}},\ }\bibfield  {title} {\bibinfo {title} {{Fractal decomposition
  of exponential operators with applications to many-body theories and Monte
  Carlo simulations}},\ }\href {https://doi.org/10.1016/0375-9601(90)90962-N}
  {\bibfield  {journal} {\bibinfo  {journal} {Phys. Lett. A}\ }\textbf
  {\bibinfo {volume} {146}},\ \bibinfo {pages} {319} (\bibinfo {year}
  {1990})}\BibitemShut {NoStop}%
\bibitem [{\citenamefont {{Suzuki}}(1991)}]{Suzuki1991}%
  \BibitemOpen
  \bibfield  {author} {\bibinfo {author} {\bibfnamefont {M.}~\bibnamefont
  {{Suzuki}}},\ }\bibfield  {title} {\bibinfo {title} {{General theory of
  fractal path integrals with applications to many-body theories and
  statistical physics}},\ }\href {https://doi.org/10.1063/1.529425} {\bibfield
  {journal} {\bibinfo  {journal} {J. Math. Phys.}\ }\textbf {\bibinfo {volume}
  {32}},\ \bibinfo {pages} {400} (\bibinfo {year} {1991})}\BibitemShut
  {NoStop}%
\bibitem [{\citenamefont {Arg\"uelles}\ and\ \citenamefont
  {Jones}(2019)}]{Arguelles:2019phs}%
  \BibitemOpen
  \bibfield  {author} {\bibinfo {author} {\bibfnamefont {C.~A.}\ \bibnamefont
  {Arg\"uelles}}\ and\ \bibinfo {author} {\bibfnamefont {B.~J.~P.}\
  \bibnamefont {Jones}},\ }\bibfield  {title} {\bibinfo {title} {{Neutrino
  Oscillations in a Quantum Processor}},\ }\href
  {https://doi.org/10.1103/PhysRevResearch.1.033176} {\bibfield  {journal}
  {\bibinfo  {journal} {Phys. Rev. Research.}\ }\textbf {\bibinfo {volume}
  {1}},\ \bibinfo {pages} {033176} (\bibinfo {year} {2019})},\ \Eprint
  {https://arxiv.org/abs/1904.10559} {arXiv:1904.10559 [quant-ph]} \BibitemShut
  {NoStop}%
\bibitem [{\citenamefont {{Vatan}}\ and\ \citenamefont
  {{Williams}}(2004)}]{Vatan:2004nmz}%
  \BibitemOpen
  \bibfield  {author} {\bibinfo {author} {\bibfnamefont {F.}~\bibnamefont
  {{Vatan}}}\ and\ \bibinfo {author} {\bibfnamefont {C.}~\bibnamefont
  {{Williams}}},\ }\bibfield  {title} {\bibinfo {title} {{Optimal quantum
  circuits for general two-qubit gates}},\ }\href
  {https://doi.org/10.1103/PhysRevA.69.032315} {\bibfield  {journal} {\bibinfo
  {journal} {Phys. Rev. A}\ }\textbf {\bibinfo {volume} {69}},\ \bibinfo
  {pages} {032315} (\bibinfo {year} {2004})},\ \Eprint
  {https://arxiv.org/abs/quant-ph/0308006} {arXiv:quant-ph/0308006}
  \BibitemShut {NoStop}%
\bibitem [{\citenamefont {Vidal}\ and\ \citenamefont
  {Dawson}(2004)}]{PhysRevA.69.010301}%
  \BibitemOpen
  \bibfield  {author} {\bibinfo {author} {\bibfnamefont {G.}~\bibnamefont
  {Vidal}}\ and\ \bibinfo {author} {\bibfnamefont {C.~M.}\ \bibnamefont
  {Dawson}},\ }\bibfield  {title} {\bibinfo {title} {Universal quantum circuit
  for two-qubit transformations with three controlled-not gates},\ }\href
  {https://doi.org/10.1103/PhysRevA.69.010301} {\bibfield  {journal} {\bibinfo
  {journal} {Phys. Rev. A}\ }\textbf {\bibinfo {volume} {69}},\ \bibinfo
  {pages} {010301} (\bibinfo {year} {2004})}\BibitemShut {NoStop}%
\bibitem [{\citenamefont {Coffey}\ \emph {et~al.}(2008)\citenamefont {Coffey},
  \citenamefont {Deiotte},\ and\ \citenamefont {Semi}}]{PhysRevA.77.066301}%
  \BibitemOpen
  \bibfield  {author} {\bibinfo {author} {\bibfnamefont {M.~W.}\ \bibnamefont
  {Coffey}}, \bibinfo {author} {\bibfnamefont {R.}~\bibnamefont {Deiotte}},\
  and\ \bibinfo {author} {\bibfnamefont {T.}~\bibnamefont {Semi}},\ }\bibfield
  {title} {\bibinfo {title} {{Comment on ``Universal quantum circuit for
  two-qubit transformations with three controlled-NOT gates'' and ``Recognizing
  small-circuit structure in two-qubit operators''}},\ }\href
  {https://doi.org/10.1103/PhysRevA.77.066301} {\bibfield  {journal} {\bibinfo
  {journal} {Phys. Rev. A}\ }\textbf {\bibinfo {volume} {77}},\ \bibinfo
  {pages} {066301} (\bibinfo {year} {2008})}\BibitemShut {NoStop}%
\bibitem [{\citenamefont {Molewski}\ and\ \citenamefont
  {Jones}(2022)}]{Molewski:2021ogs}%
  \BibitemOpen
  \bibfield  {author} {\bibinfo {author} {\bibfnamefont {M.~J.}\ \bibnamefont
  {Molewski}}\ and\ \bibinfo {author} {\bibfnamefont {B.~J.~P.}\ \bibnamefont
  {Jones}},\ }\bibfield  {title} {\bibinfo {title} {{Scalable qubit
  representations of neutrino mixing matrices}},\ }\href
  {https://doi.org/10.1103/PhysRevD.105.056024} {\bibfield  {journal} {\bibinfo
   {journal} {Phys. Rev. D}\ }\textbf {\bibinfo {volume} {105}},\ \bibinfo
  {pages} {056024} (\bibinfo {year} {2022})},\ \Eprint
  {https://arxiv.org/abs/2111.05401} {arXiv:2111.05401 [quant-ph]} \BibitemShut
  {NoStop}%
\bibitem [{\citenamefont {{Qiskit contributors}}(2024)}]{qiskit}%
  \BibitemOpen
  \bibfield  {author} {\bibinfo {author} {\bibnamefont {{Qiskit
  contributors}}},\ }\href {https://doi.org/10.5281/zenodo.10798865} {\bibinfo
  {title} {Qiskit 1.0.2: An open-source framework for quantum computing}}
  (\bibinfo {year} {2024})\BibitemShut {NoStop}%
\bibitem [{\citenamefont {Javadi-Abhari}\ \emph {et~al.}(2024)\citenamefont
  {Javadi-Abhari} \emph {et~al.}}]{Javadi-Abhari:2024kbf}%
  \BibitemOpen
  \bibfield  {author} {\bibinfo {author} {\bibfnamefont {A.}~\bibnamefont
  {Javadi-Abhari}} \emph {et~al.},\ }\href@noop {} {\bibinfo {title} {{Quantum
  computing with Qiskit}}} (\bibinfo {year} {2024}),\ \Eprint
  {https://arxiv.org/abs/2405.08810} {arXiv:2405.08810 [quant-ph]} \BibitemShut
  {NoStop}%
\bibitem [{\citenamefont {Sivarajah}\ \emph {et~al.}(2020)\citenamefont
  {Sivarajah}, \citenamefont {Dilkes}, \citenamefont {Cowtan}, \citenamefont
  {Simmons}, \citenamefont {Edgington},\ and\ \citenamefont
  {Duncan}}]{Sivarajah:2020lfo}%
  \BibitemOpen
  \bibfield  {author} {\bibinfo {author} {\bibfnamefont {S.}~\bibnamefont
  {Sivarajah}}, \bibinfo {author} {\bibfnamefont {S.}~\bibnamefont {Dilkes}},
  \bibinfo {author} {\bibfnamefont {A.}~\bibnamefont {Cowtan}}, \bibinfo
  {author} {\bibfnamefont {W.}~\bibnamefont {Simmons}}, \bibinfo {author}
  {\bibfnamefont {A.}~\bibnamefont {Edgington}},\ and\ \bibinfo {author}
  {\bibfnamefont {R.}~\bibnamefont {Duncan}},\ }\bibfield  {title} {\bibinfo
  {title} {{t$|$ket\ensuremath{\rangle}: a retargetable compiler for NISQ
  devices}},\ }\href {https://doi.org/10.1088/2058-9565/ab8e92} {\bibfield
  {journal} {\bibinfo  {journal} {Quantum Sci. Technol.}\ }\textbf {\bibinfo
  {volume} {6}},\ \bibinfo {pages} {014003} (\bibinfo {year} {2020})},\ \Eprint
  {https://arxiv.org/abs/2003.10611} {arXiv:2003.10611 [quant-ph]} \BibitemShut
  {NoStop}%
\bibitem [{\citenamefont {Shende}\ \emph {et~al.}(2006)\citenamefont {Shende},
  \citenamefont {Bullock},\ and\ \citenamefont {Markov}}]{Shende:2006}%
  \BibitemOpen
  \bibfield  {author} {\bibinfo {author} {\bibfnamefont {V.}~\bibnamefont
  {Shende}}, \bibinfo {author} {\bibfnamefont {S.}~\bibnamefont {Bullock}},\
  and\ \bibinfo {author} {\bibfnamefont {I.}~\bibnamefont {Markov}},\
  }\bibfield  {title} {\bibinfo {title} {Synthesis of quantum-logic circuits},\
  }\href {https://doi.org/10.1109/TCAD.2005.855930} {\bibfield  {journal}
  {\bibinfo  {journal} {IEEE Trans. on Computer-Aided Design}\ }\textbf
  {\bibinfo {volume} {25}},\ \bibinfo {pages} {1000} (\bibinfo {year}
  {2006})},\ \Eprint {https://arxiv.org/abs/quant-ph/0406176}
  {arXiv:quant-ph/0406176} \BibitemShut {NoStop}%
\bibitem [{\citenamefont {Mottonen}\ and\ \citenamefont
  {Vartiainen}(2006)}]{Mottonen:2006}%
  \BibitemOpen
  \bibfield  {author} {\bibinfo {author} {\bibfnamefont {M.}~\bibnamefont
  {Mottonen}}\ and\ \bibinfo {author} {\bibfnamefont {J.~J.}\ \bibnamefont
  {Vartiainen}},\ }\bibfield  {title} {\bibinfo {title} {Decompositions of
  general quantum gates},\ }in\ \href@noop {} {\emph {\bibinfo {booktitle}
  {{Trends in Quantum Computing Research}}}},\ \bibinfo {editor} {edited by\
  \bibinfo {editor} {\bibfnamefont {S.}~\bibnamefont {Shannon}}}\ (\bibinfo
  {publisher} {{NOVA Science Publishers, Inc.}},\ \bibinfo {address} {USA},\
  \bibinfo {year} {2006})\ Chap.~\bibinfo {chapter} {7},\ \Eprint
  {https://arxiv.org/abs/quant-ph/0504100} {arXiv:quant-ph/0504100}
  \BibitemShut {NoStop}%
\bibitem [{\citenamefont {Mansky}\ \emph {et~al.}(2023)\citenamefont {Mansky},
  \citenamefont {Castillo}, \citenamefont {Puigvert},\ and\ \citenamefont
  {Linnhoff-Popien}}]{Mansky:2022bai}%
  \BibitemOpen
  \bibfield  {author} {\bibinfo {author} {\bibfnamefont {M.~B.}\ \bibnamefont
  {Mansky}}, \bibinfo {author} {\bibfnamefont {S.~L.~n.}\ \bibnamefont
  {Castillo}}, \bibinfo {author} {\bibfnamefont {V.~R.}\ \bibnamefont
  {Puigvert}},\ and\ \bibinfo {author} {\bibfnamefont {C.}~\bibnamefont
  {Linnhoff-Popien}},\ }\bibfield  {title} {\bibinfo {title} {{Near-optimal
  quantum circuit construction via Cartan decomposition}},\ }\href
  {https://doi.org/10.1103/PhysRevA.108.052607} {\bibfield  {journal} {\bibinfo
   {journal} {Phys. Rev. A}\ }\textbf {\bibinfo {volume} {108}},\ \bibinfo
  {pages} {052607} (\bibinfo {year} {2023})},\ \Eprint
  {https://arxiv.org/abs/2212.12934} {arXiv:2212.12934 [quant-ph]} \BibitemShut
  {NoStop}%
\bibitem [{\citenamefont {Krol}\ and\ \citenamefont
  {Al-Ars}(2024)}]{Krol:2024taf}%
  \BibitemOpen
  \bibfield  {author} {\bibinfo {author} {\bibfnamefont {A.~M.}\ \bibnamefont
  {Krol}}\ and\ \bibinfo {author} {\bibfnamefont {Z.}~\bibnamefont {Al-Ars}},\
  }\href@noop {} {\bibinfo {title} {{Beyond Quantum Shannon: Circuit
  Construction for General n-Qubit Gates Based on Block ZXZ-Decomposition}}}
  (\bibinfo {year} {2024}),\ \Eprint {https://arxiv.org/abs/2403.13692}
  {arXiv:2403.13692 [quant-ph]} \BibitemShut {NoStop}%
\bibitem [{\citenamefont {Preskill}(2018)}]{Preskill:2018jim}%
  \BibitemOpen
  \bibfield  {author} {\bibinfo {author} {\bibfnamefont {J.}~\bibnamefont
  {Preskill}},\ }\bibfield  {title} {\bibinfo {title} {{Quantum Computing in
  the NISQ era and beyond}},\ }\href {https://doi.org/10.22331/q-2018-08-06-79}
  {\bibfield  {journal} {\bibinfo  {journal} {Quantum}\ }\textbf {\bibinfo
  {volume} {2}},\ \bibinfo {pages} {79} (\bibinfo {year} {2018})},\ \Eprint
  {https://arxiv.org/abs/1801.00862} {arXiv:1801.00862 [quant-ph]} \BibitemShut
  {NoStop}%
\bibitem [{\citenamefont {Li}\ and\ \citenamefont
  {Benjamin}(2017)}]{Li:2016vmf}%
  \BibitemOpen
  \bibfield  {author} {\bibinfo {author} {\bibfnamefont {Y.}~\bibnamefont
  {Li}}\ and\ \bibinfo {author} {\bibfnamefont {S.~C.}\ \bibnamefont
  {Benjamin}},\ }\bibfield  {title} {\bibinfo {title} {{Efficient Variational
  Quantum Simulator Incorporating Active Error Minimization}},\ }\href
  {https://doi.org/10.1103/physrevx.7.021050} {\bibfield  {journal} {\bibinfo
  {journal} {Phys. Rev. X}\ }\textbf {\bibinfo {volume} {7}},\ \bibinfo {pages}
  {021050} (\bibinfo {year} {2017})},\ \Eprint
  {https://arxiv.org/abs/1611.09301} {arXiv:1611.09301 [quant-ph]} \BibitemShut
  {NoStop}%
\bibitem [{\citenamefont {Temme}\ \emph {et~al.}(2017)\citenamefont {Temme},
  \citenamefont {Bravyi},\ and\ \citenamefont {Gambetta}}]{Temme:2016vkz}%
  \BibitemOpen
  \bibfield  {author} {\bibinfo {author} {\bibfnamefont {K.}~\bibnamefont
  {Temme}}, \bibinfo {author} {\bibfnamefont {S.}~\bibnamefont {Bravyi}},\ and\
  \bibinfo {author} {\bibfnamefont {J.~M.}\ \bibnamefont {Gambetta}},\
  }\bibfield  {title} {\bibinfo {title} {{Error Mitigation for Short-Depth
  Quantum Circuits}},\ }\href {https://doi.org/10.1103/physrevlett.119.180509}
  {\bibfield  {journal} {\bibinfo  {journal} {Phys. Rev. Lett.}\ }\textbf
  {\bibinfo {volume} {119}},\ \bibinfo {pages} {180509} (\bibinfo {year}
  {2017})},\ \Eprint {https://arxiv.org/abs/1612.02058} {arXiv:1612.02058
  [quant-ph]} \BibitemShut {NoStop}%
\bibitem [{\citenamefont {Giurgica-Tiron}\ \emph {et~al.}(2020)\citenamefont
  {Giurgica-Tiron}, \citenamefont {Hindy}, \citenamefont {LaRose},
  \citenamefont {Mari},\ and\ \citenamefont {Zeng}}]{2020arXiv200510921G}%
  \BibitemOpen
  \bibfield  {author} {\bibinfo {author} {\bibfnamefont {T.}~\bibnamefont
  {Giurgica-Tiron}}, \bibinfo {author} {\bibfnamefont {Y.}~\bibnamefont
  {Hindy}}, \bibinfo {author} {\bibfnamefont {R.}~\bibnamefont {LaRose}},
  \bibinfo {author} {\bibfnamefont {A.}~\bibnamefont {Mari}},\ and\ \bibinfo
  {author} {\bibfnamefont {W.~J.}\ \bibnamefont {Zeng}},\ }\bibfield  {title}
  {\bibinfo {title} {Digital zero noise extrapolation for quantum error
  mitigation},\ }in\ \href {https://doi.org/10.1109/QCE49297.2020.00045} {\emph
  {\bibinfo {booktitle} {2020 IEEE International Conference on Quantum
  Computing and Engineering (QCE)}}}\ (\bibinfo {year} {2020})\ pp.\ \bibinfo
  {pages} {306--316},\ \Eprint {https://arxiv.org/abs/2005.10921}
  {arXiv:2005.10921 [quant-ph]} \BibitemShut {NoStop}%
\bibitem [{\citenamefont {van~den Berg}\ \emph {et~al.}(2023)\citenamefont
  {van~den Berg}, \citenamefont {Minev}, \citenamefont {Kandala},\ and\
  \citenamefont {Temme}}]{Berg:2022ugn}%
  \BibitemOpen
  \bibfield  {author} {\bibinfo {author} {\bibfnamefont {E.}~\bibnamefont
  {van~den Berg}}, \bibinfo {author} {\bibfnamefont {Z.~K.}\ \bibnamefont
  {Minev}}, \bibinfo {author} {\bibfnamefont {A.}~\bibnamefont {Kandala}},\
  and\ \bibinfo {author} {\bibfnamefont {K.}~\bibnamefont {Temme}},\ }\bibfield
   {title} {\bibinfo {title} {{Probabilistic error cancellation with sparse
  Pauli\textendash{}Lindblad models on noisy quantum processors}},\ }\href
  {https://doi.org/10.1038/s41567-023-02042-2} {\bibfield  {journal} {\bibinfo
  {journal} {Nat. Phys.}\ }\textbf {\bibinfo {volume} {19}},\ \bibinfo {pages}
  {1116} (\bibinfo {year} {2023})},\ \Eprint {https://arxiv.org/abs/2201.09866}
  {arXiv:2201.09866 [quant-ph]} \BibitemShut {NoStop}%
\bibitem [{\citenamefont {Urbanek}\ \emph {et~al.}(2021)\citenamefont
  {Urbanek}, \citenamefont {Nachman}, \citenamefont {Pascuzzi}, \citenamefont
  {He}, \citenamefont {Bauer},\ and\ \citenamefont
  {de~Jong}}]{Urbanek:2021oej}%
  \BibitemOpen
  \bibfield  {author} {\bibinfo {author} {\bibfnamefont {M.}~\bibnamefont
  {Urbanek}}, \bibinfo {author} {\bibfnamefont {B.}~\bibnamefont {Nachman}},
  \bibinfo {author} {\bibfnamefont {V.~R.}\ \bibnamefont {Pascuzzi}}, \bibinfo
  {author} {\bibfnamefont {A.}~\bibnamefont {He}}, \bibinfo {author}
  {\bibfnamefont {C.~W.}\ \bibnamefont {Bauer}},\ and\ \bibinfo {author}
  {\bibfnamefont {W.~A.}\ \bibnamefont {de~Jong}},\ }\bibfield  {title}
  {\bibinfo {title} {Mitigating depolarizing noise on quantum computers with
  noise-estimation circuits},\ }\href
  {https://doi.org/10.1103/PhysRevLett.127.270502} {\bibfield  {journal}
  {\bibinfo  {journal} {Phys. Rev. Lett.}\ }\textbf {\bibinfo {volume} {127}},\
  \bibinfo {pages} {270502} (\bibinfo {year} {2021})},\ \Eprint
  {https://arxiv.org/abs/2103.08591} {arXiv:2103.08591 [quant-ph]} \BibitemShut
  {NoStop}%
\bibitem [{\citenamefont {A~Rahman}\ \emph {et~al.}(2022)\citenamefont
  {A~Rahman}, \citenamefont {Lewis}, \citenamefont {Mendicelli},\ and\
  \citenamefont {Powell}}]{ARahman:2022tkr}%
  \BibitemOpen
  \bibfield  {author} {\bibinfo {author} {\bibfnamefont {S.}~\bibnamefont
  {A~Rahman}}, \bibinfo {author} {\bibfnamefont {R.}~\bibnamefont {Lewis}},
  \bibinfo {author} {\bibfnamefont {E.}~\bibnamefont {Mendicelli}},\ and\
  \bibinfo {author} {\bibfnamefont {S.}~\bibnamefont {Powell}},\ }\bibfield
  {title} {\bibinfo {title} {Self-mitigating trotter circuits for su(2) lattice
  gauge theory on a quantum computer},\ }\href
  {https://doi.org/10.1103/PhysRevD.106.074502} {\bibfield  {journal} {\bibinfo
   {journal} {Phys. Rev. D}\ }\textbf {\bibinfo {volume} {106}},\ \bibinfo
  {pages} {074502} (\bibinfo {year} {2022})}\BibitemShut {NoStop}%
\bibitem [{\citenamefont {Farrell}\ \emph {et~al.}(2023)\citenamefont
  {Farrell}, \citenamefont {Chernyshev}, \citenamefont {Powell}, \citenamefont
  {Zemlevskiy}, \citenamefont {Illa},\ and\ \citenamefont
  {Savage}}]{Farrell:2022wyt}%
  \BibitemOpen
  \bibfield  {author} {\bibinfo {author} {\bibfnamefont {R.~C.}\ \bibnamefont
  {Farrell}}, \bibinfo {author} {\bibfnamefont {I.~A.}\ \bibnamefont
  {Chernyshev}}, \bibinfo {author} {\bibfnamefont {S.~J.~M.}\ \bibnamefont
  {Powell}}, \bibinfo {author} {\bibfnamefont {N.~A.}\ \bibnamefont
  {Zemlevskiy}}, \bibinfo {author} {\bibfnamefont {M.}~\bibnamefont {Illa}},\
  and\ \bibinfo {author} {\bibfnamefont {M.~J.}\ \bibnamefont {Savage}},\
  }\bibfield  {title} {\bibinfo {title} {{Preparations for quantum simulations
  of quantum chromodynamics in 1+1 dimensions. I. Axial gauge}},\ }\href
  {https://doi.org/10.1103/PhysRevD.107.054512} {\bibfield  {journal} {\bibinfo
   {journal} {Phys. Rev. D}\ }\textbf {\bibinfo {volume} {107}},\ \bibinfo
  {pages} {054512} (\bibinfo {year} {2023})},\ \Eprint
  {https://arxiv.org/abs/2207.01731} {arXiv:2207.01731 [quant-ph]} \BibitemShut
  {NoStop}%
\bibitem [{\citenamefont {Ciavarella}(2023)}]{Ciavarella:2023mfc}%
  \BibitemOpen
  \bibfield  {author} {\bibinfo {author} {\bibfnamefont {A.~N.}\ \bibnamefont
  {Ciavarella}},\ }\bibfield  {title} {\bibinfo {title} {{Quantum simulation of
  lattice QCD with improved Hamiltonians}},\ }\href
  {https://doi.org/10.1103/PhysRevD.108.094513} {\bibfield  {journal} {\bibinfo
   {journal} {Phys. Rev. D}\ }\textbf {\bibinfo {volume} {108}},\ \bibinfo
  {pages} {094513} (\bibinfo {year} {2023})},\ \Eprint
  {https://arxiv.org/abs/2307.05593} {arXiv:2307.05593 [hep-lat]} \BibitemShut
  {NoStop}%
\bibitem [{\citenamefont {Farrell}\ \emph
  {et~al.}(2024{\natexlab{a}})\citenamefont {Farrell}, \citenamefont {Illa},
  \citenamefont {Ciavarella},\ and\ \citenamefont {Savage}}]{Farrell:2023fgd}%
  \BibitemOpen
  \bibfield  {author} {\bibinfo {author} {\bibfnamefont {R.~C.}\ \bibnamefont
  {Farrell}}, \bibinfo {author} {\bibfnamefont {M.}~\bibnamefont {Illa}},
  \bibinfo {author} {\bibfnamefont {A.~N.}\ \bibnamefont {Ciavarella}},\ and\
  \bibinfo {author} {\bibfnamefont {M.~J.}\ \bibnamefont {Savage}},\ }\bibfield
   {title} {\bibinfo {title} {{Scalable Circuits for Preparing Ground States on
  Digital Quantum Computers: The Schwinger Model Vacuum on 100 Qubits}},\
  }\href {https://doi.org/10.1103/PRXQuantum.5.020315} {\bibfield  {journal}
  {\bibinfo  {journal} {PRX Quantum}\ }\textbf {\bibinfo {volume} {5}},\
  \bibinfo {pages} {020315} (\bibinfo {year} {2024}{\natexlab{a}})},\ \Eprint
  {https://arxiv.org/abs/2308.04481} {arXiv:2308.04481 [quant-ph]} \BibitemShut
  {NoStop}%
\bibitem [{\citenamefont {Farrell}\ \emph
  {et~al.}(2024{\natexlab{b}})\citenamefont {Farrell}, \citenamefont {Illa},
  \citenamefont {Ciavarella},\ and\ \citenamefont {Savage}}]{Farrell:2024fit}%
  \BibitemOpen
  \bibfield  {author} {\bibinfo {author} {\bibfnamefont {R.~C.}\ \bibnamefont
  {Farrell}}, \bibinfo {author} {\bibfnamefont {M.}~\bibnamefont {Illa}},
  \bibinfo {author} {\bibfnamefont {A.~N.}\ \bibnamefont {Ciavarella}},\ and\
  \bibinfo {author} {\bibfnamefont {M.~J.}\ \bibnamefont {Savage}},\ }\bibfield
   {title} {\bibinfo {title} {{Quantum simulations of hadron dynamics in the
  Schwinger model using 112 qubits}},\ }\href
  {https://doi.org/10.1103/PhysRevD.109.114510} {\bibfield  {journal} {\bibinfo
   {journal} {Phys. Rev. D}\ }\textbf {\bibinfo {volume} {109}},\ \bibinfo
  {pages} {114510} (\bibinfo {year} {2024}{\natexlab{b}})},\ \Eprint
  {https://arxiv.org/abs/2401.08044} {arXiv:2401.08044 [quant-ph]} \BibitemShut
  {NoStop}%
\bibitem [{\citenamefont {Ciavarella}\ and\ \citenamefont
  {Bauer}(2024)}]{Ciavarella:2024fzw}%
  \BibitemOpen
  \bibfield  {author} {\bibinfo {author} {\bibfnamefont {A.~N.}\ \bibnamefont
  {Ciavarella}}\ and\ \bibinfo {author} {\bibfnamefont {C.~W.}\ \bibnamefont
  {Bauer}},\ }\href@noop {} {\bibinfo {title} {{Quantum Simulation of SU(3)
  Lattice Yang Mills Theory at Leading Order in Large N}}} (\bibinfo {year}
  {2024}),\ \Eprint {https://arxiv.org/abs/2402.10265} {arXiv:2402.10265
  [hep-ph]} \BibitemShut {NoStop}%
\bibitem [{\citenamefont {Nguyen}\ \emph {et~al.}(2022)\citenamefont {Nguyen},
  \citenamefont {Tran}, \citenamefont {Zhu}, \citenamefont {Green},
  \citenamefont {Alderete}, \citenamefont {Davoudi},\ and\ \citenamefont
  {Linke}}]{Nguyen:2021hyk}%
  \BibitemOpen
  \bibfield  {author} {\bibinfo {author} {\bibfnamefont {N.~H.}\ \bibnamefont
  {Nguyen}}, \bibinfo {author} {\bibfnamefont {M.~C.}\ \bibnamefont {Tran}},
  \bibinfo {author} {\bibfnamefont {Y.}~\bibnamefont {Zhu}}, \bibinfo {author}
  {\bibfnamefont {A.~M.}\ \bibnamefont {Green}}, \bibinfo {author}
  {\bibfnamefont {C.~H.}\ \bibnamefont {Alderete}}, \bibinfo {author}
  {\bibfnamefont {Z.}~\bibnamefont {Davoudi}},\ and\ \bibinfo {author}
  {\bibfnamefont {N.~M.}\ \bibnamefont {Linke}},\ }\bibfield  {title} {\bibinfo
  {title} {{Digital Quantum Simulation of the Schwinger Model and Symmetry
  Protection with Trapped Ions}},\ }\href
  {https://doi.org/10.1103/PRXQuantum.3.020324} {\bibfield  {journal} {\bibinfo
   {journal} {PRX Quantum}\ }\textbf {\bibinfo {volume} {3}},\ \bibinfo {pages}
  {020324} (\bibinfo {year} {2022})},\ \Eprint
  {https://arxiv.org/abs/2112.14262} {arXiv:2112.14262 [quant-ph]} \BibitemShut
  {NoStop}%
\bibitem [{\citenamefont {Wallman}\ and\ \citenamefont
  {Emerson}(2016)}]{Wallman:2015uzh}%
  \BibitemOpen
  \bibfield  {author} {\bibinfo {author} {\bibfnamefont {J.~J.}\ \bibnamefont
  {Wallman}}\ and\ \bibinfo {author} {\bibfnamefont {J.}~\bibnamefont
  {Emerson}},\ }\bibfield  {title} {\bibinfo {title} {{Noise tailoring for
  scalable quantum computation via randomized compiling}},\ }\href
  {https://doi.org/10.1103/PhysRevA.94.052325} {\bibfield  {journal} {\bibinfo
  {journal} {Phys. Rev. A}\ }\textbf {\bibinfo {volume} {94}},\ \bibinfo
  {pages} {052325} (\bibinfo {year} {2016})},\ \Eprint
  {https://arxiv.org/abs/1512.01098} {arXiv:1512.01098 [quant-ph]} \BibitemShut
  {NoStop}%
\bibitem [{\citenamefont {Hashim}\ \emph {et~al.}(2021)\citenamefont {Hashim}
  \emph {et~al.}}]{Hashim:2020cop}%
  \BibitemOpen
  \bibfield  {author} {\bibinfo {author} {\bibfnamefont {A.}~\bibnamefont
  {Hashim}} \emph {et~al.},\ }\bibfield  {title} {\bibinfo {title} {{Randomized
  Compiling for Scalable Quantum Computing on a Noisy Superconducting Quantum
  Processor}},\ }\href {https://doi.org/10.1103/PhysRevX.11.041039} {\bibfield
  {journal} {\bibinfo  {journal} {Phys. Rev. X}\ }\textbf {\bibinfo {volume}
  {11}},\ \bibinfo {pages} {041039} (\bibinfo {year} {2021})},\ \Eprint
  {https://arxiv.org/abs/2010.00215} {arXiv:2010.00215 [quant-ph]} \BibitemShut
  {NoStop}%
\bibitem [{\citenamefont {Viola}\ and\ \citenamefont
  {Lloyd}(1998)}]{Viola:1998dsd}%
  \BibitemOpen
  \bibfield  {author} {\bibinfo {author} {\bibfnamefont {L.}~\bibnamefont
  {Viola}}\ and\ \bibinfo {author} {\bibfnamefont {S.}~\bibnamefont {Lloyd}},\
  }\bibfield  {title} {\bibinfo {title} {Dynamical suppression of decoherence
  in two-state quantum systems},\ }\href
  {https://doi.org/10.1103/PhysRevA.58.2733} {\bibfield  {journal} {\bibinfo
  {journal} {Phys. Rev. A}\ }\textbf {\bibinfo {volume} {58}},\ \bibinfo
  {pages} {2733} (\bibinfo {year} {1998})},\ \Eprint
  {https://arxiv.org/abs/quant-ph/9803057} {arXiv:quant-ph/9803057}
  \BibitemShut {NoStop}%
\bibitem [{\citenamefont {{Souza}}\ \emph {et~al.}(2012)\citenamefont
  {{Souza}}, \citenamefont {{Álvarez}},\ and\ \citenamefont
  {{Suter}}}]{2012RSPTA.370.4748S}%
  \BibitemOpen
  \bibfield  {author} {\bibinfo {author} {\bibfnamefont {A.~M.}\ \bibnamefont
  {{Souza}}}, \bibinfo {author} {\bibfnamefont {G.~A.}\ \bibnamefont
  {{Álvarez}}},\ and\ \bibinfo {author} {\bibfnamefont {D.}~\bibnamefont
  {{Suter}}},\ }\bibfield  {title} {\bibinfo {title} {{Robust dynamical
  decoupling}},\ }\href {https://doi.org/10.1098/rsta.2011.0355} {\bibfield
  {journal} {\bibinfo  {journal} {Phil. Trans. R. Soc.}\ }\textbf {\bibinfo
  {volume} {370}},\ \bibinfo {pages} {4748} (\bibinfo {year} {2012})},\ \Eprint
  {https://arxiv.org/abs/1110.6334} {arXiv:1110.6334 [quant-ph]} \BibitemShut
  {NoStop}%
\bibitem [{\citenamefont {Ezzell}\ \emph {et~al.}(2023)\citenamefont {Ezzell},
  \citenamefont {Pokharel}, \citenamefont {Tewala}, \citenamefont {Quiroz},\
  and\ \citenamefont {Lidar}}]{Ezzell:2022uat}%
  \BibitemOpen
  \bibfield  {author} {\bibinfo {author} {\bibfnamefont {N.}~\bibnamefont
  {Ezzell}}, \bibinfo {author} {\bibfnamefont {B.}~\bibnamefont {Pokharel}},
  \bibinfo {author} {\bibfnamefont {L.}~\bibnamefont {Tewala}}, \bibinfo
  {author} {\bibfnamefont {G.}~\bibnamefont {Quiroz}},\ and\ \bibinfo {author}
  {\bibfnamefont {D.~A.}\ \bibnamefont {Lidar}},\ }\bibfield  {title} {\bibinfo
  {title} {{Dynamical decoupling for superconducting qubits: A performance
  survey}},\ }\href {https://doi.org/10.1103/PhysRevApplied.20.064027}
  {\bibfield  {journal} {\bibinfo  {journal} {Phys. Rev. Applied}\ }\textbf
  {\bibinfo {volume} {20}},\ \bibinfo {pages} {064027} (\bibinfo {year}
  {2023})},\ \Eprint {https://arxiv.org/abs/2207.03670} {arXiv:2207.03670
  [quant-ph]} \BibitemShut {NoStop}%
\bibitem [{\citenamefont {Nation}\ \emph {et~al.}(2021)\citenamefont {Nation},
  \citenamefont {Kang}, \citenamefont {Sundaresan},\ and\ \citenamefont
  {Gambetta}}]{Nation:2021kye}%
  \BibitemOpen
  \bibfield  {author} {\bibinfo {author} {\bibfnamefont {P.~D.}\ \bibnamefont
  {Nation}}, \bibinfo {author} {\bibfnamefont {H.}~\bibnamefont {Kang}},
  \bibinfo {author} {\bibfnamefont {N.}~\bibnamefont {Sundaresan}},\ and\
  \bibinfo {author} {\bibfnamefont {J.~M.}\ \bibnamefont {Gambetta}},\
  }\bibfield  {title} {\bibinfo {title} {{Scalable Mitigation of Measurement
  Errors on Quantum Computers}},\ }\href
  {https://doi.org/10.1103/PRXQuantum.2.040326} {\bibfield  {journal} {\bibinfo
   {journal} {PRX Quantum}\ }\textbf {\bibinfo {volume} {2}},\ \bibinfo {pages}
  {040326} (\bibinfo {year} {2021})},\ \Eprint
  {https://arxiv.org/abs/2108.12518} {arXiv:2108.12518 [quant-ph]} \BibitemShut
  {NoStop}%
\bibitem [{\citenamefont {Klco}\ and\ \citenamefont
  {Savage}(2020)}]{Klco:2019xro}%
  \BibitemOpen
  \bibfield  {author} {\bibinfo {author} {\bibfnamefont {N.}~\bibnamefont
  {Klco}}\ and\ \bibinfo {author} {\bibfnamefont {M.~J.}\ \bibnamefont
  {Savage}},\ }\bibfield  {title} {\bibinfo {title} {{Minimally entangled state
  preparation of localized wave functions on quantum computers}},\ }\href
  {https://doi.org/10.1103/PhysRevA.102.012612} {\bibfield  {journal} {\bibinfo
   {journal} {Phys. Rev. A}\ }\textbf {\bibinfo {volume} {102}},\ \bibinfo
  {pages} {012612} (\bibinfo {year} {2020})},\ \Eprint
  {https://arxiv.org/abs/1904.10440} {arXiv:1904.10440 [quant-ph]} \BibitemShut
  {NoStop}%
\bibitem [{\citenamefont {Smolin}\ \emph {et~al.}(2012)\citenamefont {Smolin},
  \citenamefont {Gambetta},\ and\ \citenamefont
  {Smith}}]{PhysRevLett.108.070502}%
  \BibitemOpen
  \bibfield  {author} {\bibinfo {author} {\bibfnamefont {J.~A.}\ \bibnamefont
  {Smolin}}, \bibinfo {author} {\bibfnamefont {J.~M.}\ \bibnamefont
  {Gambetta}},\ and\ \bibinfo {author} {\bibfnamefont {G.}~\bibnamefont
  {Smith}},\ }\bibfield  {title} {\bibinfo {title} {Efficient method for
  computing the maximum-likelihood quantum state from measurements with
  additive gaussian noise},\ }\href
  {https://doi.org/10.1103/PhysRevLett.108.070502} {\bibfield  {journal}
  {\bibinfo  {journal} {Phys. Rev. Lett.}\ }\textbf {\bibinfo {volume} {108}},\
  \bibinfo {pages} {070502} (\bibinfo {year} {2012})},\ \Eprint
  {https://arxiv.org/abs/1106.5458} {arXiv:1106.5458 [quant-ph]} \BibitemShut
  {NoStop}%
\bibitem [{\citenamefont {Huang}\ \emph {et~al.}(2020)\citenamefont {Huang},
  \citenamefont {Kueng},\ and\ \citenamefont {Preskill}}]{Huang:2020tih}%
  \BibitemOpen
  \bibfield  {author} {\bibinfo {author} {\bibfnamefont {H.-Y.}\ \bibnamefont
  {Huang}}, \bibinfo {author} {\bibfnamefont {R.}~\bibnamefont {Kueng}},\ and\
  \bibinfo {author} {\bibfnamefont {J.}~\bibnamefont {Preskill}},\ }\bibfield
  {title} {\bibinfo {title} {{Predicting many properties of a quantum system
  from very few measurements}},\ }\href
  {https://doi.org/10.1038/s41567-020-0932-7} {\bibfield  {journal} {\bibinfo
  {journal} {Nature Phys.}\ }\textbf {\bibinfo {volume} {16}},\ \bibinfo
  {pages} {1050} (\bibinfo {year} {2020})},\ \Eprint
  {https://arxiv.org/abs/2002.08953} {arXiv:2002.08953 [quant-ph]} \BibitemShut
  {NoStop}%
\bibitem [{\citenamefont {Acharya}\ \emph {et~al.}(2021)\citenamefont
  {Acharya}, \citenamefont {Saha},\ and\ \citenamefont
  {Sengupta}}]{Acharya:2021byb}%
  \BibitemOpen
  \bibfield  {author} {\bibinfo {author} {\bibfnamefont {A.}~\bibnamefont
  {Acharya}}, \bibinfo {author} {\bibfnamefont {S.}~\bibnamefont {Saha}},\ and\
  \bibinfo {author} {\bibfnamefont {A.~M.}\ \bibnamefont {Sengupta}},\
  }\bibfield  {title} {\bibinfo {title} {Shadow tomography based on
  informationally complete positive operator-valued measure},\ }\href
  {https://doi.org/10.1103/PhysRevA.104.052418} {\bibfield  {journal} {\bibinfo
   {journal} {Phys. Rev. A}\ }\textbf {\bibinfo {volume} {104}},\ \bibinfo
  {pages} {052418} (\bibinfo {year} {2021})},\ \Eprint
  {https://arxiv.org/abs/2105.05992} {arXiv:2105.05992 [quant-ph]} \BibitemShut
  {NoStop}%
\bibitem [{\citenamefont {Motta}\ \emph {et~al.}(2019)\citenamefont {Motta},
  \citenamefont {Sun}, \citenamefont {Tan}, \citenamefont {Rourke},
  \citenamefont {Ye}, \citenamefont {Minnich}, \citenamefont {Brand\~ao},\ and\
  \citenamefont {Chan}}]{Motta:2019yya}%
  \BibitemOpen
  \bibfield  {author} {\bibinfo {author} {\bibfnamefont {M.}~\bibnamefont
  {Motta}}, \bibinfo {author} {\bibfnamefont {C.}~\bibnamefont {Sun}}, \bibinfo
  {author} {\bibfnamefont {A.~T.~K.}\ \bibnamefont {Tan}}, \bibinfo {author}
  {\bibfnamefont {M.~J.~O.}\ \bibnamefont {Rourke}}, \bibinfo {author}
  {\bibfnamefont {E.}~\bibnamefont {Ye}}, \bibinfo {author} {\bibfnamefont
  {A.~J.}\ \bibnamefont {Minnich}}, \bibinfo {author} {\bibfnamefont {F.~G.
  S.~L.}\ \bibnamefont {Brand\~ao}},\ and\ \bibinfo {author} {\bibfnamefont
  {G.~K.-L.}\ \bibnamefont {Chan}},\ }\bibfield  {title} {\bibinfo {title}
  {{Determining eigenstates and thermal states on a quantum computer using
  quantum imaginary time evolution}},\ }\href
  {https://doi.org/10.1038/s41567-019-0704-4} {\bibfield  {journal} {\bibinfo
  {journal} {Nature Phys.}\ }\textbf {\bibinfo {volume} {16}},\ \bibinfo
  {pages} {205} (\bibinfo {year} {2019})},\ \Eprint
  {https://arxiv.org/abs/1901.07653} {arXiv:1901.07653 [quant-ph]} \BibitemShut
  {NoStop}%
\bibitem [{\citenamefont {{Sagastizabal}}\ \emph {et~al.}(2021)\citenamefont
  {{Sagastizabal}}, \citenamefont {{Premaratne}}, \citenamefont {{Klaver}},
  \citenamefont {{Rol}}, \citenamefont {{Neg{\^\i}rneac}}, \citenamefont
  {{Moreira}}, \citenamefont {{Zou}}, \citenamefont {{Johri}}, \citenamefont
  {{Muthusubramanian}}, \citenamefont {{Beekman}}, \citenamefont
  {{Zachariadis}}, \citenamefont {{Ostroukh}}, \citenamefont {{Haider}},
  \citenamefont {{Bruno}}, \citenamefont {{Matsuura}},\ and\ \citenamefont
  {{DiCarlo}}}]{sagastizabal2021variational}%
  \BibitemOpen
  \bibfield  {author} {\bibinfo {author} {\bibfnamefont {R.}~\bibnamefont
  {{Sagastizabal}}}, \bibinfo {author} {\bibfnamefont {S.~P.}\ \bibnamefont
  {{Premaratne}}}, \bibinfo {author} {\bibfnamefont {B.~A.}\ \bibnamefont
  {{Klaver}}}, \bibinfo {author} {\bibfnamefont {M.~A.}\ \bibnamefont {{Rol}}},
  \bibinfo {author} {\bibfnamefont {V.}~\bibnamefont {{Neg{\^\i}rneac}}},
  \bibinfo {author} {\bibfnamefont {M.~S.}\ \bibnamefont {{Moreira}}}, \bibinfo
  {author} {\bibfnamefont {X.}~\bibnamefont {{Zou}}}, \bibinfo {author}
  {\bibfnamefont {S.}~\bibnamefont {{Johri}}}, \bibinfo {author} {\bibfnamefont
  {N.}~\bibnamefont {{Muthusubramanian}}}, \bibinfo {author} {\bibfnamefont
  {M.}~\bibnamefont {{Beekman}}}, \bibinfo {author} {\bibfnamefont
  {C.}~\bibnamefont {{Zachariadis}}}, \bibinfo {author} {\bibfnamefont {V.~P.}\
  \bibnamefont {{Ostroukh}}}, \bibinfo {author} {\bibfnamefont
  {N.}~\bibnamefont {{Haider}}}, \bibinfo {author} {\bibfnamefont
  {A.}~\bibnamefont {{Bruno}}}, \bibinfo {author} {\bibfnamefont {A.~Y.}\
  \bibnamefont {{Matsuura}}},\ and\ \bibinfo {author} {\bibfnamefont
  {L.}~\bibnamefont {{DiCarlo}}},\ }\bibfield  {title} {\bibinfo {title}
  {{Variational preparation of finite-temperature states on a quantum
  computer}},\ }\href {https://doi.org/10.1038/s41534-021-00468-1} {\bibfield
  {journal} {\bibinfo  {journal} {npj Quantum Inf.}\ }\textbf {\bibinfo
  {volume} {7}},\ \bibinfo {eid} {130} (\bibinfo {year} {2021})},\ \Eprint
  {https://arxiv.org/abs/2012.03895} {arXiv:2012.03895 [quant-ph]} \BibitemShut
  {NoStop}%
\bibitem [{\citenamefont {Turro}(2023)}]{turro2023quantum}%
  \BibitemOpen
  \bibfield  {author} {\bibinfo {author} {\bibfnamefont {F.}~\bibnamefont
  {Turro}},\ }\href@noop {} {\bibinfo {title} {Quantum imaginary time
  propagation algorithm for preparing thermal states}} (\bibinfo {year}
  {2023}),\ \Eprint {https://arxiv.org/abs/2306.16580} {arXiv:2306.16580
  [quant-ph]} \BibitemShut {NoStop}%
\bibitem [{\citenamefont {Davoudi}\ \emph {et~al.}(2023)\citenamefont
  {Davoudi}, \citenamefont {Mueller},\ and\ \citenamefont
  {Powers}}]{Davoudi:2022uzo}%
  \BibitemOpen
  \bibfield  {author} {\bibinfo {author} {\bibfnamefont {Z.}~\bibnamefont
  {Davoudi}}, \bibinfo {author} {\bibfnamefont {N.}~\bibnamefont {Mueller}},\
  and\ \bibinfo {author} {\bibfnamefont {C.}~\bibnamefont {Powers}},\
  }\bibfield  {title} {\bibinfo {title} {{Towards Quantum Computing Phase
  Diagrams of Gauge Theories with Thermal Pure Quantum States}},\ }\href
  {https://doi.org/10.1103/PhysRevLett.131.081901} {\bibfield  {journal}
  {\bibinfo  {journal} {Phys. Rev. Lett.}\ }\textbf {\bibinfo {volume} {131}},\
  \bibinfo {pages} {081901} (\bibinfo {year} {2023})},\ \Eprint
  {https://arxiv.org/abs/2208.13112} {arXiv:2208.13112 [hep-lat]} \BibitemShut
  {NoStop}%
\bibitem [{\citenamefont {Rajput}\ \emph {et~al.}(2022)\citenamefont {Rajput},
  \citenamefont {Roggero},\ and\ \citenamefont {Wiebe}}]{Rajput:2021khs}%
  \BibitemOpen
  \bibfield  {author} {\bibinfo {author} {\bibfnamefont {A.}~\bibnamefont
  {Rajput}}, \bibinfo {author} {\bibfnamefont {A.}~\bibnamefont {Roggero}},\
  and\ \bibinfo {author} {\bibfnamefont {N.}~\bibnamefont {Wiebe}},\ }\bibfield
   {title} {\bibinfo {title} {{Hybridized Methods for Quantum Simulation in the
  Interaction Picture}},\ }\href {https://doi.org/10.22331/q-2022-08-17-780}
  {\bibfield  {journal} {\bibinfo  {journal} {Quantum}\ }\textbf {\bibinfo
  {volume} {6}},\ \bibinfo {pages} {780} (\bibinfo {year} {2022})},\ \Eprint
  {https://arxiv.org/abs/2109.03308} {arXiv:2109.03308 [quant-ph]} \BibitemShut
  {NoStop}%
\bibitem [{\citenamefont {{Wang}}\ and\ \citenamefont
  {{Carreira-Perpi{\~n}{\'a}n}}(2013)}]{Wang:2013}%
  \BibitemOpen
  \bibfield  {author} {\bibinfo {author} {\bibfnamefont {W.}~\bibnamefont
  {{Wang}}}\ and\ \bibinfo {author} {\bibfnamefont {M.~{\'A}.}\ \bibnamefont
  {{Carreira-Perpi{\~n}{\'a}n}}},\ }\href@noop {} {\bibinfo {title}
  {{Projection onto the probability simplex: An efficient algorithm with a
  simple proof, and an application}}} (\bibinfo {year} {2013}),\ \Eprint
  {https://arxiv.org/abs/1309.1541} {arXiv:1309.1541 [cs.LG]} \BibitemShut
  {NoStop}%
\end{thebibliography}%

\end{document}